%% file: Nf12StepScaling.tex
\documentclass[aps,prd,preprintnumbers,nofootinbib,superscriptaddress,showpacs,letterpaper,twocolumn,floatfix]{revtex4-1}
\pdfoutput=1
\usepackage{amsmath,amssymb,amsbsy,longtable,tabularx}
\usepackage[utf8]{inputenc}
\usepackage{graphicx,color,ulem}
\usepackage[colorlinks=true, linkcolor=blue, filecolor=blue, urlcolor=blue, citecolor=blue, pdftex, plainpages=false]{hyperref}

\newcommand\buaff{Department of Physics and Center for Computational Science, Boston University, Boston, Massachusetts 02215, USA}
\newcommand\coaff{Department of Physics, University of Colorado, Boulder, Colorado 80309, USA}

\newcommand{\be}{\ensuremath{\beta} }

\newcommand{\gc}{\ensuremath{g_c^2} }

\newcommand{\vev}[1]{\ensuremath{\left\langle #1 \right\rangle} }

\newcommand\setrow[1]{\gdef\rowmac{#1}#1\ignorespaces}
\newcommand\clearrow{\global\let\rowmac\relax}
\clearrow

\begin{document}

%%%%%%%%%%%%%%%%%%%%%%%%%%%%%% TITLEPAGE %%%%%%%%%%%%%%%%%%%%%%%%%%
\title{Gradient flow step-scaling function for SU(3) with twelve flavors }
\author{A.~Hasenfratz}\affiliation\coaff
\author{C.~Rebbi}\affiliation\buaff
\author{O.~Witzel}\email{oliver.witzel@colorado.edu}\affiliation\coaff

\date{\today}

%%%%%%%%%%%%%%%%%%%%%%%%%%%%%% ABSTRACT %%%%%%%%%%%%%%%%%%%%%%%%%%%%%%%
\begin{abstract}
We calculate the step scaling function, the lattice analog of the renormalization group $\beta$-function, for an SU(3) gauge theory with twelve flavors. The gauge coupling of this system runs very slowly, which is reflected in a small step scaling function, making numerical simulations particularly challenging. We present a detailed  analysis including the study of systematic effects of our extensive data set generated with twelve dynamical flavors using the Symanzik gauge action and three times stout smeared M\"obius domain wall fermions. Using up to $32^4$ volumes, we calculate renormalized couplings for different gradient flow schemes and determine the step-scaling $\beta$ function for a scale change $s=2$ on up to five different lattice volume pairs. Our preferred analysis is fully $O(a^2)$ Symanzik improved and uses Zeuthen flow combined with the Symanzik operator. We find an infrared fixed point  within the range $5.2 \le g_c^2 \le 6.4$ in the $c=0.250$ finite volume gradient flow scheme. We account for systematic effects by calculating the step-scaling function based on alternative flows (Wilson or Symanzik)  as well as operators (Wilson plaquette, clover) and also explore the effects of the perturbative tree-level improvement. 
\end{abstract}
\maketitle

%=================================================
\section{Introduction}
%=================================================
The renormalization group $\beta$-function characterizes the nature of gauge-fermion systems with gauge group ${\cal G}$ and $N_f$ fermion flavors in representation ${\cal R}$. Choosing e.g.~fermions in the fundamental representation and the SU(3) gauge group, the system is chirally broken with a fast running coupling for a small number of flavors. A particularly well studied case is Quantum Chromodynamics (QCD). This is a chirally broken theory which exhibits only the Gaussian fixed point (GFP) at bare coupling $g_0^2=0$. When increasing the number of flavors, predictions based on perturbation theory suggest that a second, infrared fixed point (IRFP) develops at some definite $g_\text{IRFP}^2$ \cite{Banks:1981nn}. For an even larger number of flavors, the theory becomes IR free. Theories exhibiting an IRFP are conformal. Of special interest is the lowest number of flavors, $N_f^c$, for which a gauge-fermion system (${\cal G}, {\cal R})$ is conformal because this denotes the onset of the conformal window.\footnote{Theories just below the conformal window are promising candidates to describe physics beyond the Standard Model e.g.~composite Higgs scenarios as discussed in Refs.~\cite{DeGrand:2015zxa,Nogradi:2016qek,Witzel:2019jbe} and references within.} Perturbative methods to determine the nature of a system with $N_f$ flavors may not be reliable because the IRFP can occur at large $g_\text{IRFP}^2$ and may lie outside the trustworthy region of perturbation theory. Inside the conformal window, the IRFP is expected to move to stronger couplings as the number flavors approaches $N_f^c$. Thus a perturbative estimate of $N_f^c$ is particularly troublesome and warrants us to use nonperturbative methods \cite{Baikov:2016tgj,Ryttov:2010iz,Ryttov:2016ner,Ryttov:2016hal}. 

In this work we focus on the SU(3) gauge system with twelve flavors in the fundamental representation and present details of our nonperturbative, large-scale investigation using lattice field theory techniques. We evaluate the gradient flow (GF) step-scaling function, the lattice analogue of the renormalization group (RG) $\beta$-function in the infinite volume continuum limit. Perturbation theory up to the 4-loop level predicts that the system is conformal. The 5-loop $\beta$-function, however, does not predict a fixed point \cite{Baikov:2016tgj}. Since the 5-loop correction is very large, Refs.~\cite{Ryttov:2016ner}  suggested to improve the convergence of the perturbative series by using Pad\'e approximation. Various forms of the Pad\'e series  predict that the system is conformal. The scheme-independent form \cite{Ryttov:2016hal} also finds an IRFP. There are several nonperturbative lattice studies of this system based on staggered fermions ~\cite{DeGrand:2011cu,Fodor:2011tu,Fodor:2012et, Aoki:2012eq,Itou:2013ofa,Cheng:2013eu,Cheng:2013xha,Cheng:2014jba,Lin:2015zpa,Hasenfratz:2016dou,Fodor:2016zil,Fodor:2017gtj,Fodor:2017nlp}. The results of lattice studies are controversial as Ref.~\cite{Hasenfratz:2016dou}  finds a FP in the gradient flow $c=0.25$  scheme at $g^2\approx 7.3$,  while  Refs.~\cite{Fodor:2016zil,Fodor:2017gtj,Fodor:2017nlp} do not observe any sign of a FP. The origin of this disagreement  is still to be resolved.  Our calculation is the first based on simulations with domain wall (DW) fermions \cite{Kaplan:1992bt,Shamir:1993zy,Furman:1994ky,Brower:2012vk}. DW fermions are chiral and exhibit continuum-like flavor symmetry, a property which may be crucial to investigate near-conformal or conformal gauge fermion-systems. Furthermore, DW fermions are simulated with a Pauli-Villars term which reduces effective gauge terms generated by the many flavors, resulting in  smoother gauge field configurations and reduced cutoff effects.  

In our calculation we generate gauge field configuration with stout-smeared \cite{Morningstar:2003gk} M\"obius domain wall fermions \cite{Brower:2012vk} and Symanzik gauge action \cite{Luscher:1984xn,Luscher:1985zq} on which we perform gradient flow measurements. Our preferred determination of the step scaling function is based on Zeuthen flow \cite{Ramos:2014kka,Ramos:2015baa} combined with the Symanzik operator to estimate the energy density. Other gradient flows (Wilson,  Symanzik) as well as different operators (Wilson plaquette, clover) are considered, too. Our preferred combination of action, flow, and operator is fully $O(a^2)$ Symanzik improved \cite{Ramos:2014kka,Ramos:2015baa} and discretization effects are further suppressed by the perturbative tree-level normalization introduced in Ref.~\cite{Fodor:2014cpa}. We demonstrate that this combination has indeed small discretization errors which result in mild continuum limit extrapolations. Using GF renormalization schemes $c=0.250$, 0.275, and 0.300 we present continuum-limit results which show that SU(3) with twelve fundamental flavors exhibits an IRFP,  implying that this theory is conformal. The IRFP may exhibit a mild dependence on the renormalization scheme which, however,  is not statistically resolved by our data. In the $c=0.250$ scheme we observe the IRFP in the range of $5.2 \le g_c^2 \le 6.4$.

Early results of this project have been presented in Refs.~\cite{Hasenfratz:2017mdh,Hasenfratz:2017qyr,Hasenfratz:2018wpq,Hasenfratz:2019puu}. The remainder of this paper is organized as follows: after introducing the gradient flow step-scaling function and the perturbative improvement in Sec.~\ref{Sec.GradFlow}, we discuss the details of our numerical simulations in Sec.~\ref{Sec.NumSim}. In Section \ref{Sec.nZS_ZS} we present the results for the step-scaling $\beta$-function using different renormalization schemes $c$ for our preferred combination of Zeuthen flow with Symanzik operator with and without tree-level normalization (short: nZS and ZS).  Alternative determinations using different gradient flows, operators with and without tree-level normalization are presented in Sec.~\ref{Sec.Alt}. Additional details are collected in the appendices: Appendix \ref{Sec.tree-level} lists the tree-level normalization factors for Symanzik gauge action and our three gradient flows with the three different operators, in Appendix \ref{Sec.RenCouplings} we present the renormalized couplings for our preferred (n)ZS analyses and in Appendix \ref{Sec.ContLimitExtra} we show comparison plots including additional continuum extrapolations. Finally we summarize our results and conclude in Sec.~\ref{Sec.Conclusion}.

%=================================================
\section{The continuum limit of the step-scaling function}
\label{Sec.GradFlow}

\begin{figure}[tb]
  \centering
  \includegraphics[width=0.99\columnwidth]{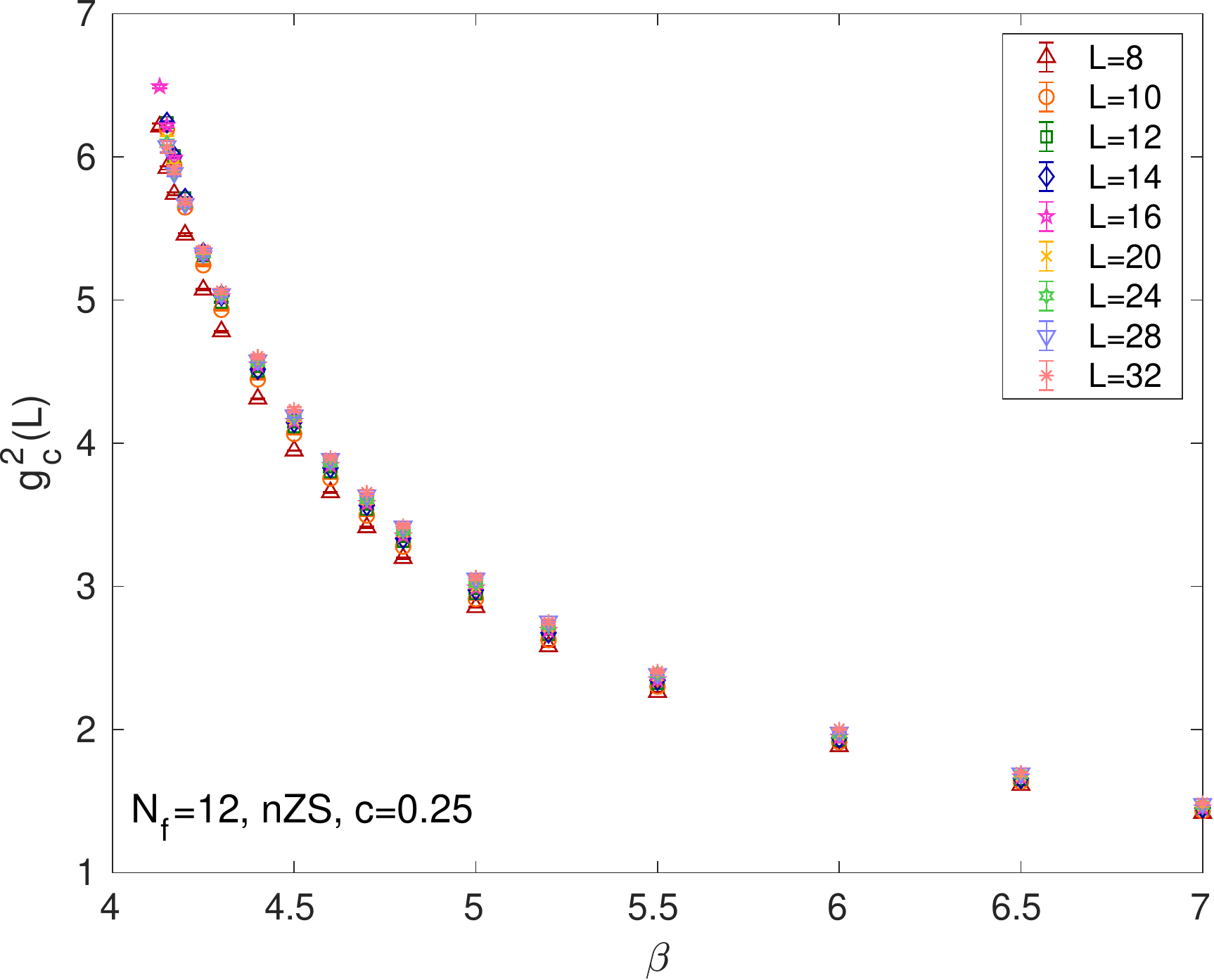}
  \caption{Renormalized couplings $g_c^2$ in the $c=0.250$ scheme  for bare couplings $4.13\le \beta\le 7.00$ on hypercubic $(L/a)^4$ volumes with $L/a$ ranging from 8 to 32. The values shown are obtained using Zeuthen flow with the Symanzik operator and  the tree-level normalization (nZS). Statistical errors are shown but barely visible.}
  \label{Fig.gcSq_nZS_c0.25}
\end{figure}  

The finite volume step scaling function, or discrete $\beta$ function  of scale change  $s$ is defined by
\begin{equation}
  \label{eq:beta}
  \be_{c,s}(\gc; L) = \frac{\gc(sL; a) - \gc(L; a)}{\log(s^2)},
\end{equation}
where $\gc(L;a)$ is a renormalized coupling at the energy scale set by the volume $\mu = (c L )^{-1}$ which is related to the gradient flow time $t$ by $\mu=1/\sqrt{8 t}$. The parameter $c$ denotes the renormalization scheme corresponding to a renormalized  gradient flow coupling given by
\begin{align}
    \label{eq:pert_g2}
    \gc(L;a) = \frac{128\pi^2}{3(N^2 - 1)} \frac{1}{C(c,L/a)} \vev{t^2 E(t)},
\end{align}
with $N=3$ for SU(3), $E(t)$ the energy density at gradient flow time $t$, and $C(c,L/a)$  a perturbatively computed tree-level improvement term\footnote{Numerical values for $C(c,L/a)$ are listed in Table \ref{Tab.tln}.} \cite{Fodor:2014cpa}. Without tree-level improvement $C(c,L/a)$ is replaced by  the term $1/(1+\delta(t/L^2))$ that compensates for the zero modes of the gauge fields in periodic volumes~\cite{Fodor:2012td}. 

The dependence on the lattice spacing $a$ reflects cutoff effects that arise because simulations are not performed with a ``perfect action,'' i.e.~along the renormalized trajectory (RT) emerging from the Gaussian FP.  As the gradient flow time $t/a^2$ increases,  irrelevant operators die out and cutoff effects are reduced.  Thus the continuum limit is approached by increasing $L/a$ while keeping $c$  and the renormalized coupling $g^2_c$ fixed. In asymptotically free theories this forces the bare coupling toward zero (the Gaussian fixed point) or $\beta \equiv 6/g_0^2 \to \infty$. If all but one irrelevant operators are negligible the remaining cutoff effects are proportional to  $(\sqrt{t}/a)^{\alpha}$ or $(L/a)^\alpha$ where $\alpha$ is the scaling exponent of the least irrelevant operator. In the vicinity of the Gaussian FP $\alpha = -2$, thus an extrapolation in $a^2/L^2$ at fixed $g_c^2$ predicts the continuum limit.

In practice simulations are performed at many bare coupling values, ``daisy-chaining'' them to cover the investigated renormalized coupling range. In slowly running (``walking'') systems a very large scale change is required to cover even a moderate change in the renormalized coupling. Figure \ref{Fig.gcSq_nZS_c0.25} shows $g^2_c$ evaluated with Zeuthen flow, Symanzik operator and tree-level normalization in the $c=0.250$ scheme as the function of the bare coupling on lattice volumes ranging from $8^4$ to $32^4$. The 16 bare coupling values cover the range $g_c^2 \in (1.5, 6.5)$ in roughly uniform $\Delta g^2_c \approx 0.4$ increments.  Since the $N_f=12$ coupling runs very slowly,  prohibitively large volumes would be needed to reach the strong coupling regime as the bare coupling is tuned toward the GFP.  When approaching an IRFP, $\beta_{c,s} \to 0$ and lattice volumes would have to grow without bounds. Nevertheless the $L/a \to \infty$ limit predicts the infinite-cutoff step scaling function correctly,  as long as the GF flow time is large enough to approach the vicinity of the RT. This reflects the fact that the renormalized trajectory describes continuum physics.

Cutoff effects scale with the scaling exponent of the least irrelevant operator. In the vicinity of the Gaussian FP $\alpha = -2$, thus an extrapolation in $a^2/L^2$ removes the cutoff effects at weak coupling. Along the RT $\alpha$ can however change and the $a^2/L^2$ continuum extrapolation can become incorrect at stronger gauge couplings.  Since it is very difficult to determine the unknown exponent $\alpha$ numerically,  an elegant way out is to chose an improved  setup where cutoff effects are suppressed and the infinite volume extrapolation is mild enough not to depend on the exact extrapolation form. We will show that our favored setup of Symanzik action, Zeuthen flow, and Symanzik operator has this property in the range of couplings covered by our simulations.

After  extrapolating the discrete $\beta$-function $\beta_{c,s}(g^2_c;L)$ to the continuum limit (i.e.~$L/a\to \infty$) at fixed $g_c^2$, the continuum $\beta$-function $\beta_{c,s}(g_c^2)$ depends only on the renormalized coupling $g^2_c$. Therefore it is expected that $\beta_{c,s}(g_c^2)$ is free of effects from irrelevant operators introduced by the lattice regularization and depends only on the renormalization scheme  $c$ and scale change $s$.

%%%%%%%%%%%%%%%%%%%%%%%%%%%%%%%%%%%%%%%
\section{Numerical simulation details }
\label{Sec.NumSim}
\begin{figure}[tb]
  \includegraphics[width=0.99\columnwidth]{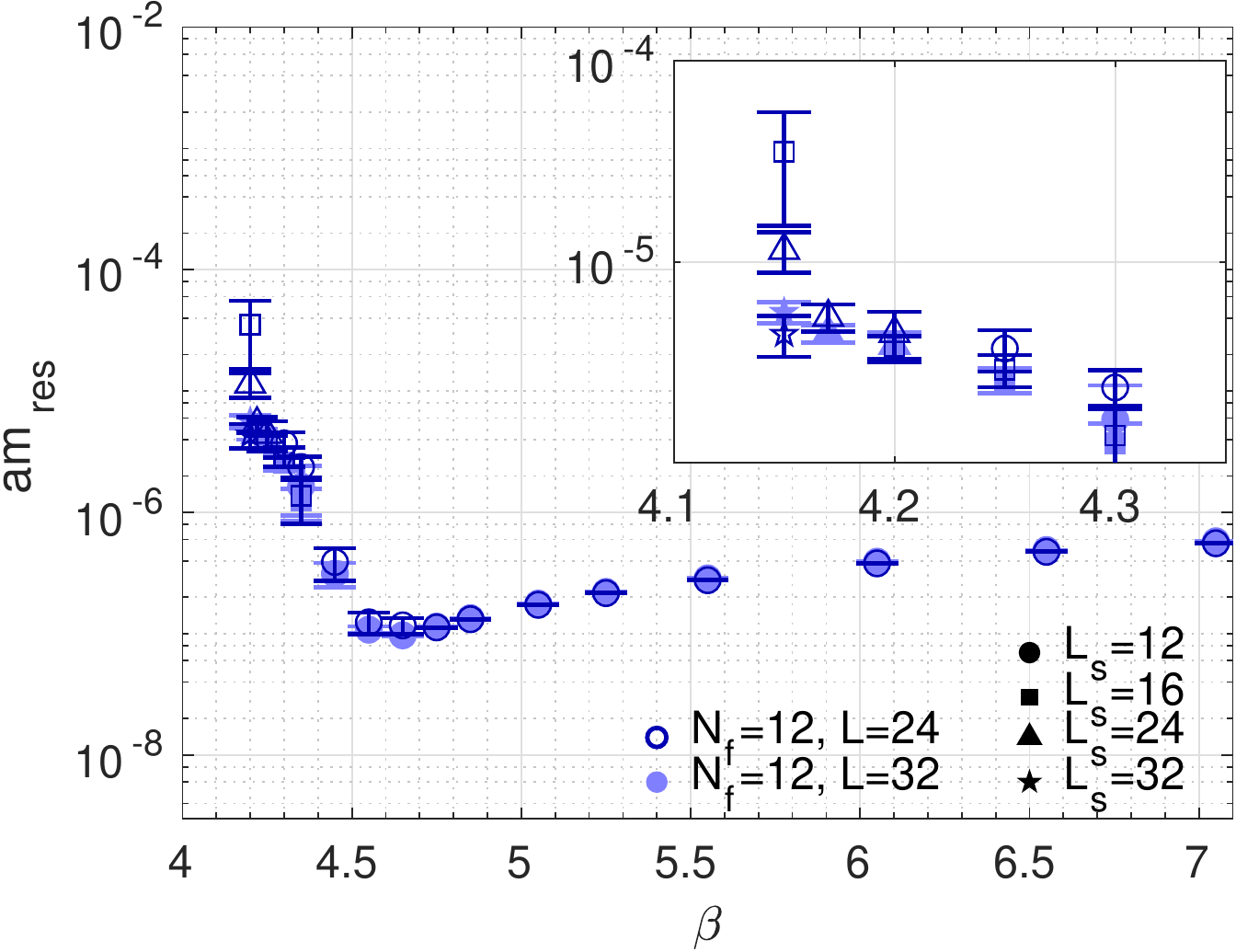}
  \caption{Residual chiral symmetry breaking, measured in terms of the residual mass $am_\text{res}$,  as function of the bare gauge coupling $\beta$ using $(L/a)^4$ volumes with $L/a=24$ and 32. Only statistical errors are shown.}
  \label{Fig.mres_vs_beta}
\end{figure}

%%%%%%%%%%%%%%%%%%%%%%%%%%%%%%%%%%%%%%%
Our nonperturbative determination of the gradient flow $\beta$ function requires the generation of dynamical gauge field configurations on $(L/a)^4$ hypercubic volumes. Choosing tree-level improved Symanzik (L\"uscher-Weisz) gauge action \cite{Luscher:1984xn,Luscher:1985zq} and M\"obius domain wall fermions (MDWF) \cite{Brower:2012vk} with three levels of stout-smearing \cite{Morningstar:2003gk} ($\varrho=0.1$) for the fermion action,\footnote{This combination of actions has already demonstrated its good properties for simulations in QCD \cite{Kaneko:2013jla,Noaki:2015xpx}.} we generate ensembles of gauge field configurations using the hybrid Monte Carlo (HMC) update algorithm \cite{Duane:1987de} with six massless two-flavor fermion fields and trajectories of length $\tau=2$ in molecular dynamics time units (MDTU). MDWF provide a prescription to simulate chiral fermions with continuum-like flavor symmetries by adding a fifth dimension to separate the chiral, physical modes of four dimensional space-time. In practice, the extent of the fifth dimension, $L_s$, is finite which results in a residual chiral symmetry breaking, conventionally parametrized by an additive mass term $am_\text{res}$. To study the $\beta$ function for  twelve flavors in the chiral limit, we set the input quark mass to zero and choose $L_s$ such that $am_\text{res} <5\cdot 10^{-6}$. As can be seen in Fig.~\ref{Fig.mres_vs_beta}, the residual mass increases when approaching strong coupling but the numerically determined values of $am_\text{res}$ are largely independent of the four dimensional volume. In our step-scaling calculation we therefore increase $L_s$ from 12 at weak coupling up to 32 for simulations at our strongest couplings to ensure that any effect from nonzero $am_\text{res}$ is negligible. The specific values of $L_s$ for each value of $\beta$ and $L/a$ are listed in Table \ref{Tab.Ls}.  At the strongest gauge couplings we performed additional simulations with alternative choices of $L_s$ to verify that the $L_s$ values listed in Table \ref{Tab.Ls} are sufficient for the flow times ($c$ values) used in this analysis. Values of the renormalized couplings on these additional ensembles are listed in Appendix \ref{Sec.RenCouplings} in Table \ref{Tab.nZS_ZS_Ls}. In all cases we find that renormalized couplings for the same choice of $L/a$ and $\beta$ but different choices of $L_s$ agree at the 1$\sigma$ level. We further found consistent results for $\beta_{c,s}(g_c^2,L)$ when substituting ensembles with different $L_s$.
\begin{table}[t]
  \caption{Values of the extent of the fifth dimension $L_s$ for the ensembles entering our main analysis as function of the bare gauge coupling $\beta$ and the length $L/a$ of the four dimensional $(L/a)^4$ volume.}
  \label{Tab.Ls}
  \begin{tabular}{cccccccccc}
    \hline\hline
            & \multicolumn{9}{c}{$L/a$}\\
    $\beta$      & 8  & 10 & 12 & 14 & 16 & 20 & 24 & 28 & 32\\
    \hline
    7.00--4.40   & 12 & 12 & 12 & 12 & 12 & 12 & 12 & 12 & 12\\
    4.30, 4.25   & 12 & 12 & 12 & 12 & 12 & 12 & 12 & 16 & 16\\
    4.20         & 16 & 16 & 16 & 16 & 16 & 16 & 16 & 24 & 24\\
    4.17         & 16 & 16 & 16 & 16 & 24 & 24 & 24 & 24 & 24\\
    4.15         & 16 & 16 & 16 & 24 & 24 & 24 & 24 & 32 & 32\\    
    4.13         & 32 & -- & -- & -- & 32 & -- & -- & -- & --\\
    \hline\hline
    \end{tabular}
\end{table}  

\begin{figure}[tb]
  \includegraphics[width=0.99\columnwidth]{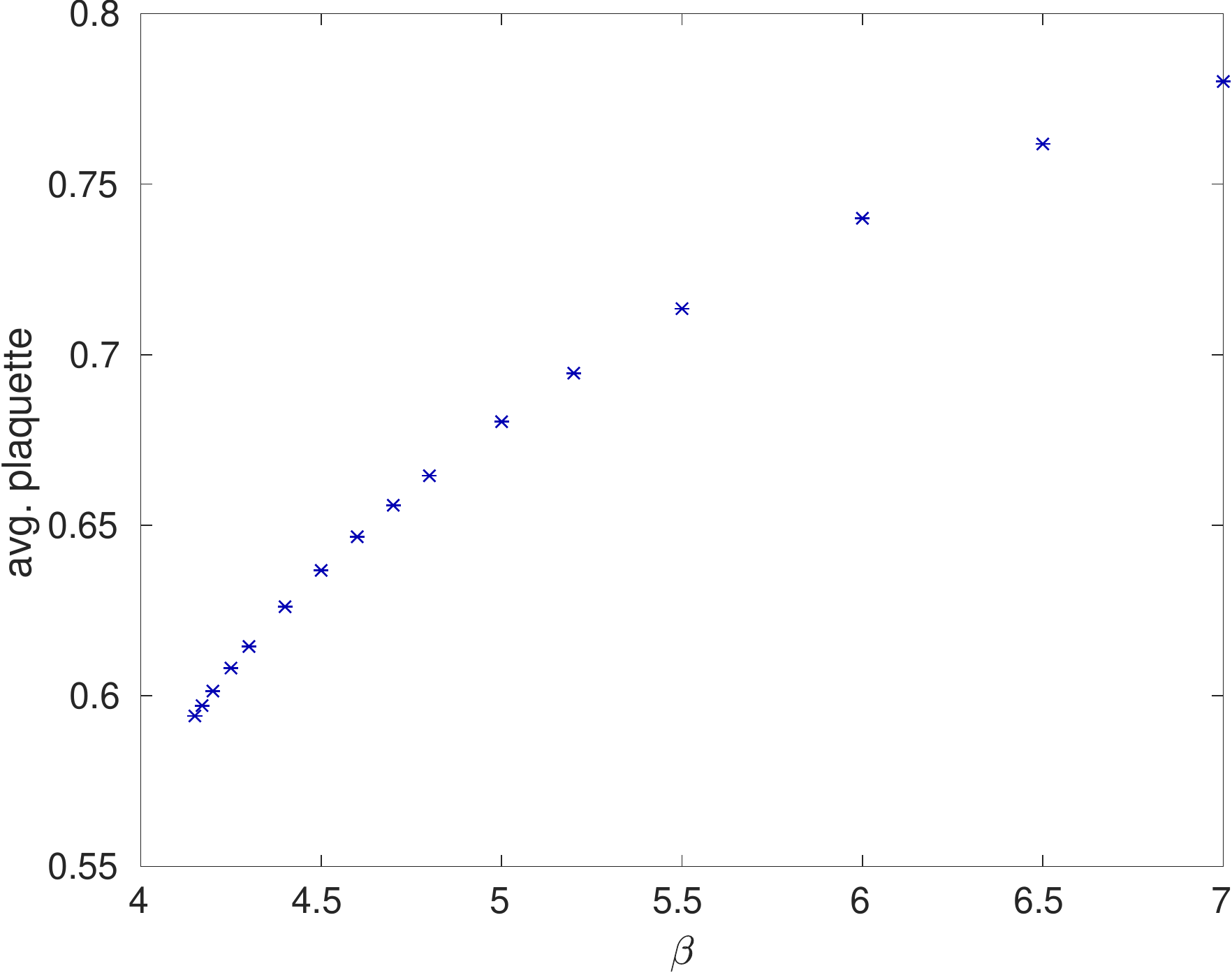}
  \caption{Average plaquette (normalized to 1) as the function of the bare gauge coupling $\beta$ on $32^4$ volumes. Statistical errors are shown but barely visible.}
  \label{Fig.plaq_vs_beta}
\end{figure}

We set the domain wall height $M_5=1$ and have Pauli-Villars terms of mass one.  Simulations are performed with the same boundary conditions (BC) in all four directions: periodic BC for the gauge field and antiperiodic BC for the fermion fields. The latter triggers a gap in the eigenvalues of the Dirac operator and thus allows simulations with zero input quark mass. In order to explore renormalized couplings up to $g_c^2\sim 6.5$, we create ensembles for a set of bare couplings starting in the weak coupling limit with $\beta=7.00$ and going down to $\beta=4.15$ as our strongest coupling.\footnote{For $L/a=8$ and 16 we also simulated at $\beta=4.13$.} We choose hypercubic $(L/a)^4$ volumes with $L/a=8$, 10, 12, 14, 16, 20, 24, 28, and 32 and typically generate 6-10k (2-4k) thermalized MDTU for most  ensembles with $L/a\le 24$ ($L/a=28,$ 32). All ensembles are generated using \texttt{GRID} \cite{Grid, Boyle:2015tjk} and a gauge field configuration is saved every five trajectories (10 MDTU).   In Fig.~\ref{Fig.plaq_vs_beta} we demonstrate that our combination of actions leads to sufficiently smooth gauge field configurations with the average plaquette, the smallest $1\times 1$ Wilson loop (normalized to 1), ranging from 0.78 to about 0.59.  Such values are comparable to those  in QCD simulations and hence we expect the gauge fields to be sufficiently smooth to extrapolate to the correct  continuum limit.

We perform gradient flow measurements on every fifth trajectory (10 MDTU). With the intention to investigate systematics of the gradient flow, we measure Wilson (W), Symanzik (S), and Zeuthen (Z) flow on each gauge field configuration using \texttt{QLUA} \cite{Pochinsky:2008zz,qlua}. For all three flows, we estimate the energy density using three different operators: the Wilson-plaquette (W), clover (C), and  Symanzik (S) operator. As described in Sec.~\ref{Sec.GradFlow}, the energy densities are proportional to the renormalized coupling which in turn lead to the step-scaling $\beta$ function.   In Appendix \ref{Sec.RenCouplings} we show the full details of our determinations of the energy densities for our preferred analyses based on Zeuthen flow with Symanzik operator (with and without tree-level normalization). Besides presenting $g_c^2$ values in the renormalization schemes $c=0.250$, 0.275, and 0.300,  we also list the total number of measurements for each ensemble as well as the integrated autocorrelation time determined using the $\Gamma$-method \cite{Wolff:2003sm}.

In addition we use the gradient flow measurements to determine the topological charge and confirm it vanishes as expected for massless simulations.

%=================================================
\section{\texorpdfstring{Gradient flow $\beta$ function for twelve fundamental flavors}{Gradient flow beta function for twelve fundamental flavors}}
\label{Sec.Results}

As discussed above, we have calculated energy densities using different gradient flows and operators for all of our ensembles of gauge field configurations. This allows us to determine the gradient flow $\beta$ function in multiple ways and check for possible systematic effects.
%=================================================
\subsection{Preferred (n)ZS analysis}
\label{Sec.nZS_ZS}
 Our preferred analysis is based on Zeuthen flow and the Symanzik operator. Since our ensembles are generated with Symanzik gauge action, this combination is fully $O(a^2)$ improved and we indeed find  small discretization effects. In the weak coupling limit, discretization effects can be further suppressed by applying the perturbatively calculated tree-level normalization factors (see Eq.~(\ref{eq:pert_g2})). A priori the range of validity in $g_c^2$ for perturbatively calculated coefficients is not known. Therefore, we present the results for our preferred analysis with and without tree-level normalization. We refer to the two analysis as nZS and ZS, respectively. 
Our results are presented for the renormalization schemes $c=0.250$, 0.275, and 0.300 in Figs.~\ref{Fig.beta_c250}--\ref{Fig.beta_c300} and are obtained from the renormalized couplings listed in Appendix in \ref{Sec.RenCouplings} Table \ref{Tab.nZS_ZS}.
\begin{figure*}[t]
  \begin{minipage}{0.49\textwidth}
   \flushright 
   \includegraphics[width=0.96\textwidth]{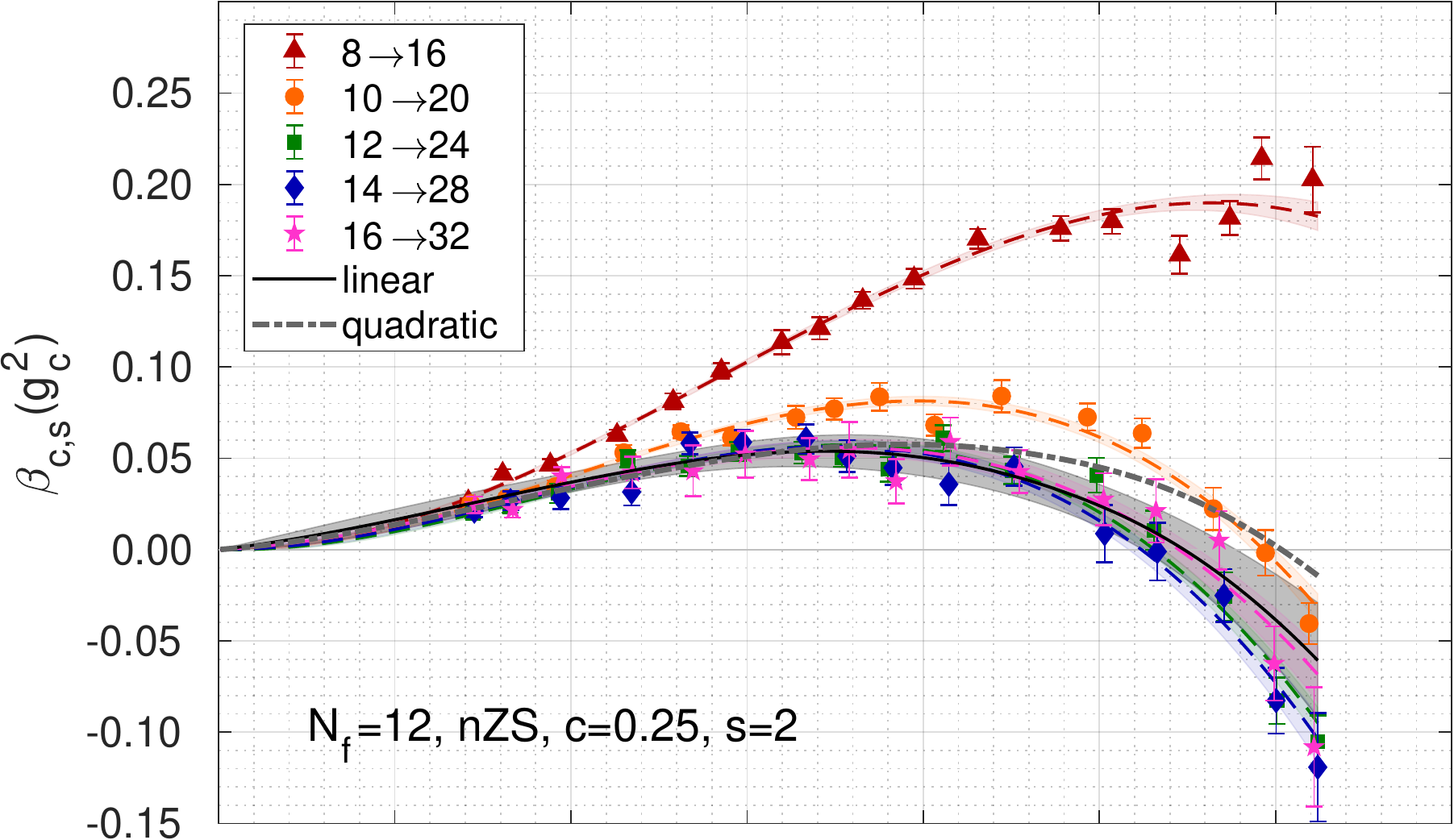}\\
   \includegraphics[width=0.924\textwidth]{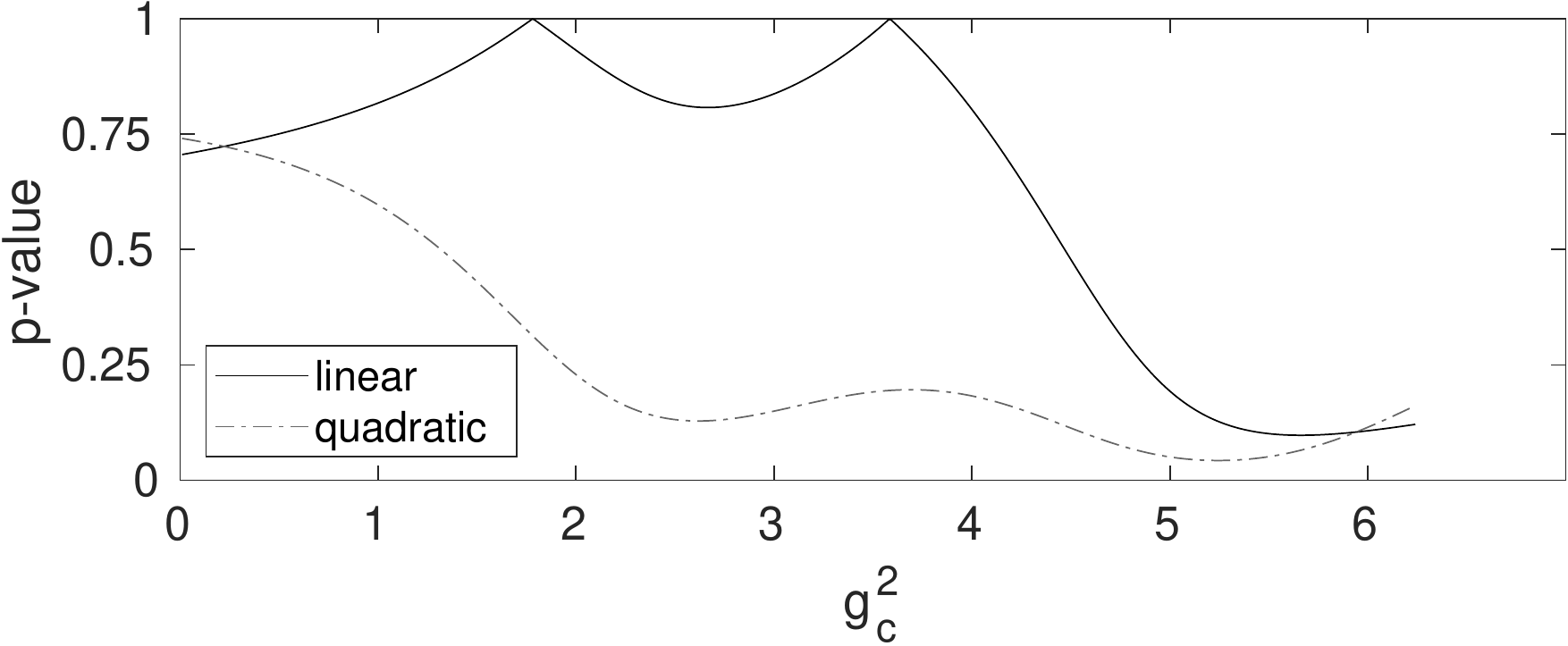}\\
   \includegraphics[width=0.96\textwidth]{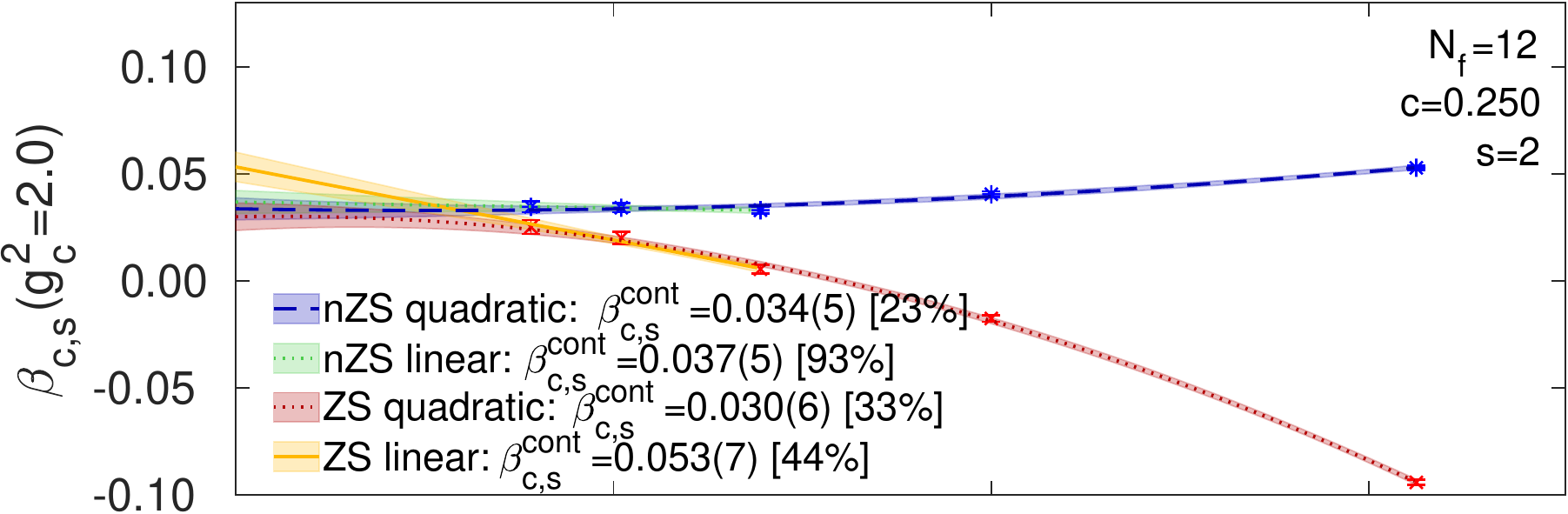}\\
   \includegraphics[width=0.96\textwidth]{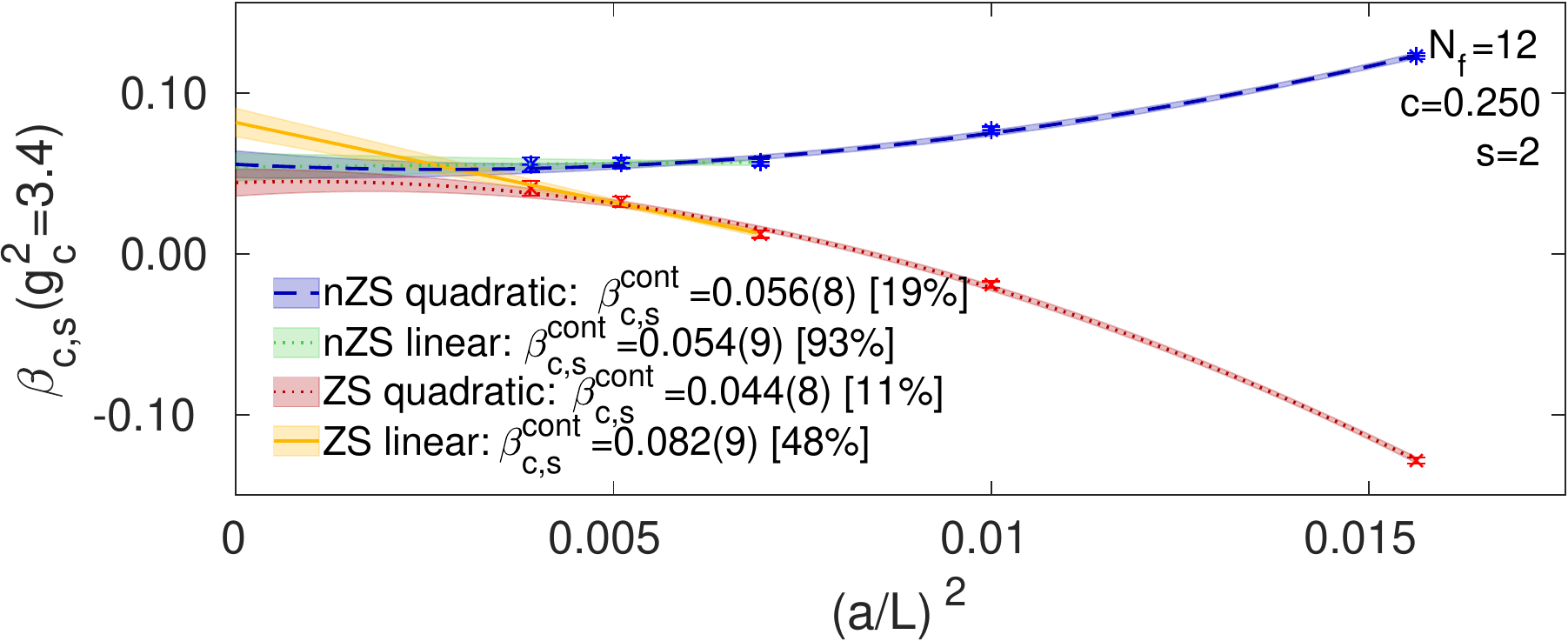}     
  \end{minipage}
  \begin{minipage}{0.49\textwidth}
    \flushright
    \includegraphics[width=0.96\textwidth]{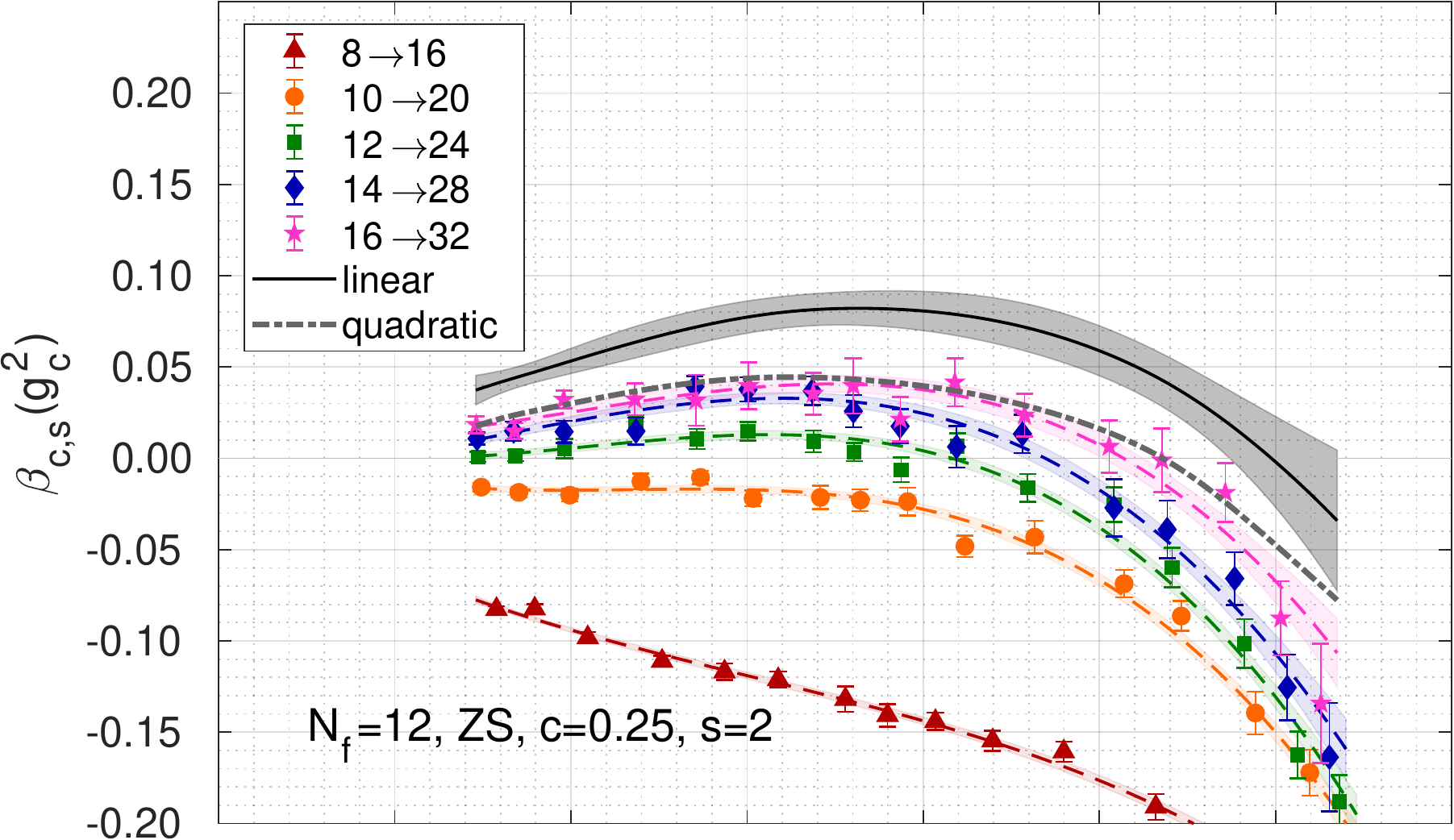}\\    
    \includegraphics[width=0.924\textwidth]{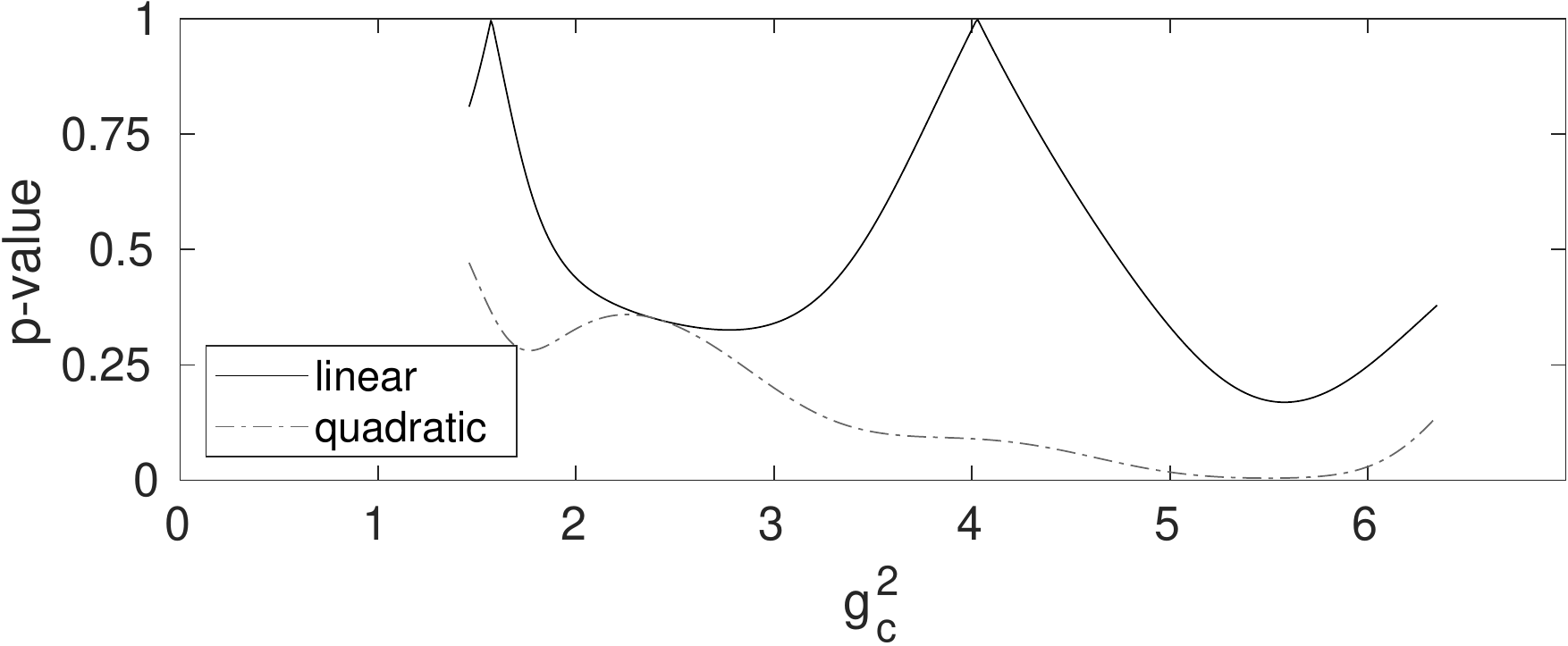}
    \includegraphics[width=0.96\textwidth]{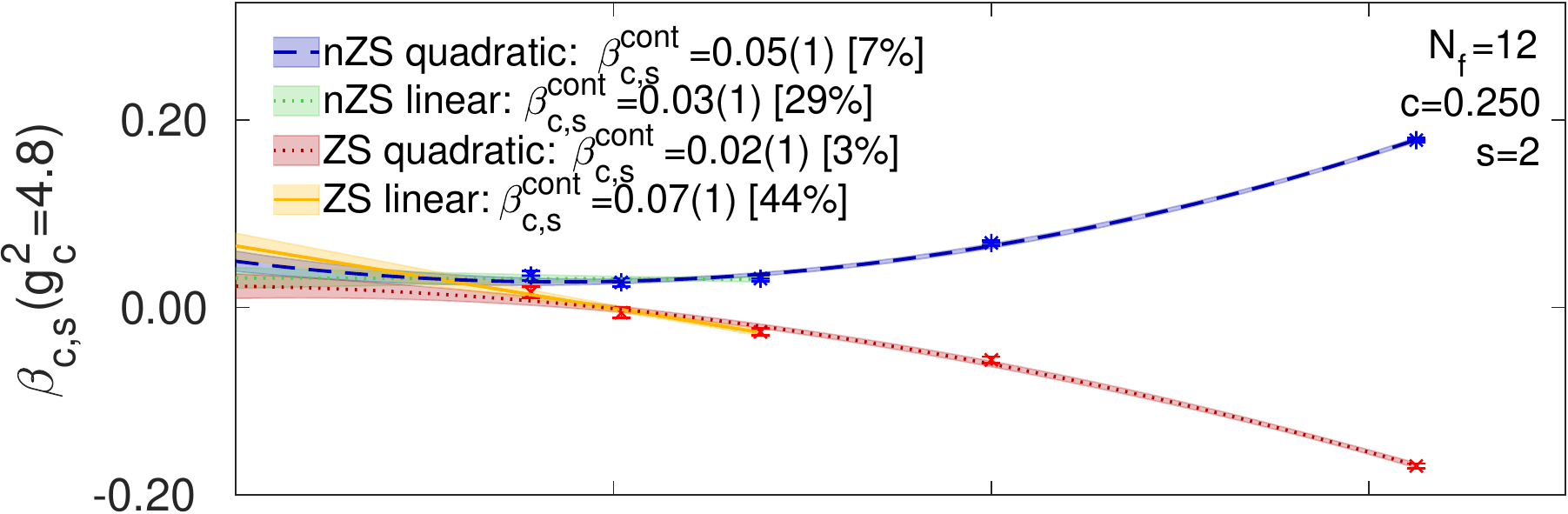}\\
    \includegraphics[width=0.96\textwidth]{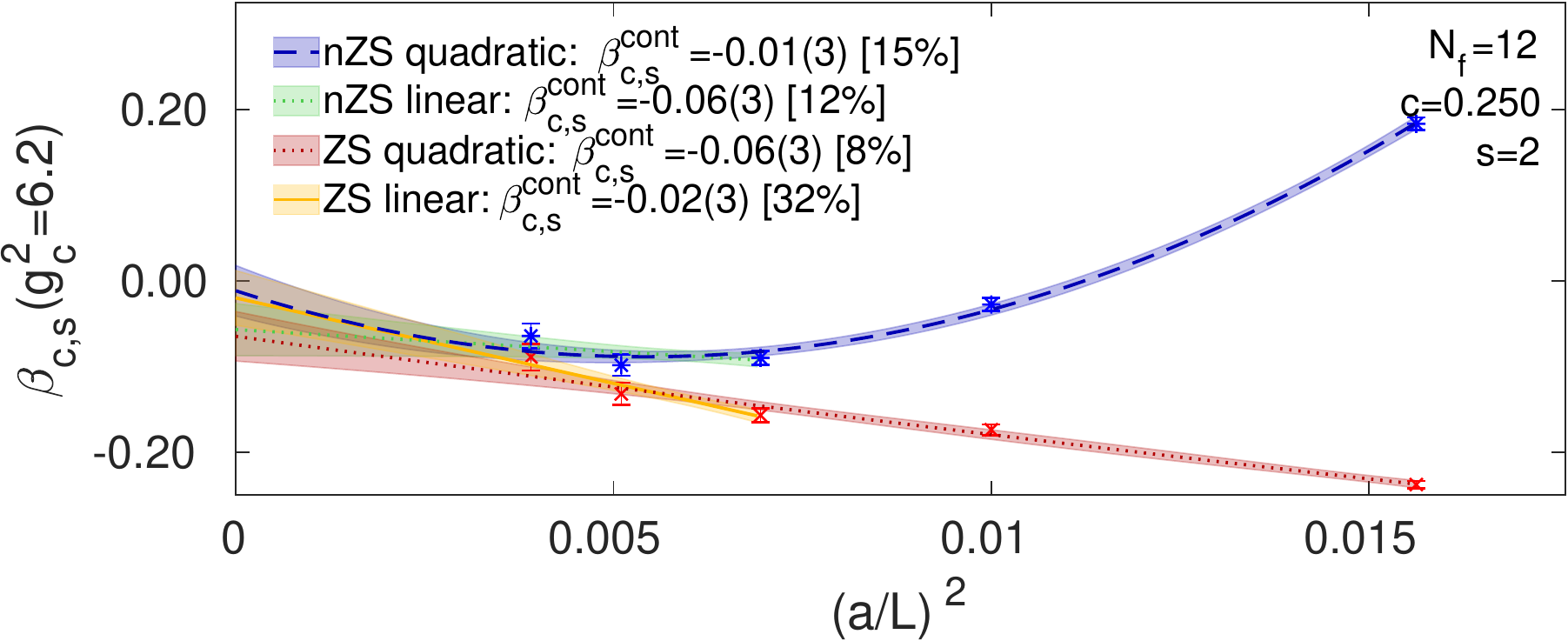}    
  \end{minipage}
  \caption{Discrete step-scaling $\beta$-function in the $c=0.250$ gradient flow scheme for our preferred nZS (left) and ZS (right) data sets. The symbols in the top row show our results for the finite volume discrete $\beta$ function with scale change $s=2$. The dashed lines with shaded error bands in the same color of the data points show the interpolating fits. We perform two continuum extrapolations: a linear fit in $a^2/L^2$ to the three largest volume pairs (black line with gray error band) and a quadratic fit in $a^2/L^2$ to all five volume pairs (gray dash-dotted line). The $p$-values of the continuum extrapolation fits are shown in the plots in the second row. Further details of the continuum extrapolation at selected $g_c^2$ values are presented in the small panels at the bottom where the legend lists the extrapolated values in the continuum limit with $p$-values in brackets. Only statistical errors are shown.}
  \label{Fig.beta_c250}
\end{figure*}  
\begin{figure*}[t]
  \begin{minipage}{0.49\textwidth}
   \flushright 
   \includegraphics[width=0.96\textwidth]{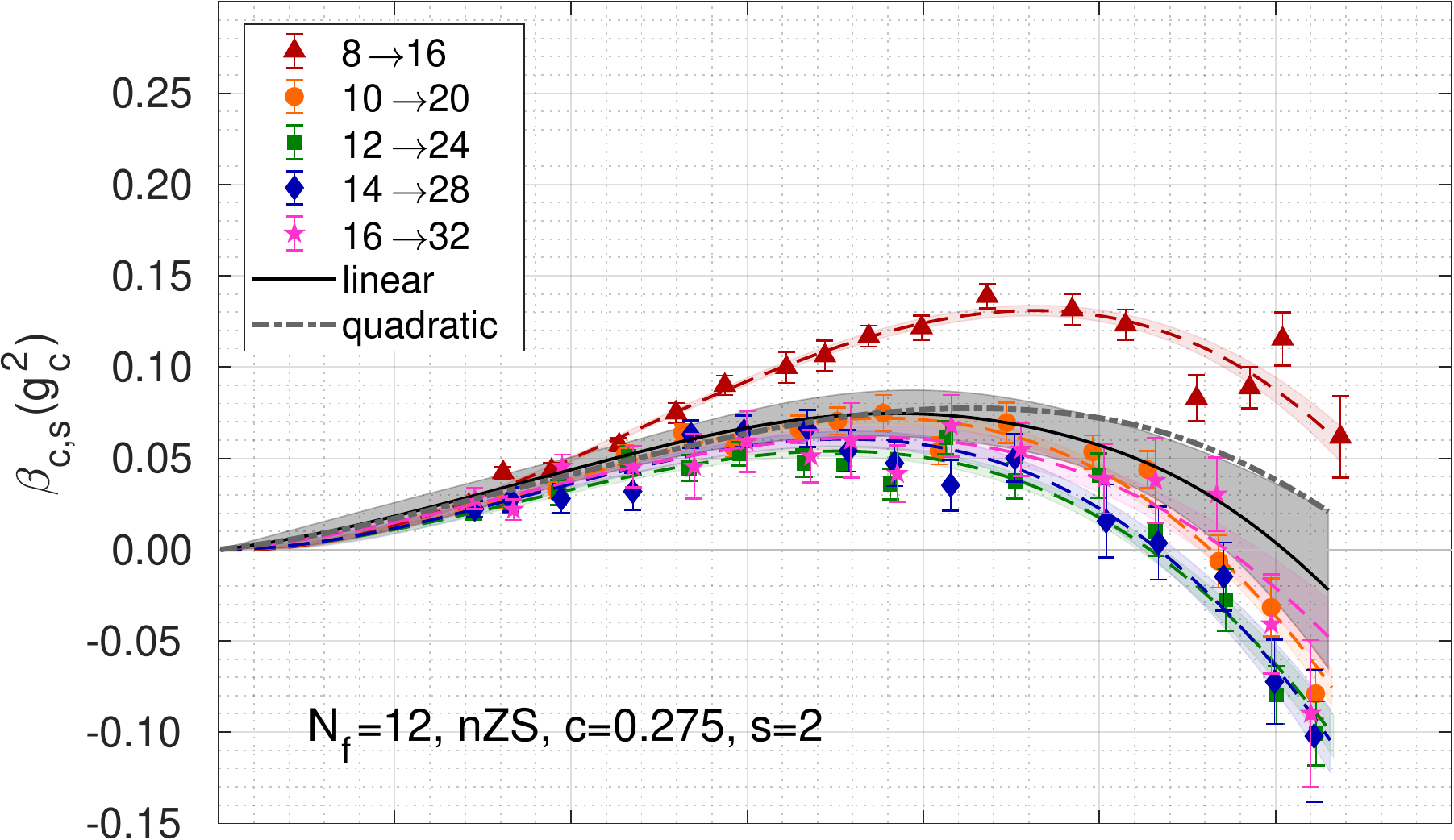}\\
   \includegraphics[width=0.924\textwidth]{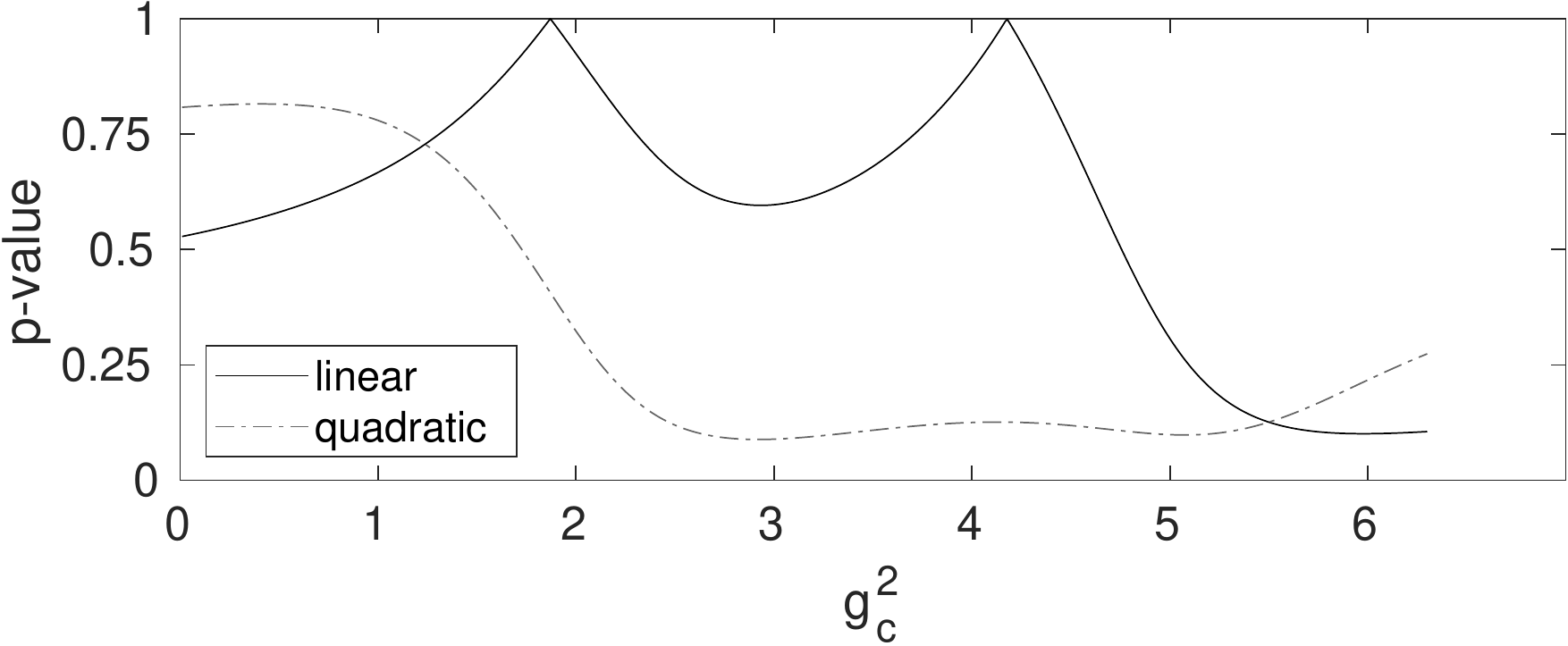}\\
   \includegraphics[width=0.96\textwidth]{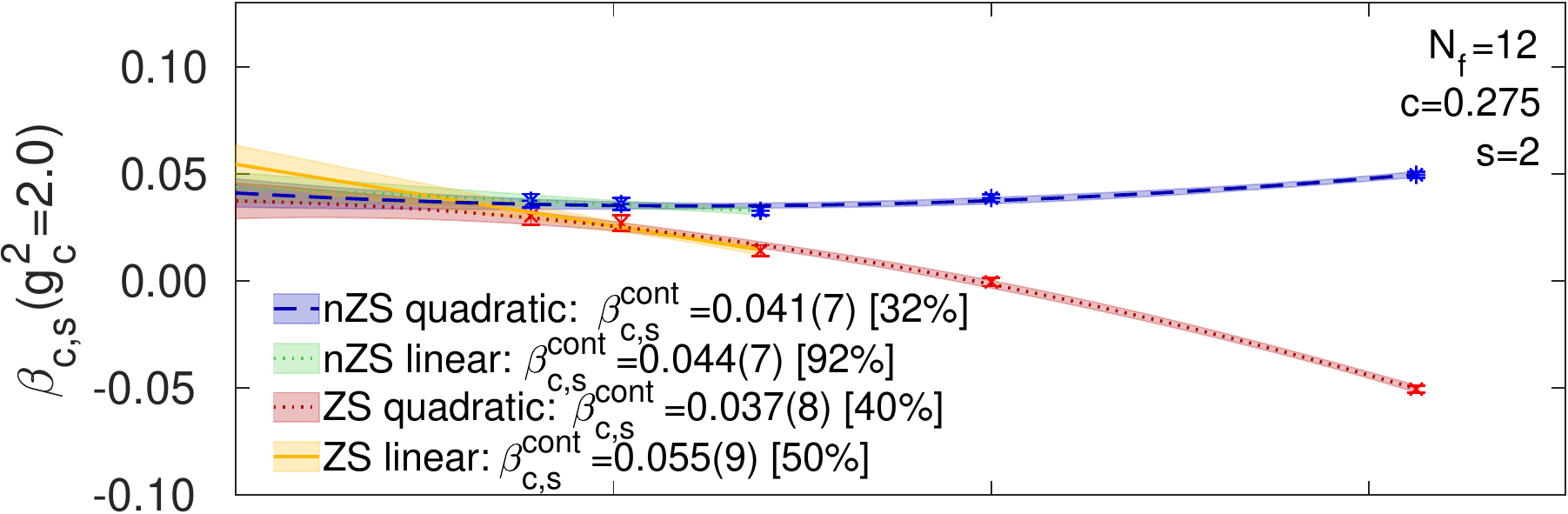}\\
   \includegraphics[width=0.96\textwidth]{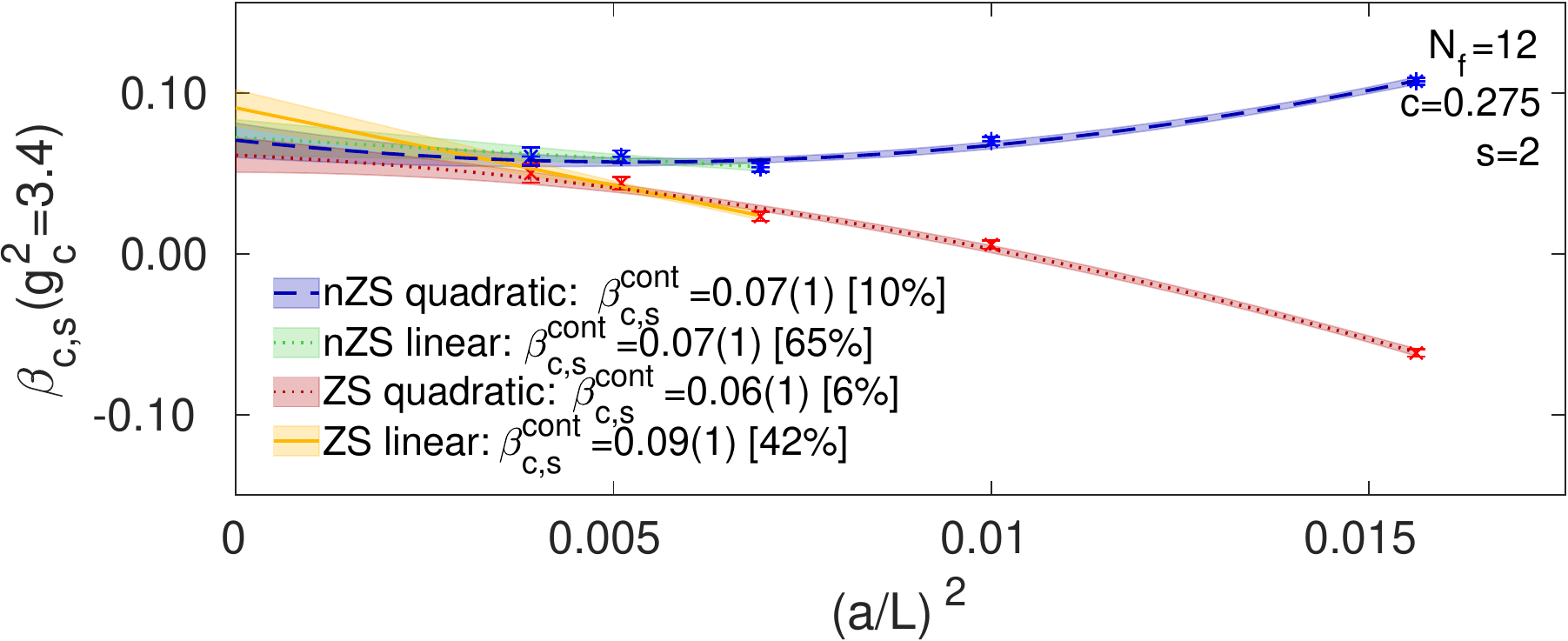}     
  \end{minipage}
  \begin{minipage}{0.49\textwidth}
    \flushright
    \includegraphics[width=0.96\textwidth]{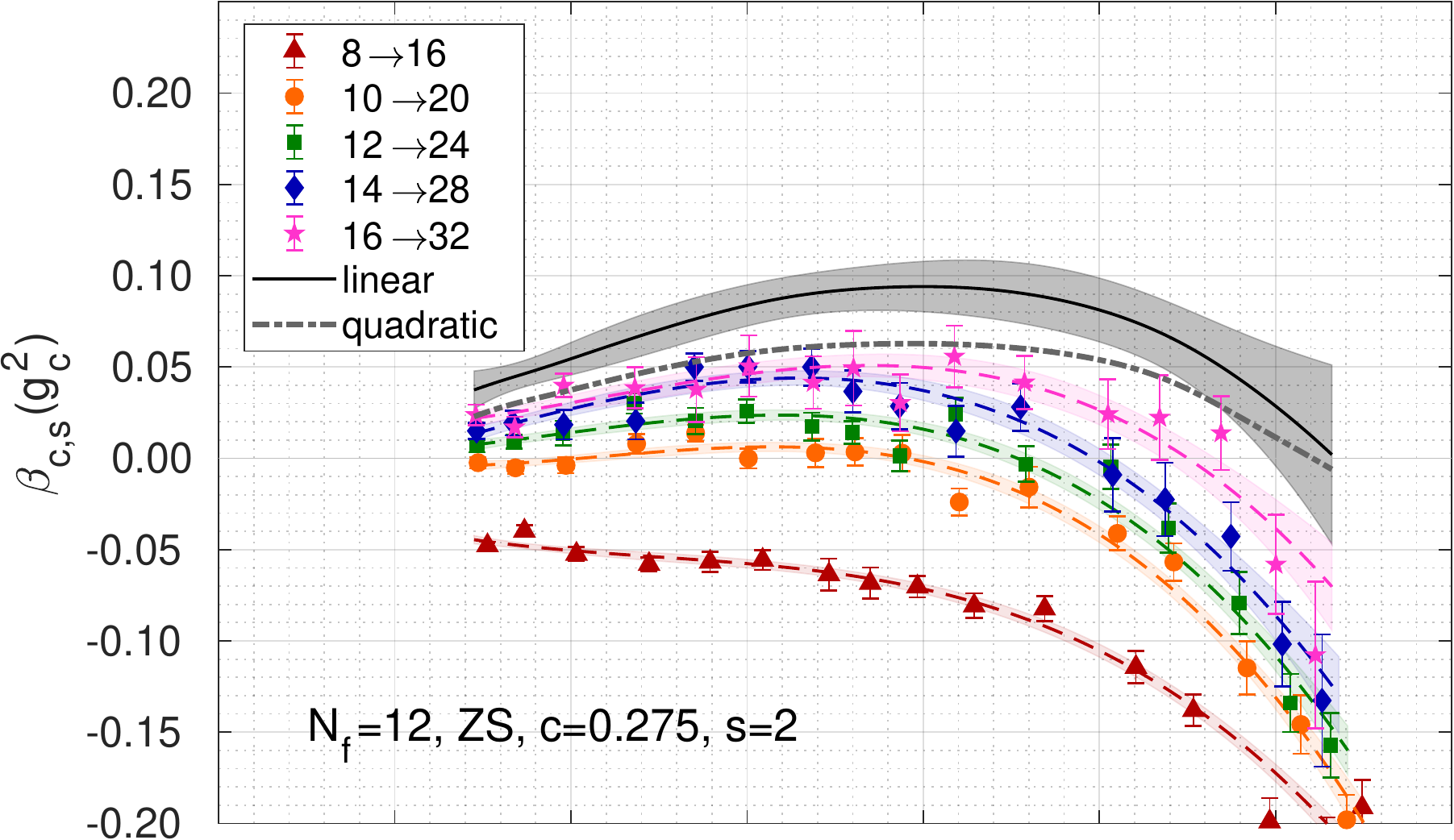}\\    
    \includegraphics[width=0.924\textwidth]{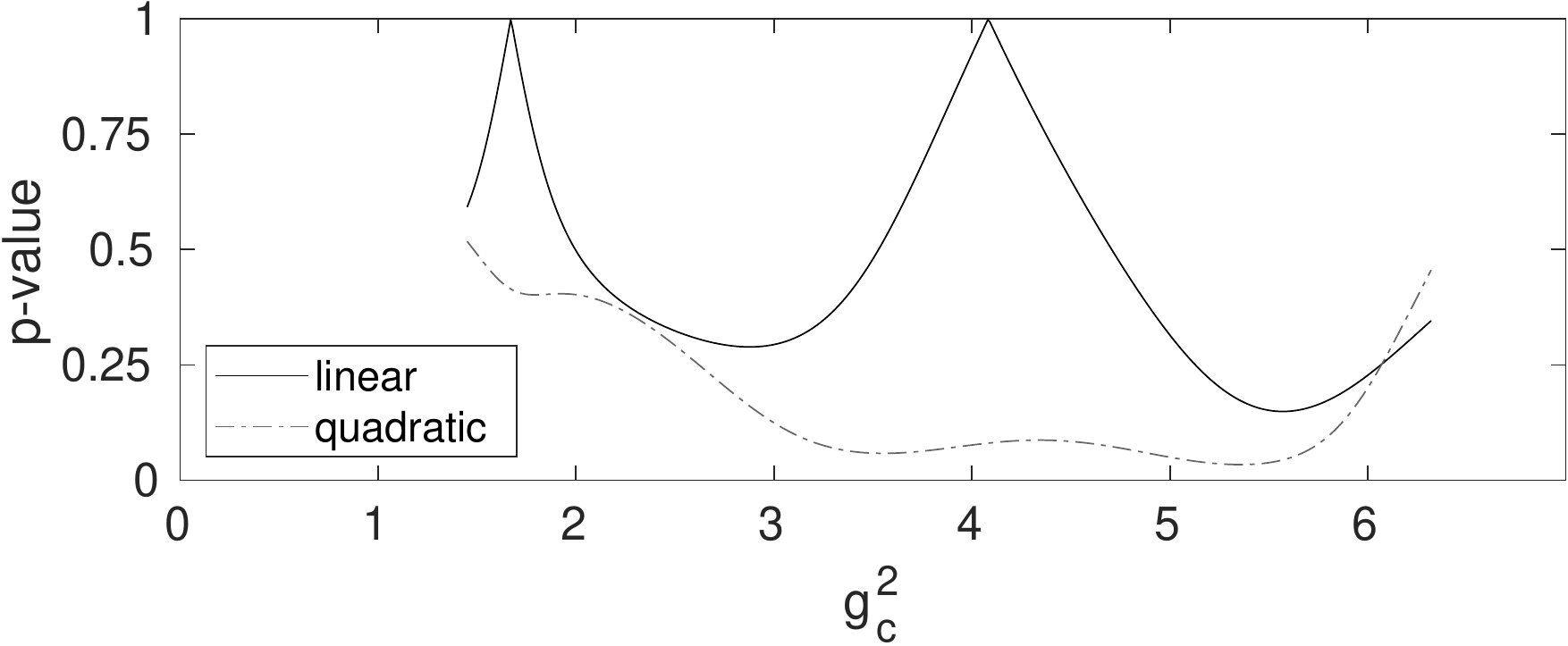}
    \includegraphics[width=0.96\textwidth]{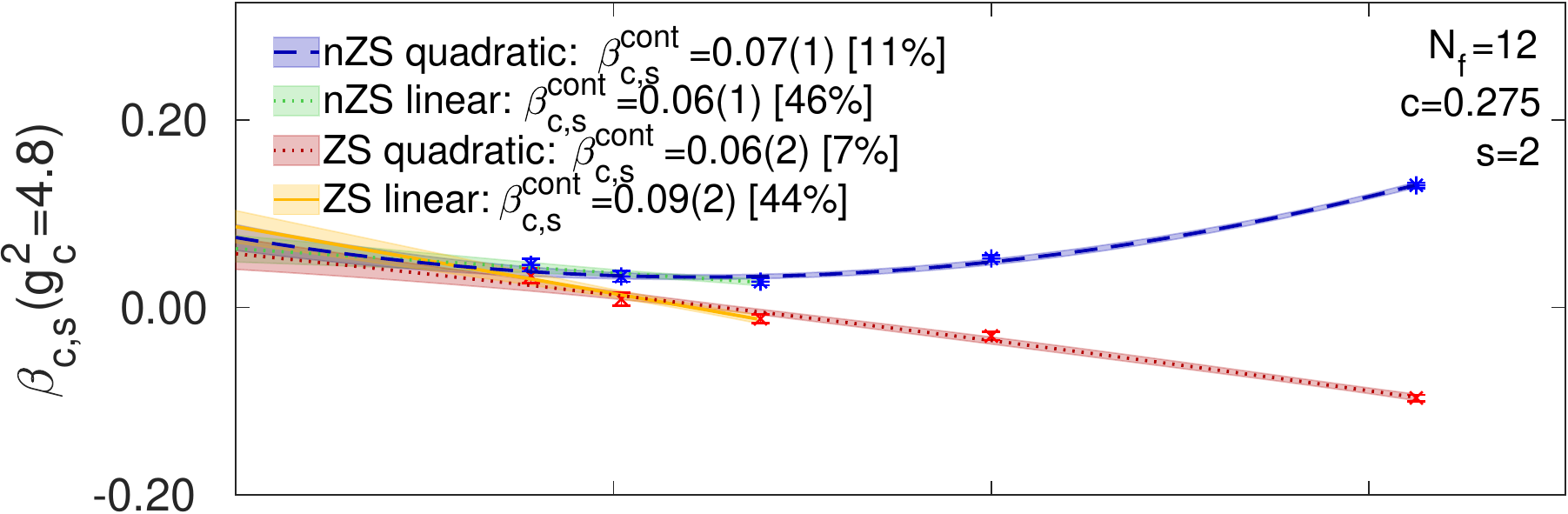}\\
    \includegraphics[width=0.96\textwidth]{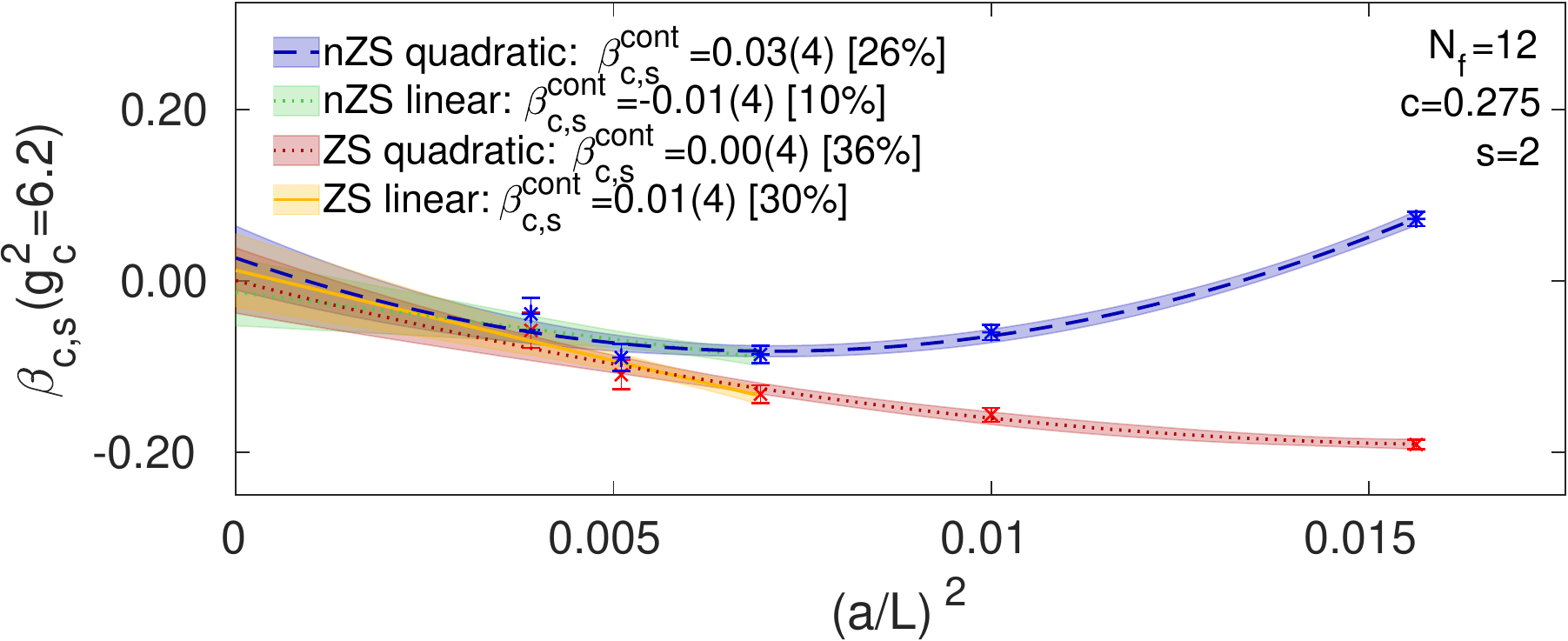}    
  \end{minipage}
  \caption{Discrete step-scaling $\beta$-function in the $c=0.275$ gradient flow scheme for our preferred nZS (left) and ZS (right) data sets. The symbols in the top row show our results for the finite volume discrete $\beta$ function with scale change  $s=2$. The dashed lines with shaded error bands in the same color of the data points show the interpolating fits. We perform two continuum extrapolations: a linear fit in $a^2/L^2$ to the three largest volume pairs (black line with gray error band) and a quadratic fit in $a^2/L^2$ to all five volume pairs (gray dash-dotted line). The $p$-values of the continuum extrapolation fits are shown in the plots in the second row. Further details of the continuum extrapolation at selected $g_c^2$ values are presented in the small panels at the bottom where the legend lists the extrapolated values in the continuum limit with $p$-values in brackets. Only statistical errors are shown.}
 \label{Fig.beta_c275}
\end{figure*}  
\begin{figure*}[t]
  \begin{minipage}{0.49\textwidth}
   \flushright 
   \includegraphics[width=0.96\textwidth]{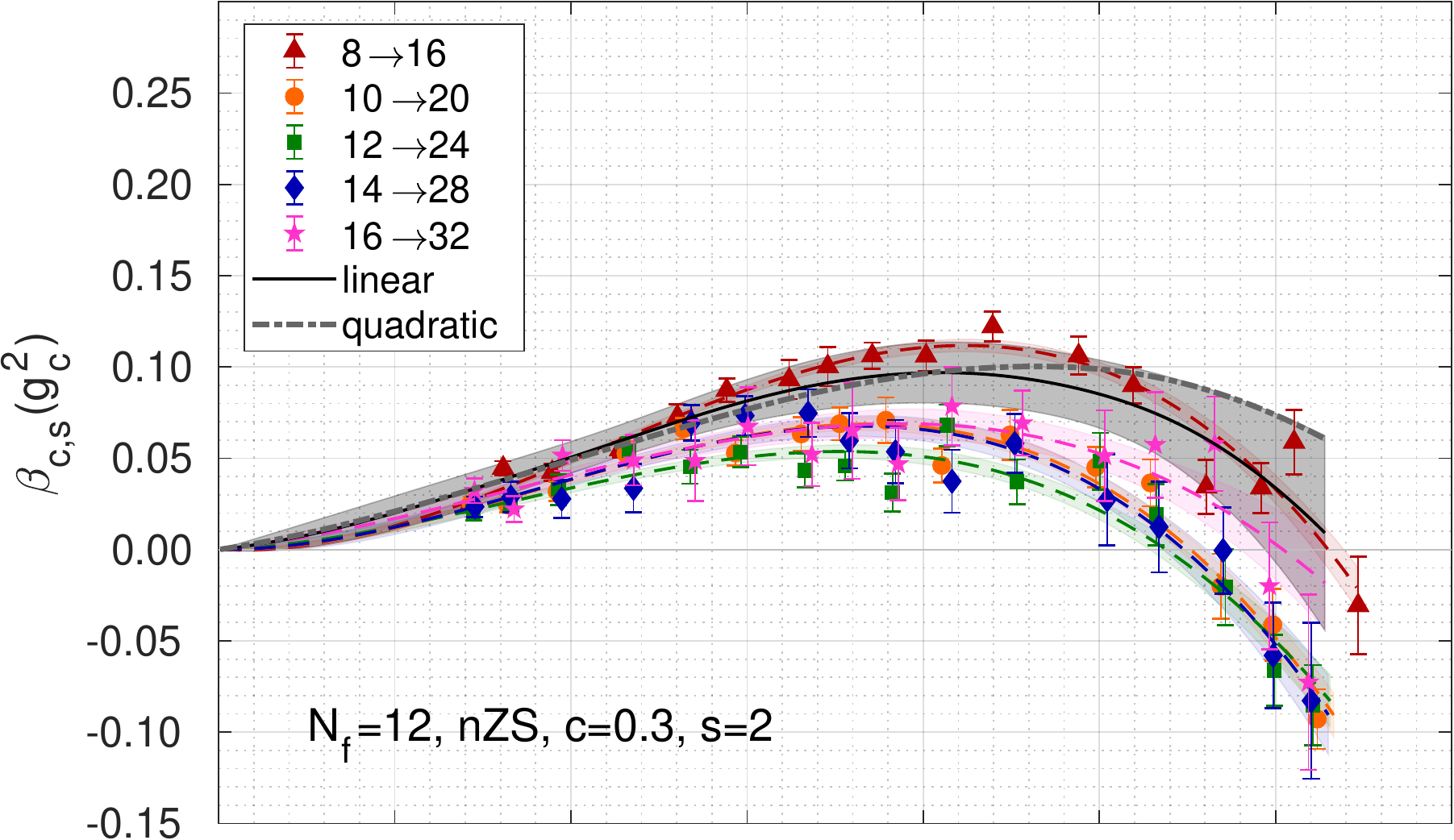}\\
   \includegraphics[width=0.924\textwidth]{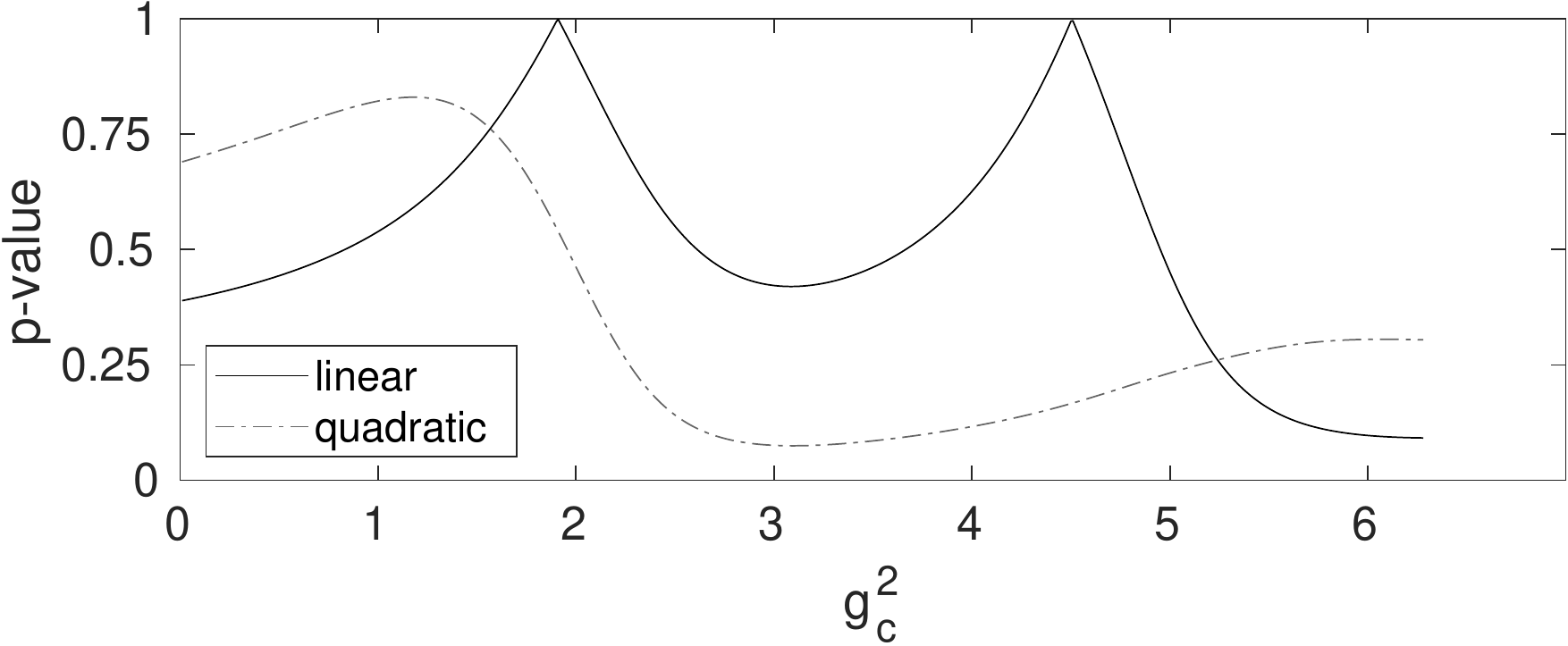}\\
   \includegraphics[width=0.96\textwidth]{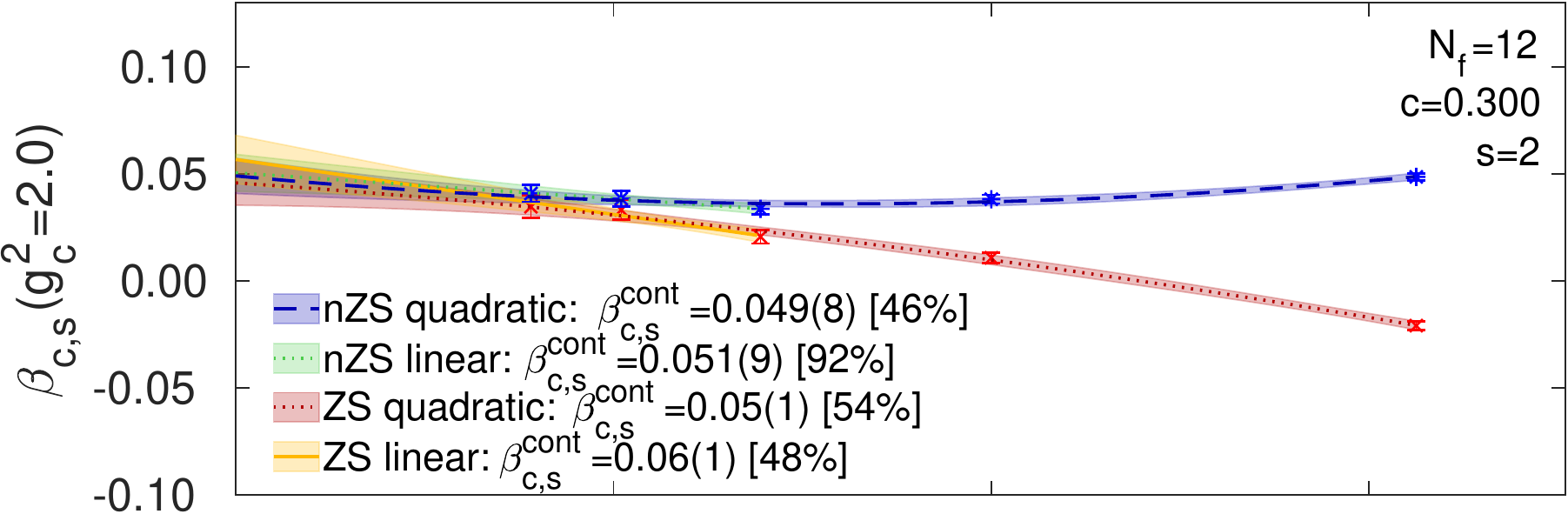}\\
   \includegraphics[width=0.96\textwidth]{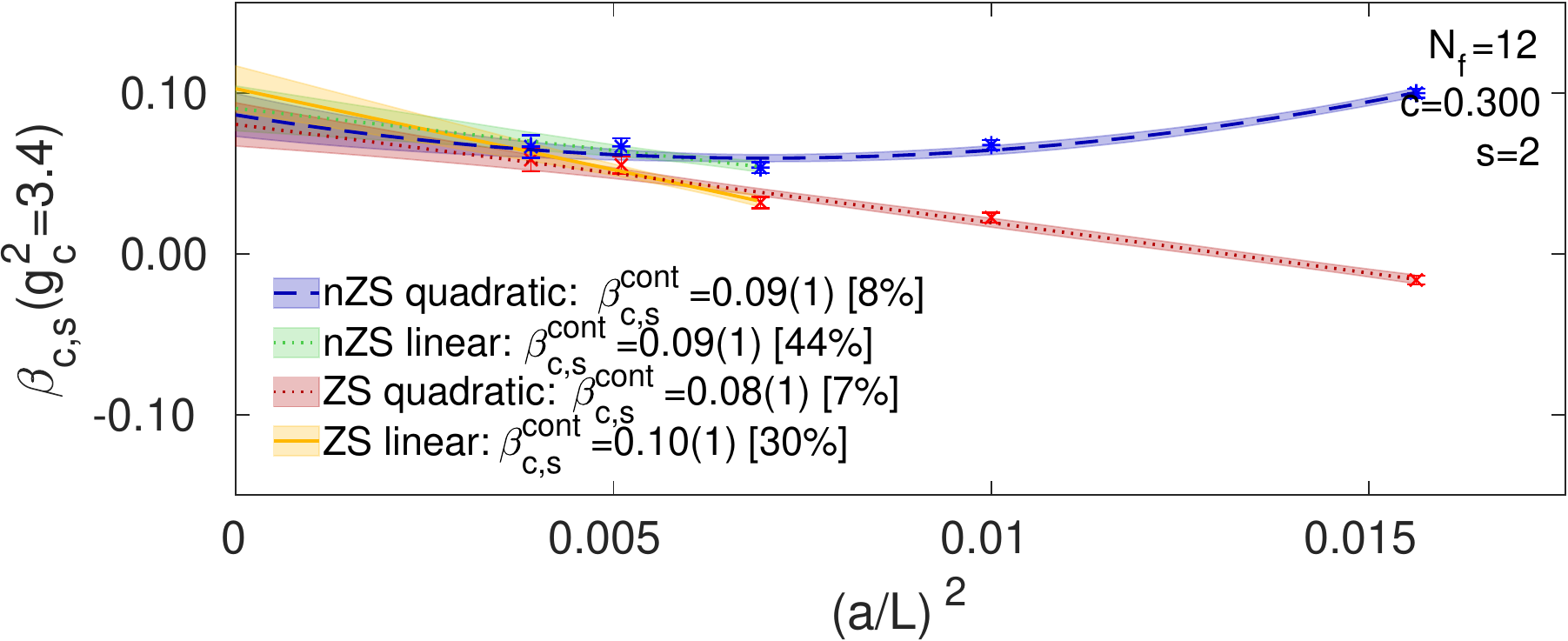}     
  \end{minipage}
  \begin{minipage}{0.49\textwidth}
    \flushright
    \includegraphics[width=0.96\textwidth]{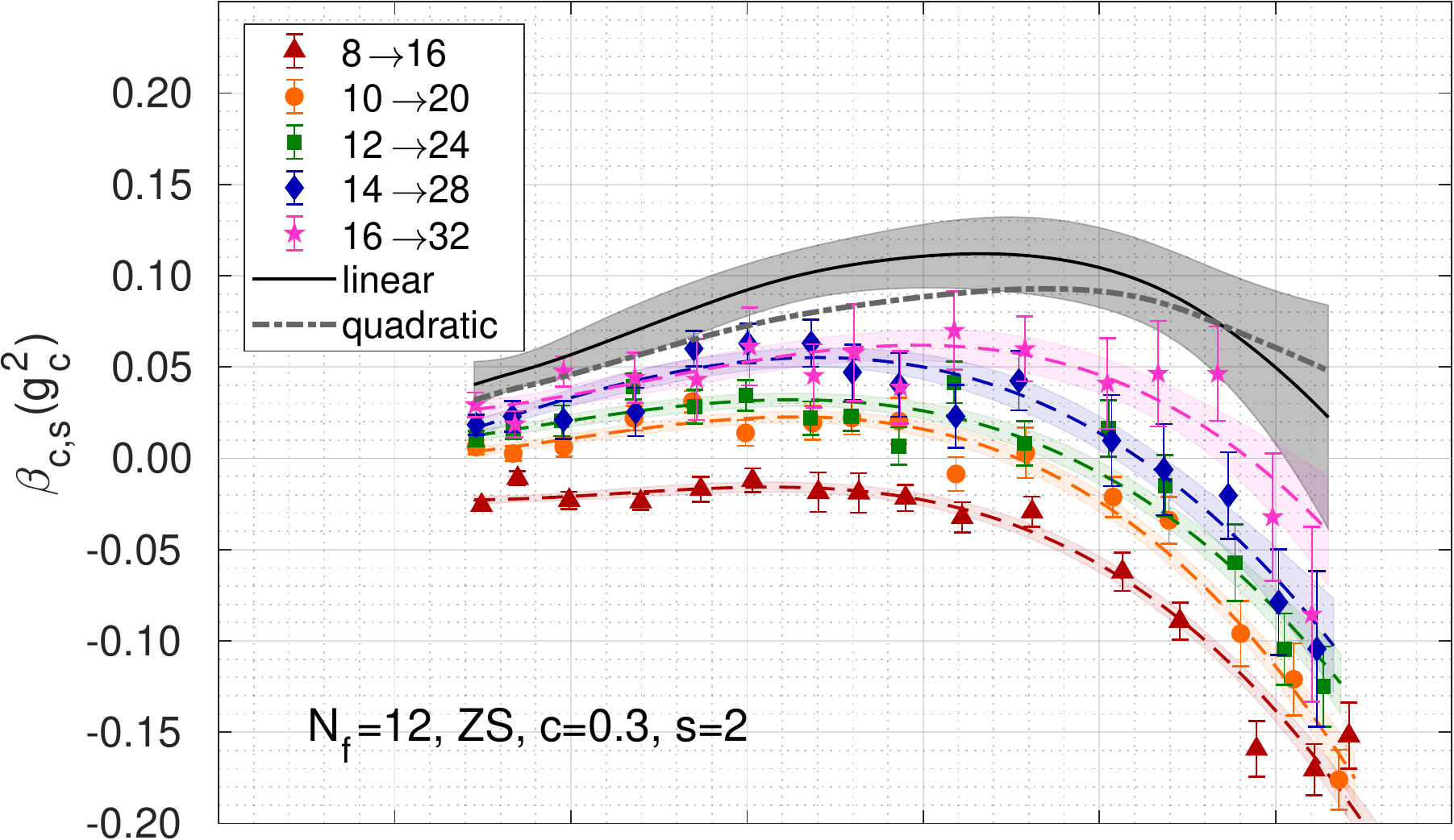}\\    
    \includegraphics[width=0.924\textwidth]{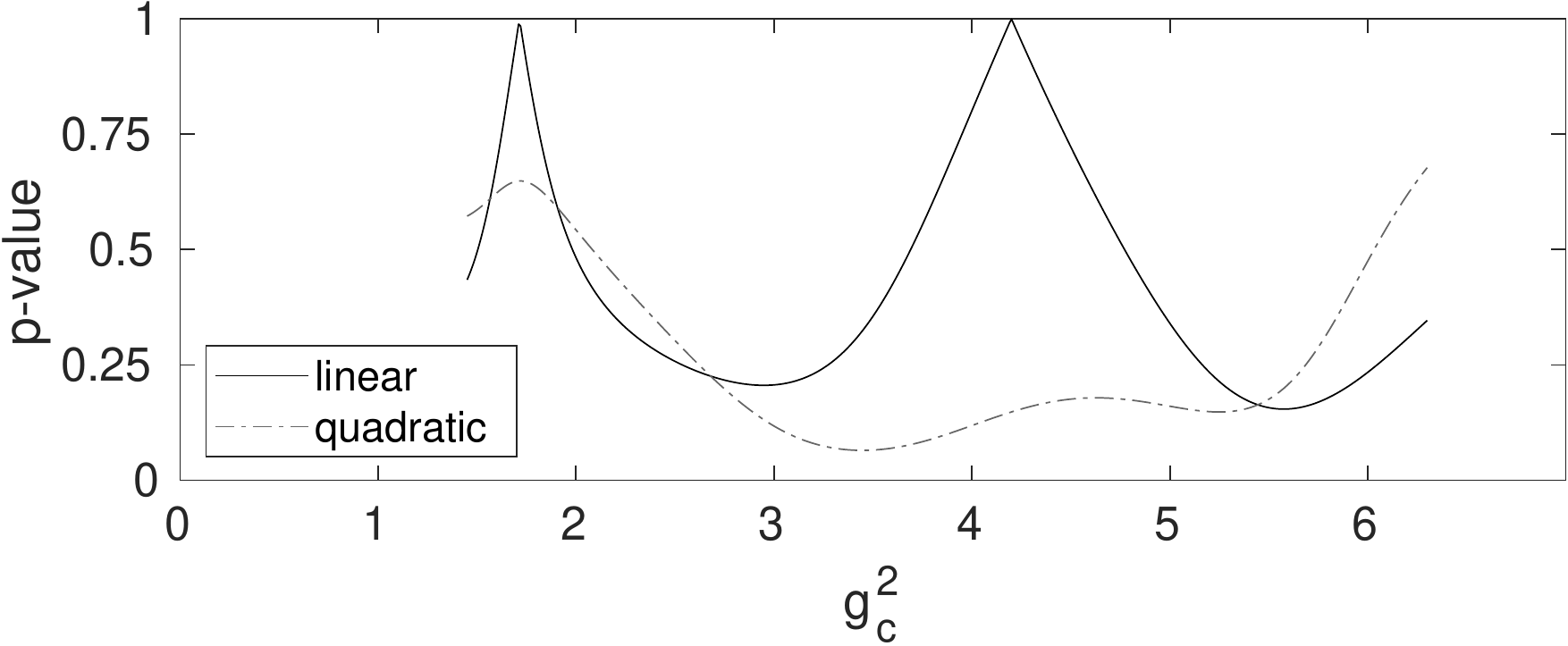}
    \includegraphics[width=0.96\textwidth]{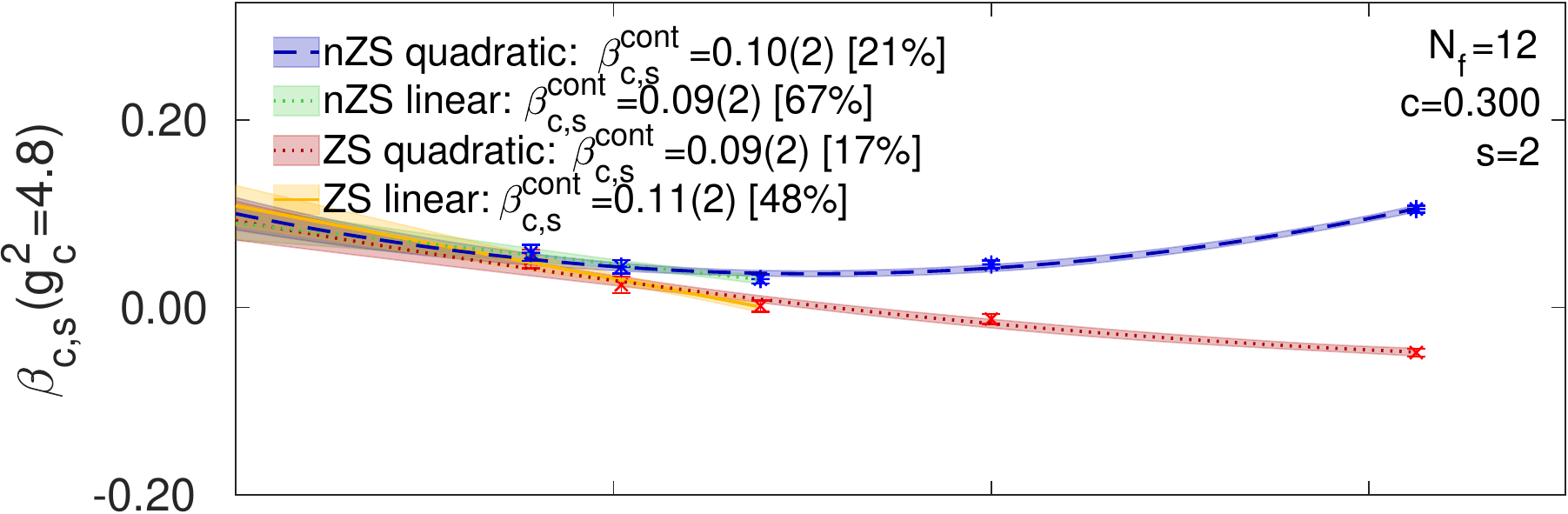}\\
    \includegraphics[width=0.96\textwidth]{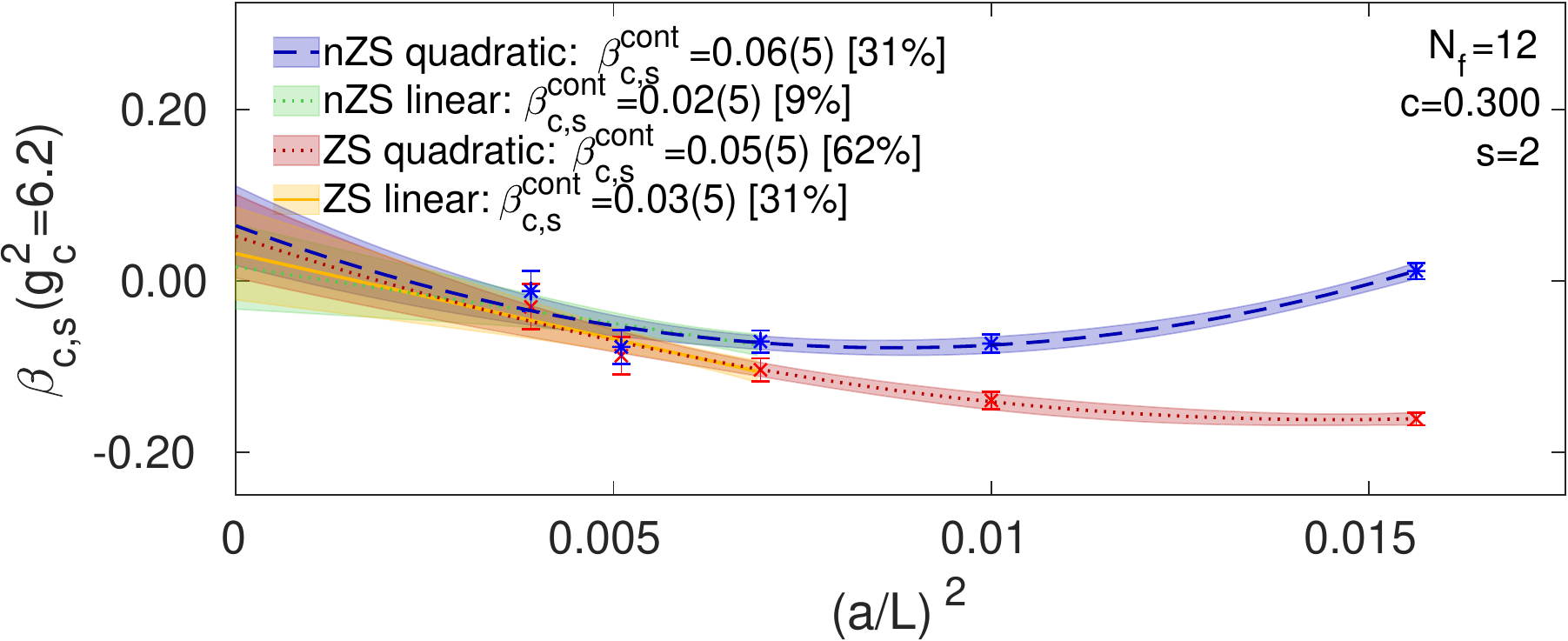}    
  \end{minipage}
  \caption{Discrete step-scaling $\beta$-function in the $c=0.300$ gradient flow scheme for our preferred nZS (left) and ZS (right) data sets. The symbols in the top row show our results for the finite volume discrete $\beta$ function with scale change  $s=2$. The dashed lines with shaded error bands in the same color of the data points show the interpolating fits. We perform two continuum extrapolations: a linear fit in $a^2/L^2$ to the three largest volume pairs (black line with gray error band) and a quadratic fit in $a^2/L^2$ to all five volume pairs (gray dash-dotted line). The $p$-values of the continuum extrapolation fits are shown in the plots in the second row. Further details of the continuum extrapolation at selected $g_c^2$ values are presented in the small panels at the bottom where the legend lists the extrapolated values in the continuum limit with $p$-values in brackets. Only statistical errors are shown.}
  \label{Fig.beta_c300}
\end{figure*}  
Following Eq.~(\ref{eq:beta}), we calculate the discrete $\beta_{c,s}$ functions for five different volume pairs with scale change $s=2$. The resulting values are denoted by the colored data symbols in the plots in the top panels of Figs.~\ref{Fig.beta_c250}--\ref{Fig.beta_c300}. By simulating a set of bare couplings $\beta$ for all volumes, we directly obtain these statistically independent data points.

We interpolate values for each lattice volume pair using a polynomial Ansatz motivated by the perturbative expansion
\begin{align}
\beta_{c,s}(g_c^2;L) = \sum_{i=0}^{n} b_i g_c^{2i}.
\label{eq:fit_form}
\end{align}
In practice we find that $n=3$ is sufficient to describe our data  and obtain fits with good $p$-value. In the case of nZS data, discretization effects are sufficiently small  that we omit the constant term with coefficient $b_0$. Looking at the data, it does not seem justified to force a zero intercept for ZS although numerically our fit does not resolve $b_0$. We list the results of our interpolation in Table \ref{interpolations} where we also quote the $\chi^2$/dof as well as the $p$-value of the fits. The $p$-values reflect that the fitted lines including statistical uncertainties pass through almost all data points. Hence the fit Ansatz Eq.~(\ref{eq:fit_form}) leads to a good description of our data. Only the $8\to 16$ volume pair corresponding to the smallest volumes used exhibits a low $p$-value around 5\%. This data set suffers most from discretization effects and also exhibits a few ``outliers'' in the strong coupling limit which cause a larger $\chi^2$ (lower $p$-value).

\begin{table*}[t]
  \caption{Results of the interpolation fits for the five lattice volume pairs for our preferred (n)ZS analysis using renormalization schemes $c=0.250$ (top panel), 0.275 (middle panel), and 0.300 (bottom panel). Since discretization effects are sufficiently small for nZS, we constrain the constant term $b_0=0$ in Eq.~(\ref{eq:fit_form}) and perform fits with 13 degrees of freedom (dof) or in case of the $8 \to 16$ volume pair with 14 dof. For ZS the intercept $b_0$ is fitted and we have 12 and 13 dof, respectively. In addition we list the $\chi^2/\text{dof}$ as well as the $p$-value. }
  \label{interpolations}
  \begin{tabular}{c@{~~~~}cccccc@{~~~~}cccccc}
    \hline \hline
    & \multicolumn{5}{c}{nZS} && \multicolumn{6}{c}{ZS} \\
    \cline{2-6} \cline{8-13}
       & $\chi^2$/dof & $p$-val. & $b_3$ & $b_2$ & $b_1$ && $\chi^2$/dof & $p$-val. &  $b_3$ & $b_2$ & $b_1$ & $b_0$\\ \hline
    $8\to 16$  & 1.738 & 0.042 & -0.00220(20) & 0.0188(14) & -0.0024(24) &&1.687 &0.056 &-0.00136(42) &0.0124(47) &-0.061(16) &-0.011(17)\\
    $10\to 20$ & 1.385 & 0.157 & -0.00269(22) & 0.0162(17) & -0.0013(29) &&1.407 &0.154 &-0.00281(55) &0.0198(60) &-0.045(20) &0.016(19) \\
    $12\to 24$ & 1.507 & 0.106 & -0.00269(23) & 0.0147(18) & -0.0021(32) &&1.591 &0.086 &-0.00290(67) &0.0181(72) &-0.028(24) &0.013(23) \\
    $14\to 28$ & 0.905 & 0.547 & -0.00267(34) & 0.0140(26) & -0.0002(45) &&0.971 &0.474 &-0.0023(10)  &0.011(11)  &0.002(34)  &-0.009(31) \\
    $16\to 32$ & 0.875 & 0.579 & -0.00209(43) & 0.0106(32) &  0.0045(53) &&0.933 &0.512 &-0.0025(12)  &0.015(13)  &-0.016(43) &0.016(41)\\
    \hline
    $8\to 16$  &1.620 & 0.066 &-0.00283(23) &0.0201(17) &-0.0040(29) &&1.660 &0.062 &-0.00221(53) &0.0165(59) &-0.048(20) &-0.003(20)\\
    $10\to 20$ &1.344 & 0.179 &-0.00278(27) &0.0159(21) &-0.0013(35) &&1.410 &0.153 &-0.00291(70) &0.0190(75) &-0.033(25) &0.013(24) \\
    $12\to 24$ &1.439 & 0.132 &-0.00242(29) &0.0126(23) &0.0008(40) &&1.547 &0.100 &-0.00248(86) &0.0140(93) &-0.014(30) &0.005(29)\\
    $14\to 28$ &0.815 & 0.644 &-0.00269(45) &0.0143(34) &0.0001(59) &&0.857 &0.592 &-0.0019(13) &0.007(14)  &0.019(44) &-0.023(41) \\
    $16\to 32$ &0.863 & 0.593 &-0.00188(56) &0.0094(41) &0.0073(68) &&0.920 &0.525 &-0.0024(16) &0.016(17) &-0.017(54) &0.021(52) \\
    \hline
    $8\to 16$  &1.709 & 0.047 &-0.00319(27) &0.0208(20) &-0.0046(34) && 1.787 &0.039 &-0.00273(64) &0.0186(71) &-0.036(24) &-0.001(23) \\
    $10\to 20$ &1.274 & 0.220 &-0.00282(33) &0.0158(25) &-0.0011(43) && 1.365 &0.174 &-0.00279(87) &0.0166(94) &-0.019(31) &0.005(30)\\
    $12\to 24$ &1.423 & 0.140 &-0.00212(37) &0.0106(28) &0.0040(50) && 1.535 &0.103 &-0.0020(11) &0.010(12) &-0.002(38) &-0.000(37)\\
    $14\to 28$ &0.732 & 0.732 &-0.00278(56) &0.0153(43) &-0.0002(75) &&0.754 &0.699 &-0.0017(17) &0.004(18) &0.034(56) &-0.035(52) \\
    $16\to 32$ &0.884 & 0.570 &-0.00168(70) &0.0085(52) &0.0103(84) &&0.930 &0.515 &-0.0027(20) &0.020(21) &-0.030(67) &0.036(64)\\
    \hline \hline
  \end{tabular}
\end{table*}

The interpolating fits  predict finite volume discrete step-scaling functions $\beta_{c,s}(g_c^2; L)$. These discrete step-scaling functions are shown in the top row panels of Figs.~\ref{Fig.beta_c250}--\ref{Fig.beta_c300} by dashed lines with shaded error band in the same color as the corresponding data points. The interpolation for nZS starts at $g_c^2=0$ due to the constraint $b_0=0$ of the interpolating polynomial,  whereas for ZS it begins with our data at the weakest coupling. The upper end of the interpolation depends on the range where we have data and varies slightly depending on the chosen scheme $c$.

Subsequently, we perform infinite volume continuum limit extrapolations using the interpolated finite-volume $\beta$-functions $\beta_{c,s}(g_c^2; L)$, which are continuous in $g_c^2$. We consider two Ans\"atze for the extrapolation
  \begin{itemize}
  \item a linear fit in $a^2/L^2$ to the three largest volume pairs,
  \item a quadratic fit in $a^2/L^2$ to all five volume pairs,
  \end{itemize}
  which are motivated by the expected ${\cal O}(a^2)$ discretization effects in the weak coupling limit. Similar to the use of tree-level normalization, the validity of an extrapolation proportional to $a^2/L^2$ is limited to sufficiently weak couplings. Moreover correction terms of higher order may only be resolved if the statistical uncertainties are small enough. We therefore monitor the $p$-value of the extrapolation fits as a criteria to judge the validity of the continuum limit extrapolations. $p$-values are shown as a function of $g_c^2$ in the plots in the second row panels of Figs.~\ref{Fig.beta_c250}--\ref{Fig.beta_c300}. In addition we show details of the continuum limit extrapolations at $g_c^2=2.0$, 3.4, 4.8, and 6.2 in the four smaller plots at the bottom of the three figures. The resulting continuum step-scaling functions are shown in the top row plots by the solid black line with gray error band (linear fit) and the gray dash-dotted line (quadratic fit).

Overall most extrapolation exhibit excellent $p$-values over the full range in $g_c^2$ covered by our simulations and corrections to an $a^2/L^2$ extrapolation seem to be small. Performing a linear extrapolation in $a^2/L^2$ to our three largest volume pairs is well justified for all data sets although a noticeable drop of the $p$-value can be observed for nZS data around $g_c^2\sim 4.5$ for all three $c$ values considered. However, even for our strongest couplings the linear fit still gives a good $p$-value greater than 10\%. In case of the ZS data, the $p$-values for linear extrapolations are mostly well above the 20\% level. They do however show a similar behavior like in case of the nZS fits with lower values around $g_c^2\sim 2.9$ and 5.5.  The quality of the quadratic extrapolation based on all five lattice volume pairs is strongly affected by our smallest $8\to 16$ volume pair.  Discretization effects for the $8\to 16$ data set are quite significant for $c=0.25$ but decrease for $c=0.275$ and 0.3. This corresponds to the well-known property that smaller $c$ values require larger lattice volumes~\cite{Fritzsch:2013je}. Using tree-level normalization, this effect is attenuated. For nZS, the $p$-value of the quadratic extrapolation even at $c=0.250$ remains at the acceptable 4--5\% level for the full range in $g_c^2$ covered, whereas the quadratic extrapolation for ZS at $c=0.250$ exhibits vanishing $p$-values around $g_c^2\sim 5.5$.

Looking more closely at the details of the extrapolation plots presented at the bottom of Figs.~\ref{Fig.beta_c250}--\ref{Fig.beta_c300}, we observe that the nZS data for the three largest lattice volume pairs exhibit very little $a^2/L^2$ dependence for all $c$ values and in fact can also be described by fitting only a constant \cite{Hasenfratz:2017qyr}. For the smaller volumes at stronger couplings, corrections to $a^2/L^2$ arise and the data show an upward curvature. In contrast to that the large volume ZS data clearly exhibit a slope in $a^2/L^2$ for all $c$-values. An explanation could be that corrections at weak coupling are caused by higher order terms which can essentially be removed by the tree-level normalization, whereas the corrections at strong coupling indicate that perturbative improvement and extrapolations in $a^2/L^2$ are subject to nonperturbative corrections.

%=================================================
\subsection{Alternative flow/operators}
\label{Sec.Alt}

\begin{figure*}[t]
  \includegraphics[width=0.99\columnwidth]{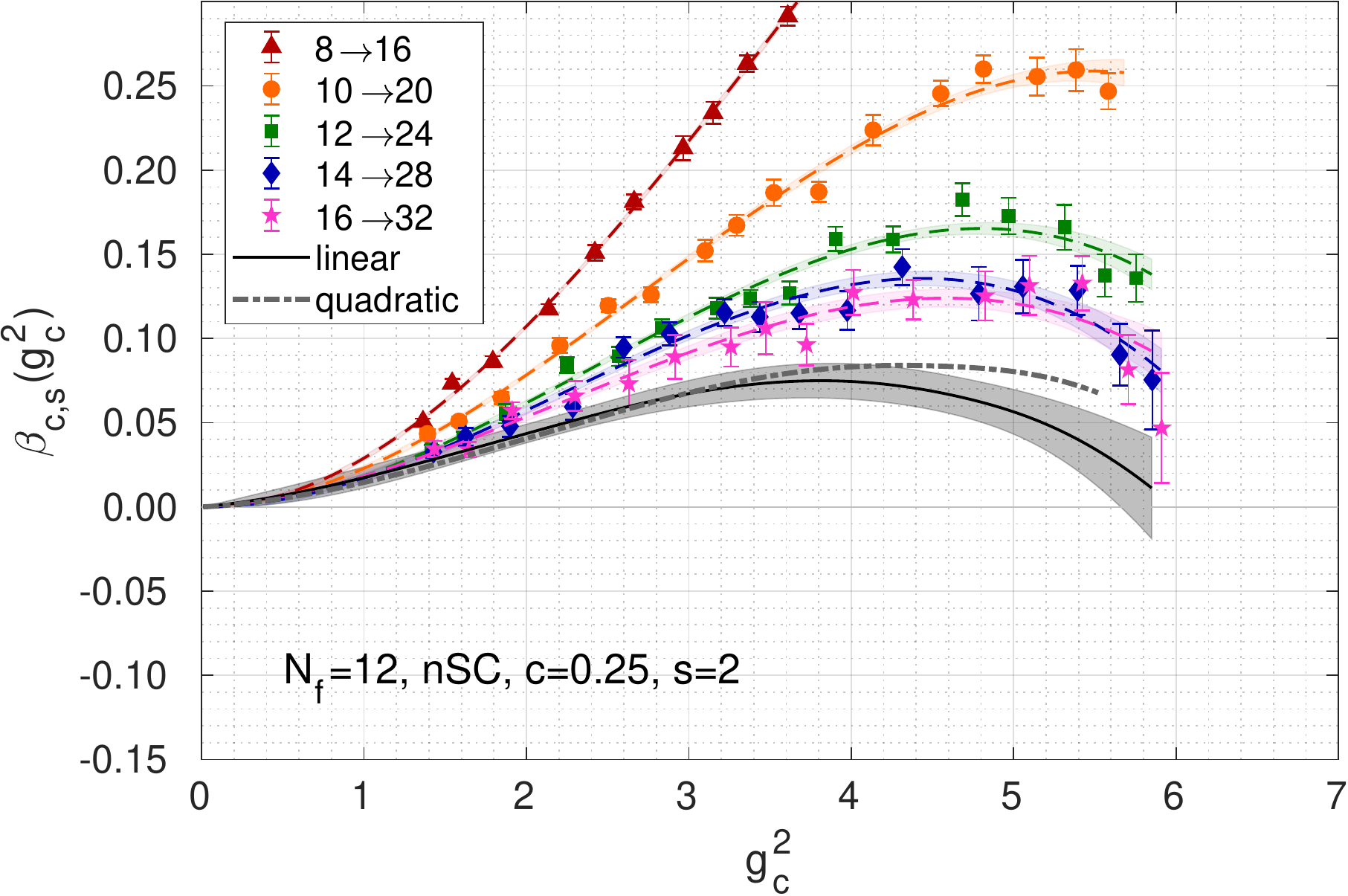}\hfill
  \includegraphics[width=0.99\columnwidth]{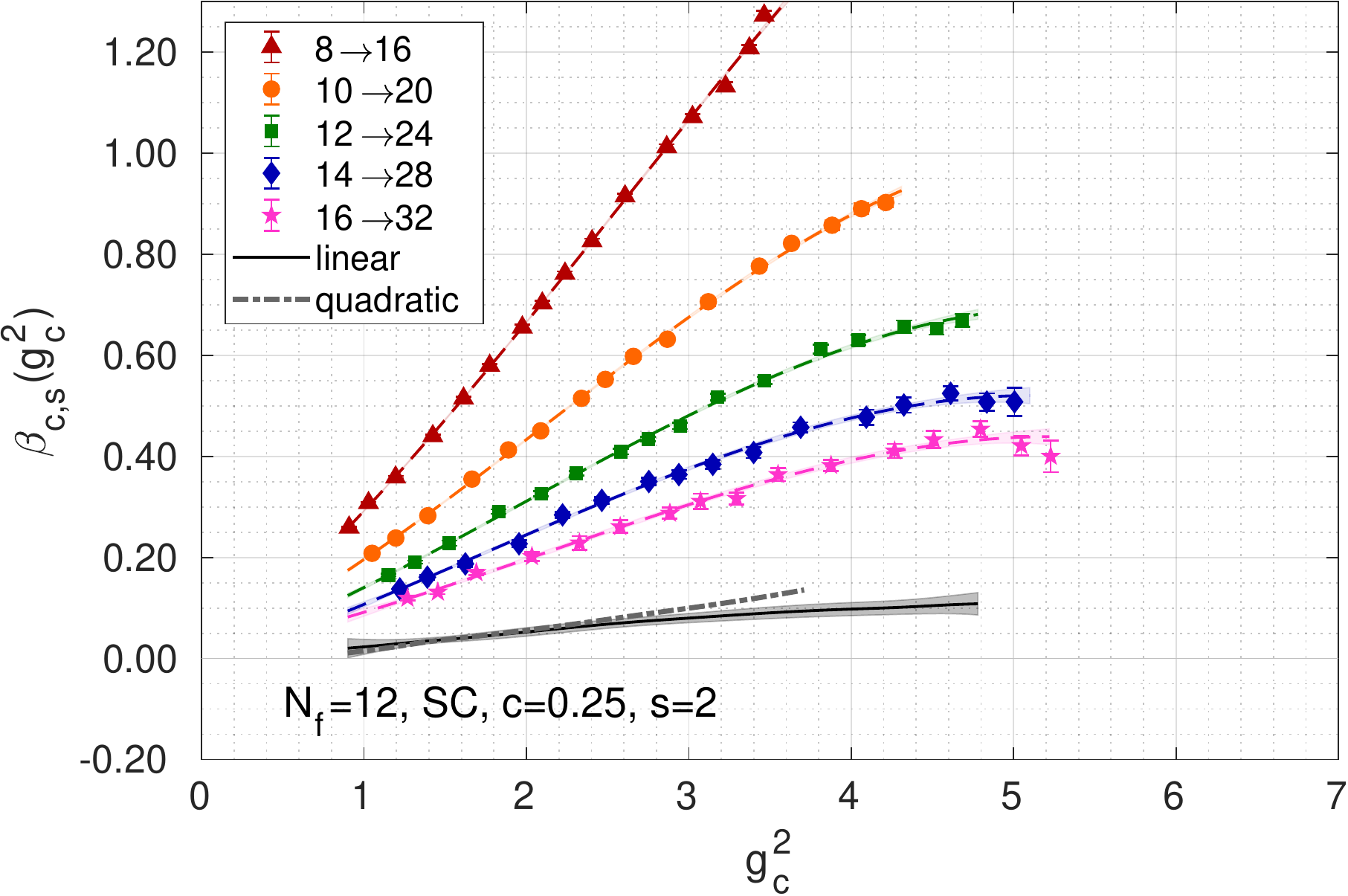}\\
  \includegraphics[width=0.99\columnwidth]{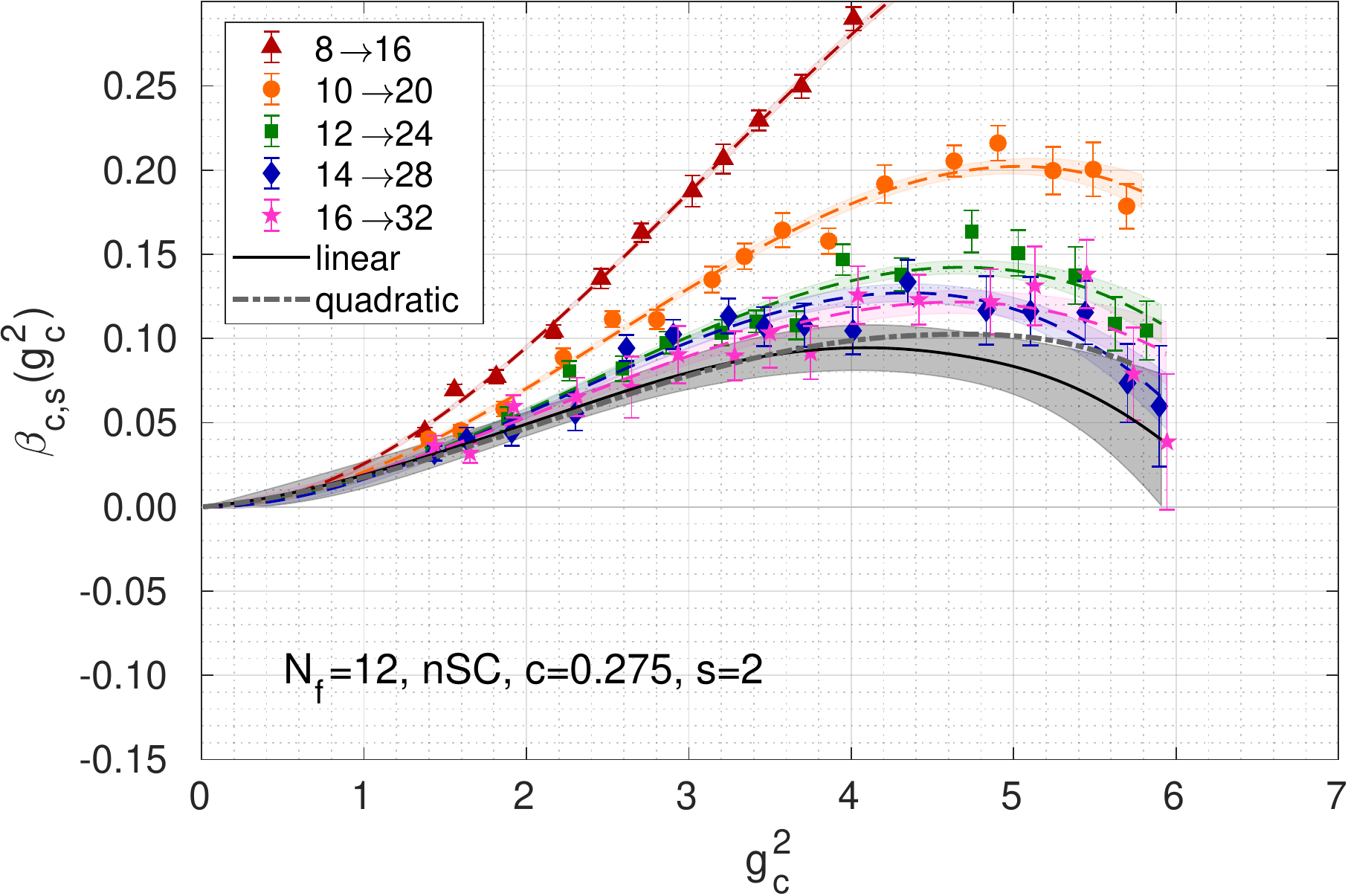}\hfill
  \includegraphics[width=0.99\columnwidth]{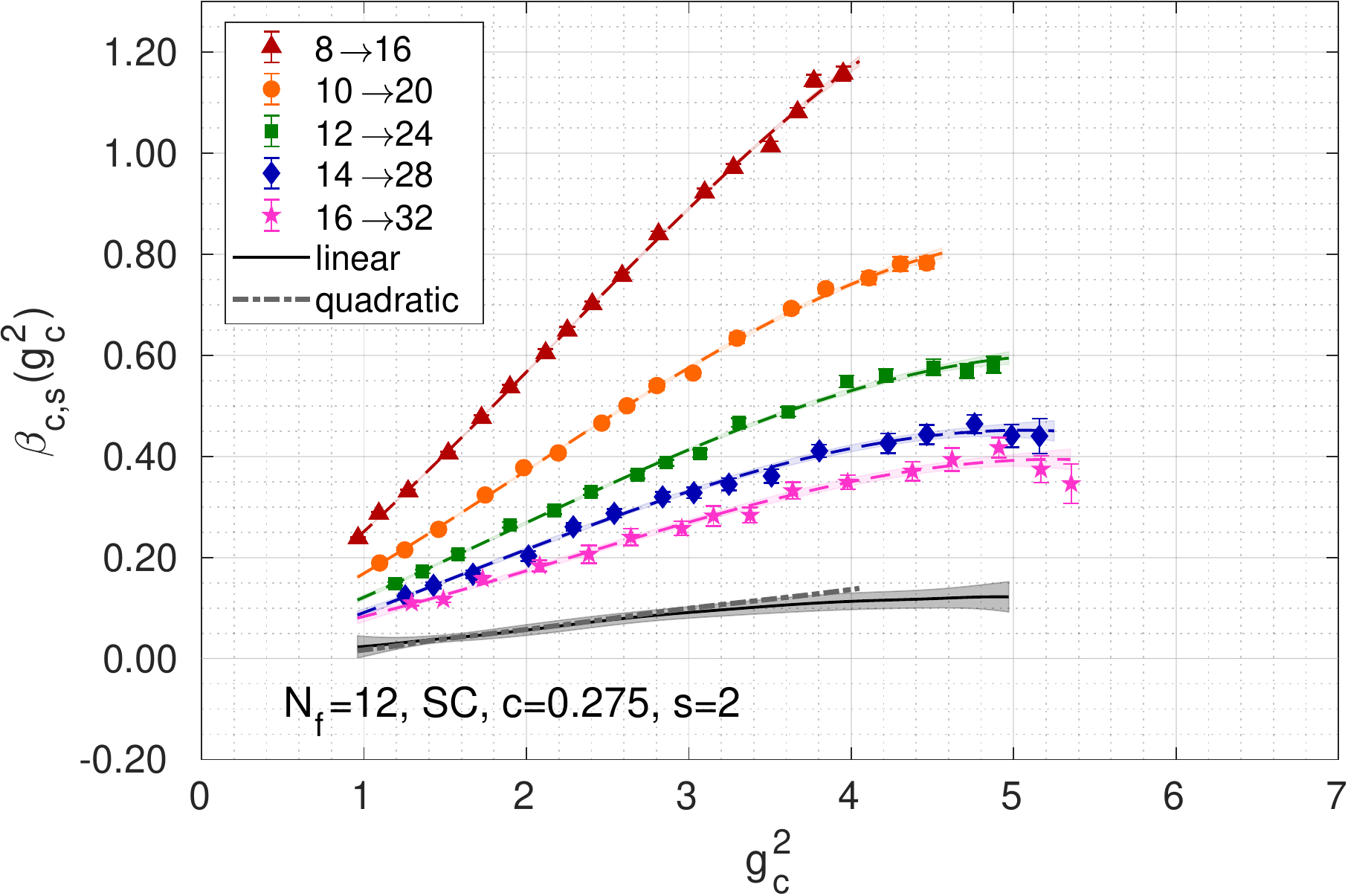}\\
  \includegraphics[width=0.99\columnwidth]{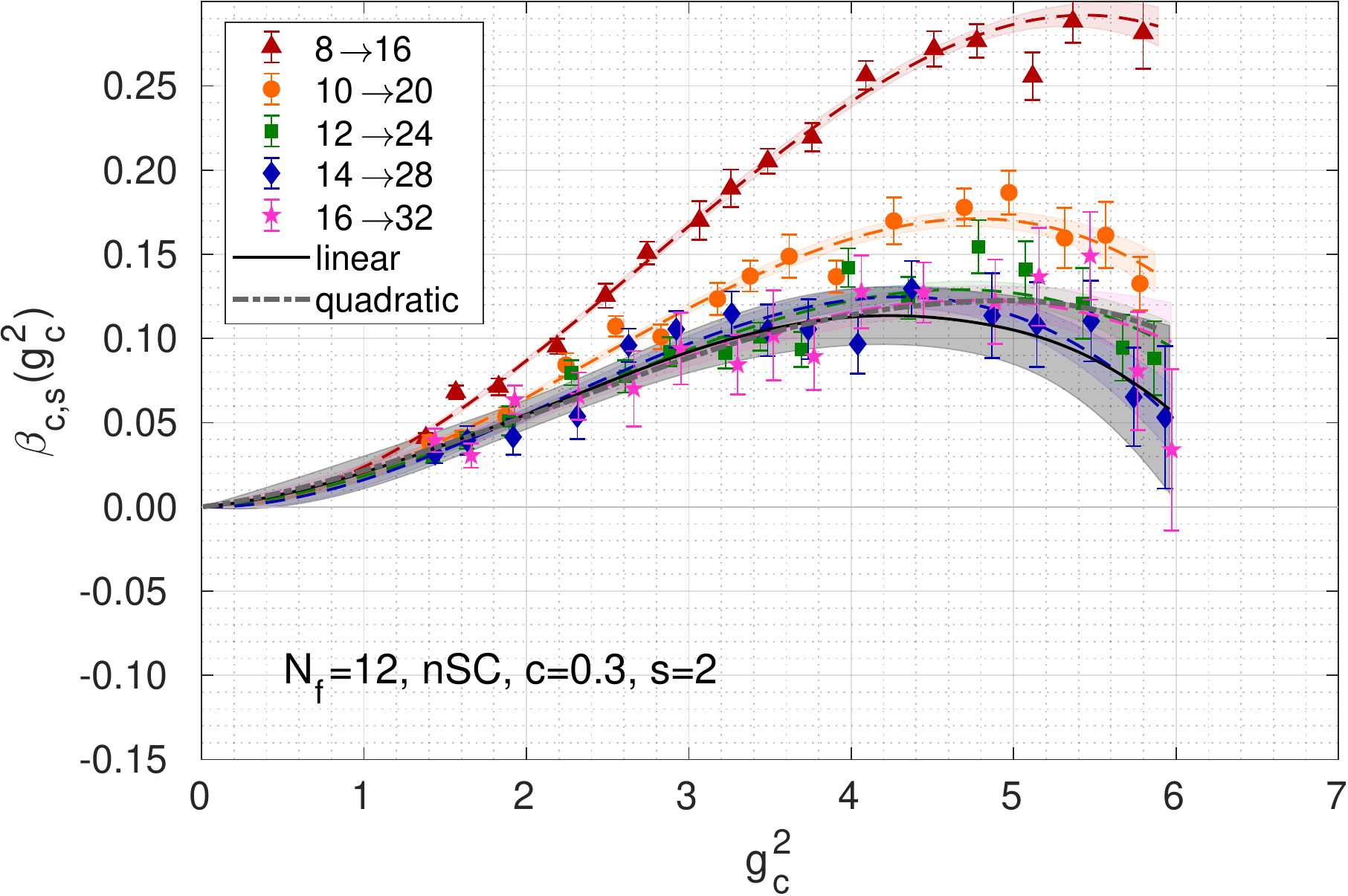}\hfill
  \includegraphics[width=0.99\columnwidth]{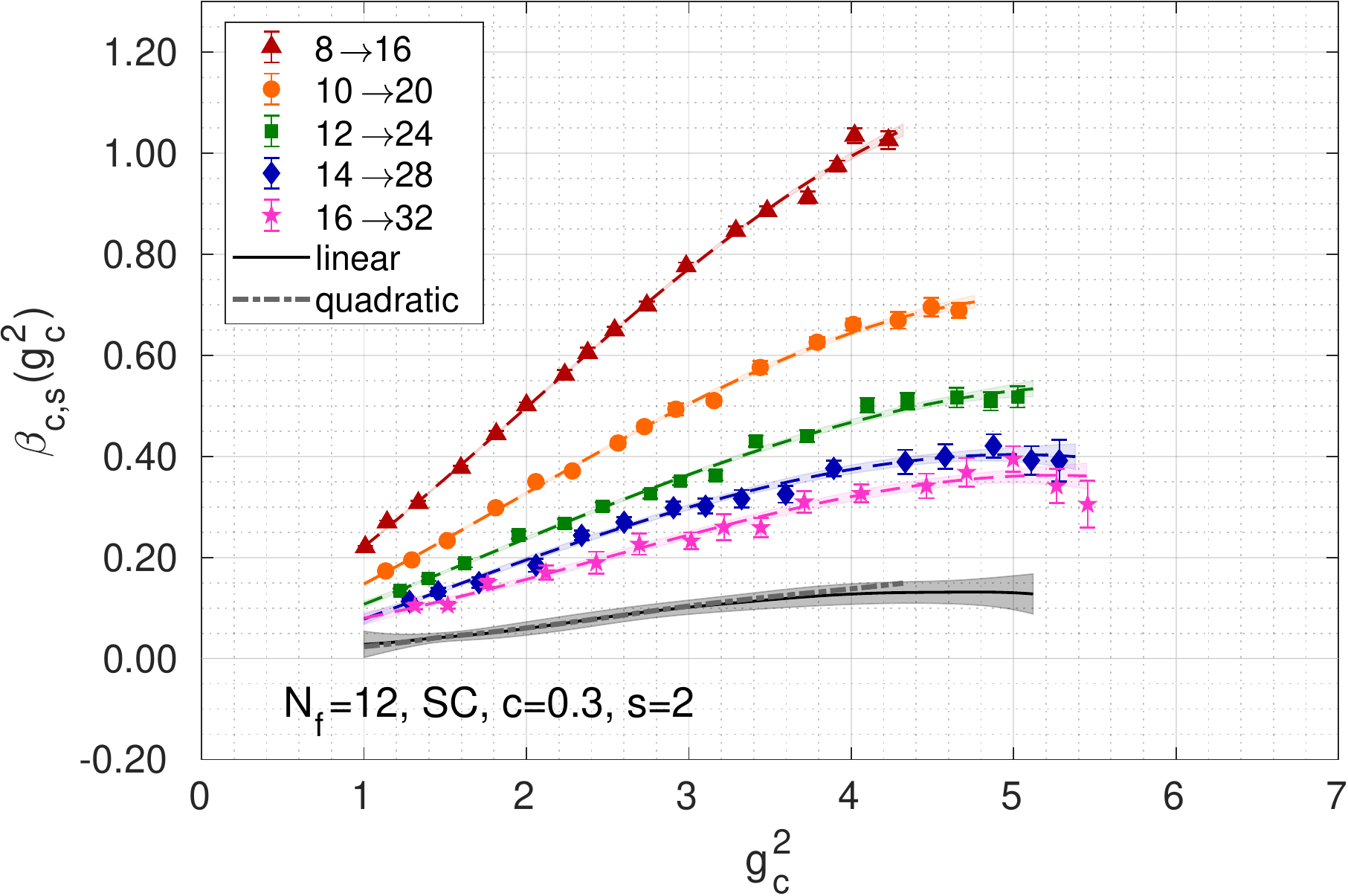}  
  \caption{Alternative determination of the discrete $\beta$ function using Symanzik flow with clover operator. Plots on the left show the analysis including the tree-level improvement (nSC), plots on the right without (SC). From top to bottom we present results for the renormalization scheme $c=0.250$, 0.275, and 0.30. Only statistical errors are shown.}
  \label{Fig.beta_alt_SC}
\end{figure*}

\begin{figure*}[t]
  \includegraphics[width=0.99\columnwidth]{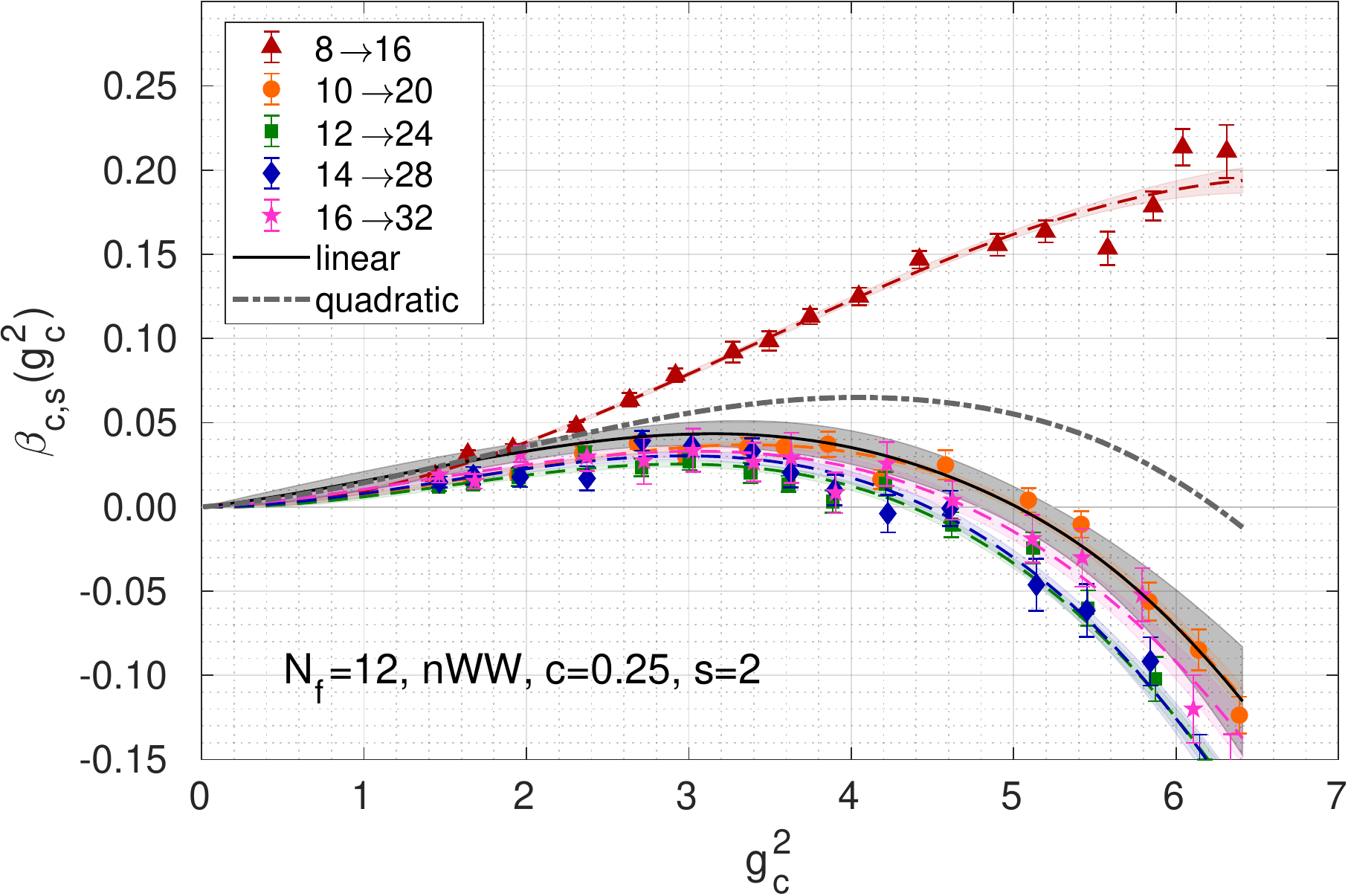}\hfill
  \includegraphics[width=0.99\columnwidth]{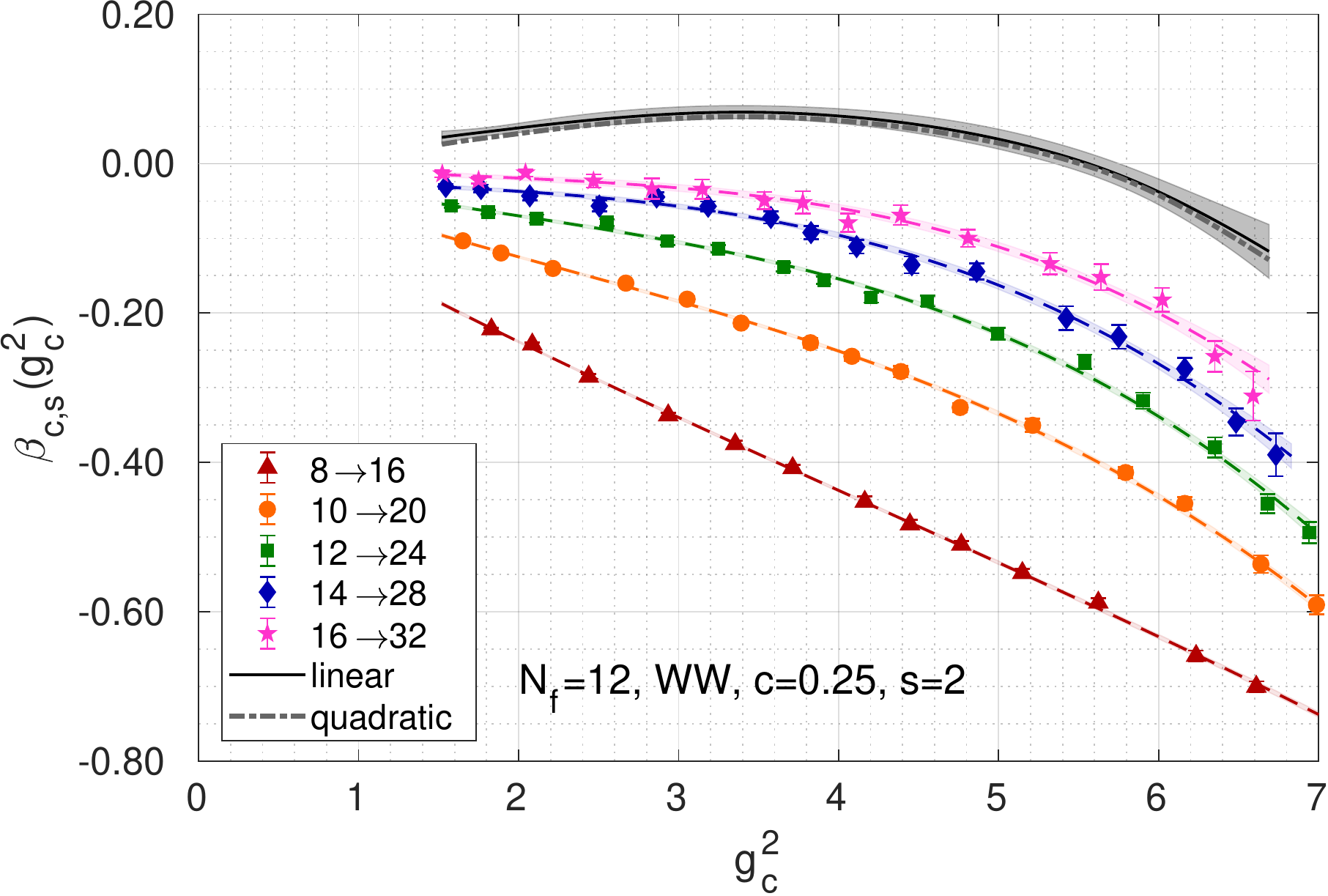}\\
  \includegraphics[width=0.99\columnwidth]{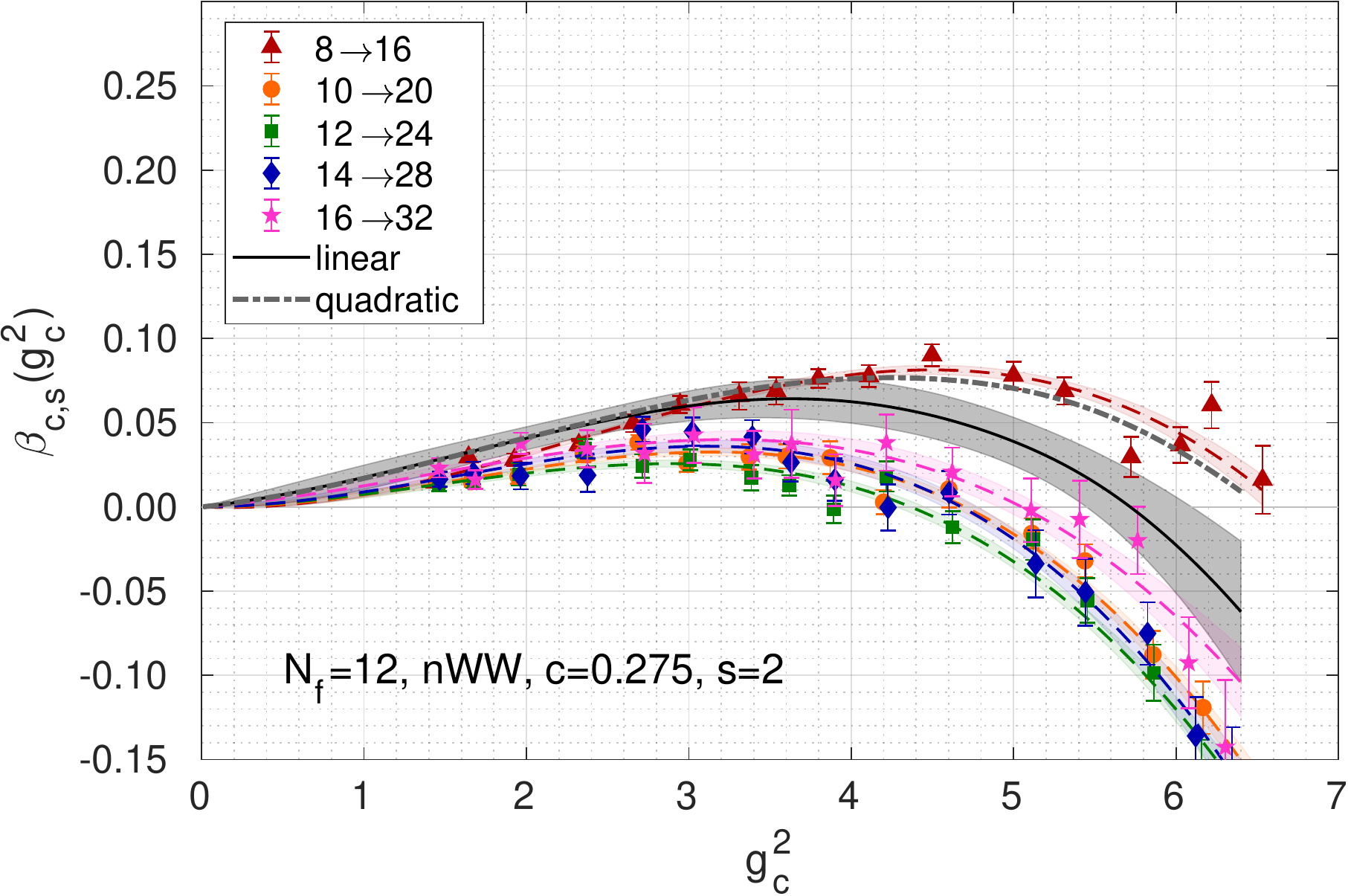}\hfill
  \includegraphics[width=0.99\columnwidth]{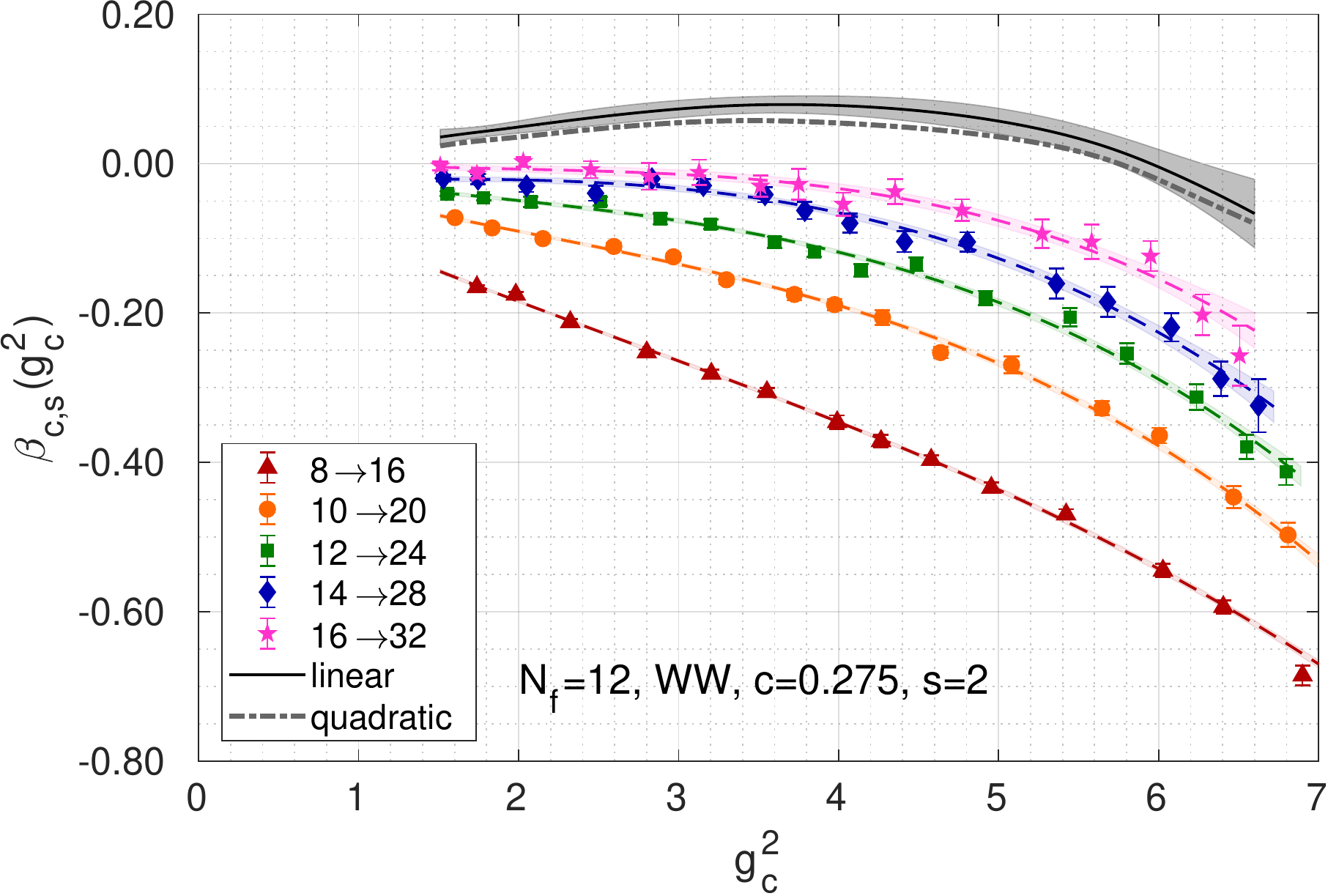}\\
  \includegraphics[width=0.99\columnwidth]{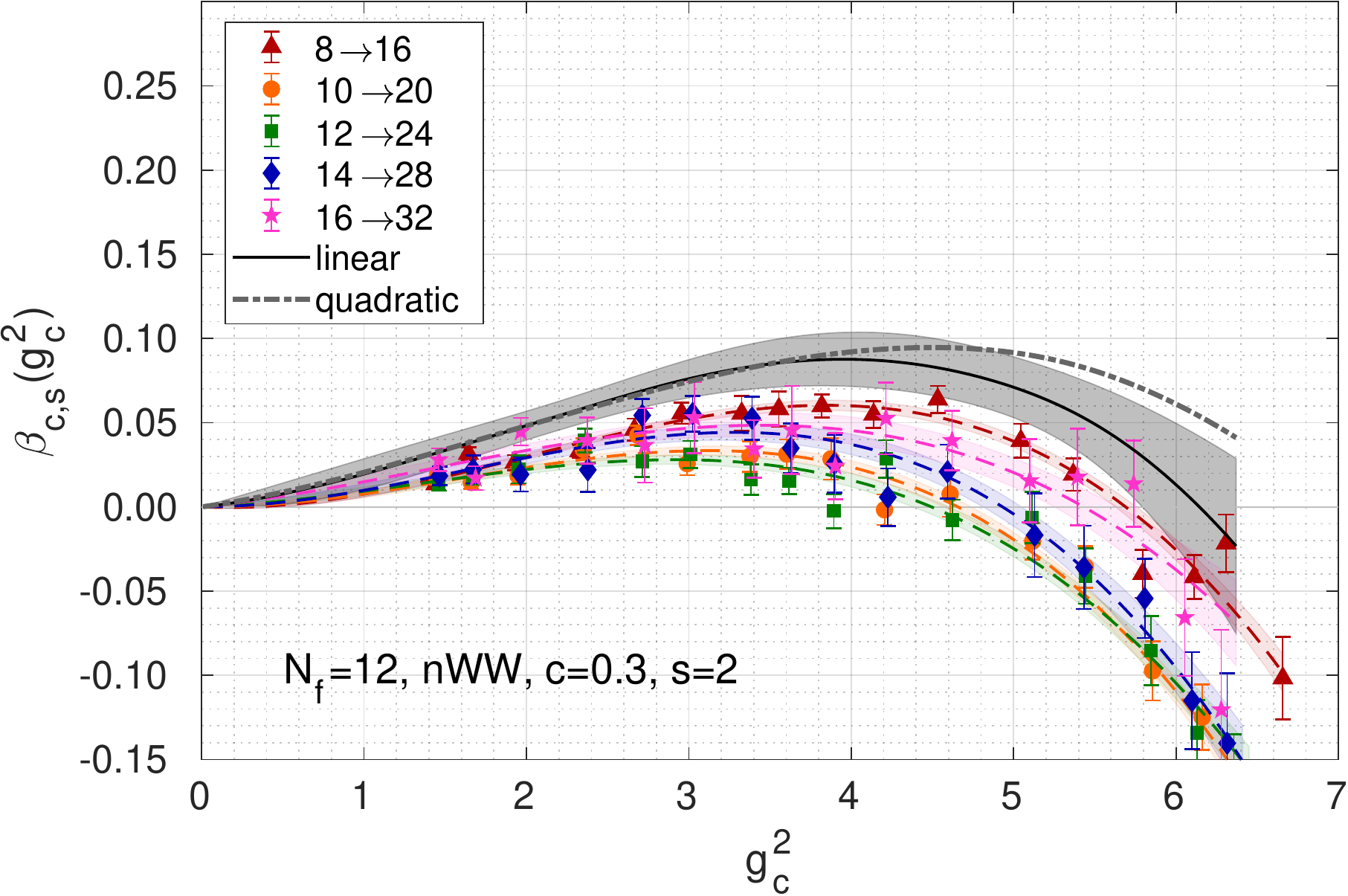}\hfill
  \includegraphics[width=0.99\columnwidth]{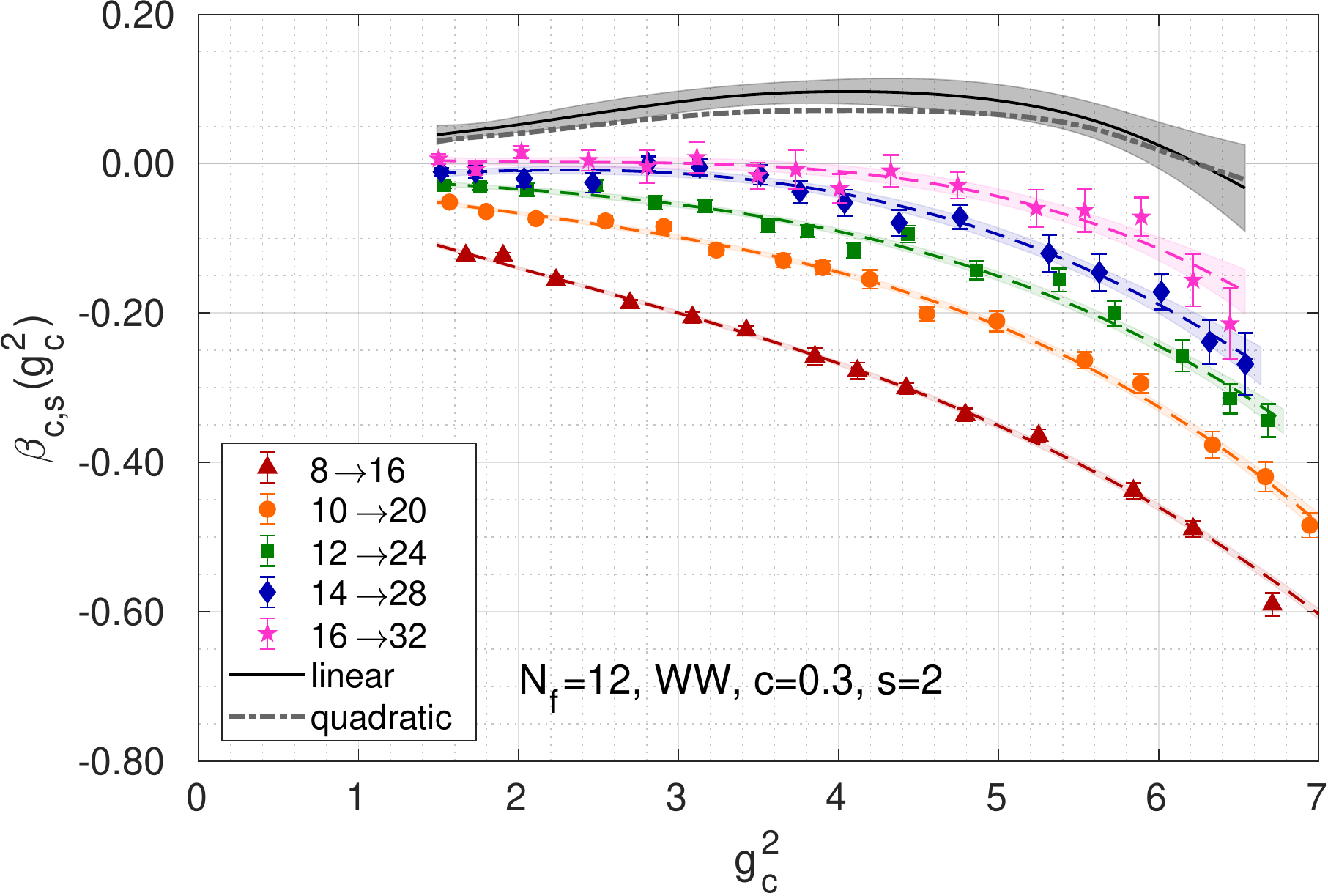}  
  \caption{Alternative determination of the discrete $\beta$ function using Wilson flow with Wilson plaquette operator. Plots on the left show the analysis including the tree-level improvement (nWW), plots on the right without (WW). From top to bottom we present results for the renormalization scheme $c=0.250$, 0.275, and 0.30. Only statistical errors are shown.}
  \label{Fig.beta_alt_WW}
\end{figure*}

In order to estimate systematic effects, we consider the step-scaling $\beta$ function obtained with different flows (Wilson and Symanzik flow) as well as different operators (Wilson plaquette and clover). We repeat the same steps of the analysis as for our preferred (n)ZS data i.e.~we first obtain data points for the discrete $\beta$ functions for our five lattice volume pairs and next perform an interpolating fit. Again we constrain $b_0$ to be zero when using tree-level improvement but fit $b_0$ otherwise.  Finally we again carry out an infinite volume continuum limit extrapolation using a linear Ansatz for the three largest volume pairs and a quadratic Ansatz using all five volume pairs. As examples for our alternative determinations, we present in Fig.~\ref{Fig.beta_alt_SC} the determination for the step-scaling $\beta$ function using Symanzik flow and the clover operator and in Fig.~\ref{Fig.beta_alt_WW} Wilson flow with the Wilson plaquette operator. The plots on the left show the analysis with tree-level normalization factors, the plots on the right the analysis without tree-level improvement. Again we show results (top to bottom) for $c=0.250$, 0.275, and 0.300. When using tree-level improvement the discretization effects between the different lattice volume pairs are significantly reduced resulting also in a much smaller spread (smaller range of the ordinate) of the data sets. The continuum limit extrapolations are again shown by the solid line with gray error band (linear fit to the three largest volume pairs) and the gray dash-dotted line (quadratic fit to all five volume pairs). Details on the continuum limit extrapolations are presented in Appendix \ref{Sec.ContLimitExtra} in Figs.~\ref{Fig.cont_extra_c0250}--\ref{Fig.cont_extra_c0300} where we show extrapolations for our preferred (n)ZS data in comparison to (n)SC and (n)WW.

%=================================================

%=================================================

While continuum limit predictions for the renormalized $\beta_{c,s}(g_c^2)$ obtained from different flow/operator combinations are expected to agree, the finite volume predictions of $\beta_{c,s}(g_c^2,L)$ are subject to discretization effects and hence  depend on the specific flow/operator combination. At finite volume, different flow/operator combinations predict different renormalized couplings even on the same set of gauge field ensembles evaluated using the same scheme $c$. Only after taking the continuum limit, the obtained $\beta_{c,s}(g_c^2)$  are expected to be free of discretization effects and can be meaningfully compared. The accessible range of $g_c^2$ and $\beta_{c,s}(g_c^2)$ varies for each flow/operator combination at each bare coupling $\beta\equiv 6/g_0^2$ and $(L/a)^4$ volume. How different flow/operator combinations approach the same continuum limit  can be seen by comparing Figs.~\ref{Fig.beta_c250}--\ref{Fig.beta_alt_WW}.\footnote{
Another way to understand this effect is to consider the gradient flow  as a continuous renormalization group (RG) transformation where the continuum limit is reached as the flow approaches the renormalized trajectory (RT)~\cite{Carosso:2018bmz,Hasenfratz:2019hpg}. Since the RT in bare parameter space depends on the RG transformation, different flow/operator combinations approach their corresponding RT differently, predicting different $g^2_c$ on the same ensembles.}

\begin{figure*}[p]
  \includegraphics[height=0.304\textheight]{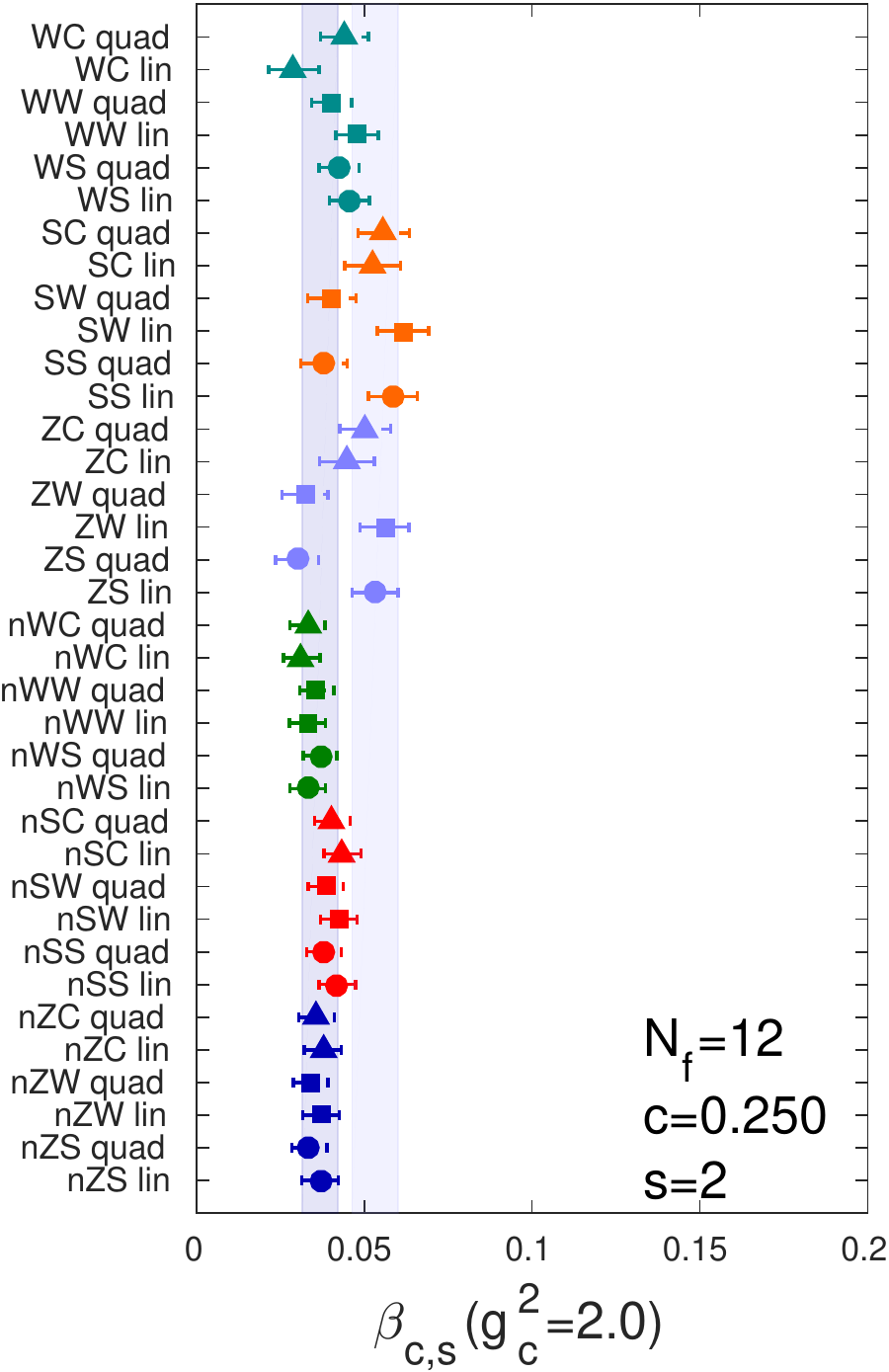}
  \includegraphics[height=0.304\textheight]{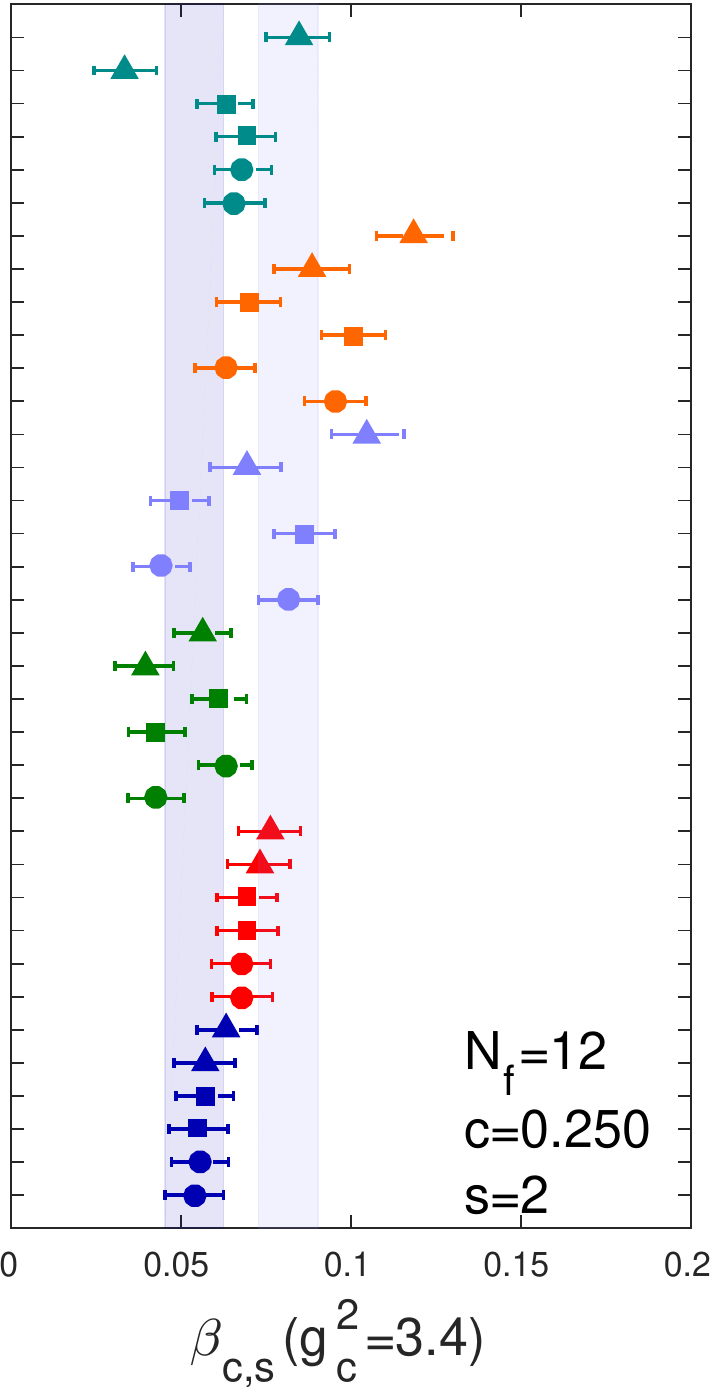}
  \includegraphics[height=0.304\textheight]{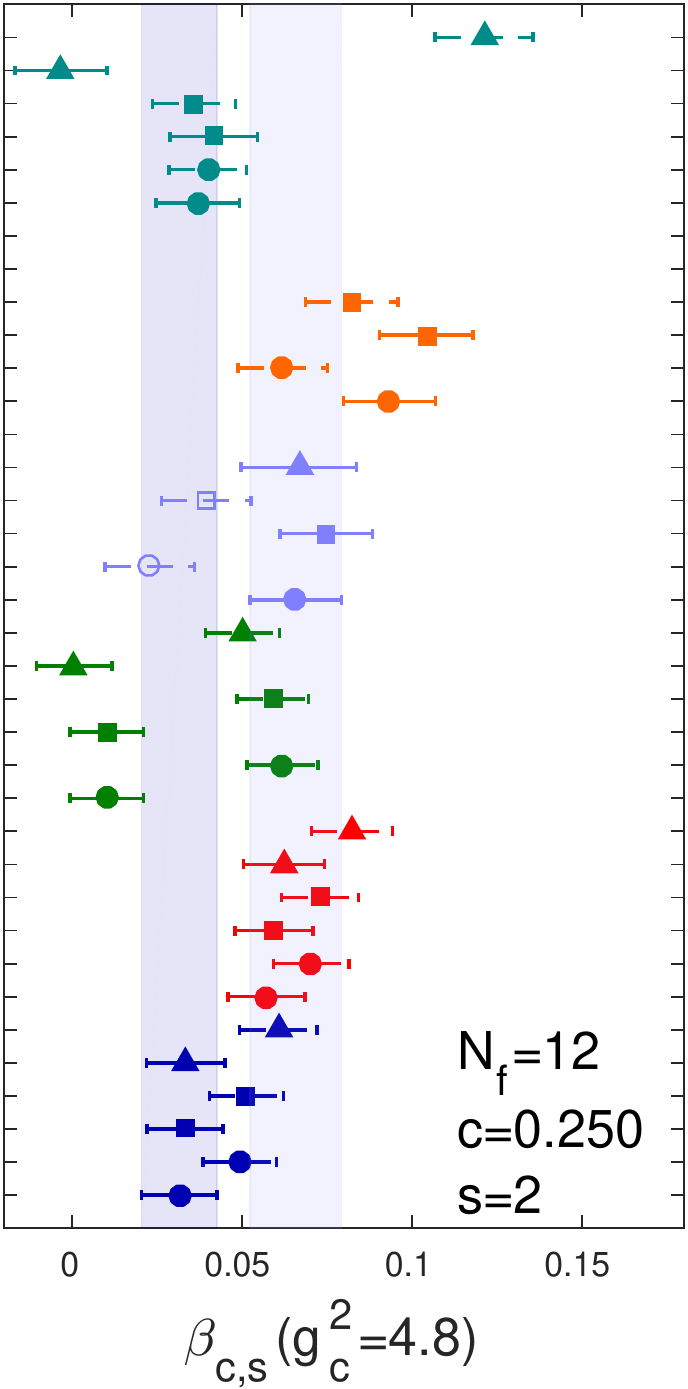}
  \includegraphics[height=0.304\textheight]{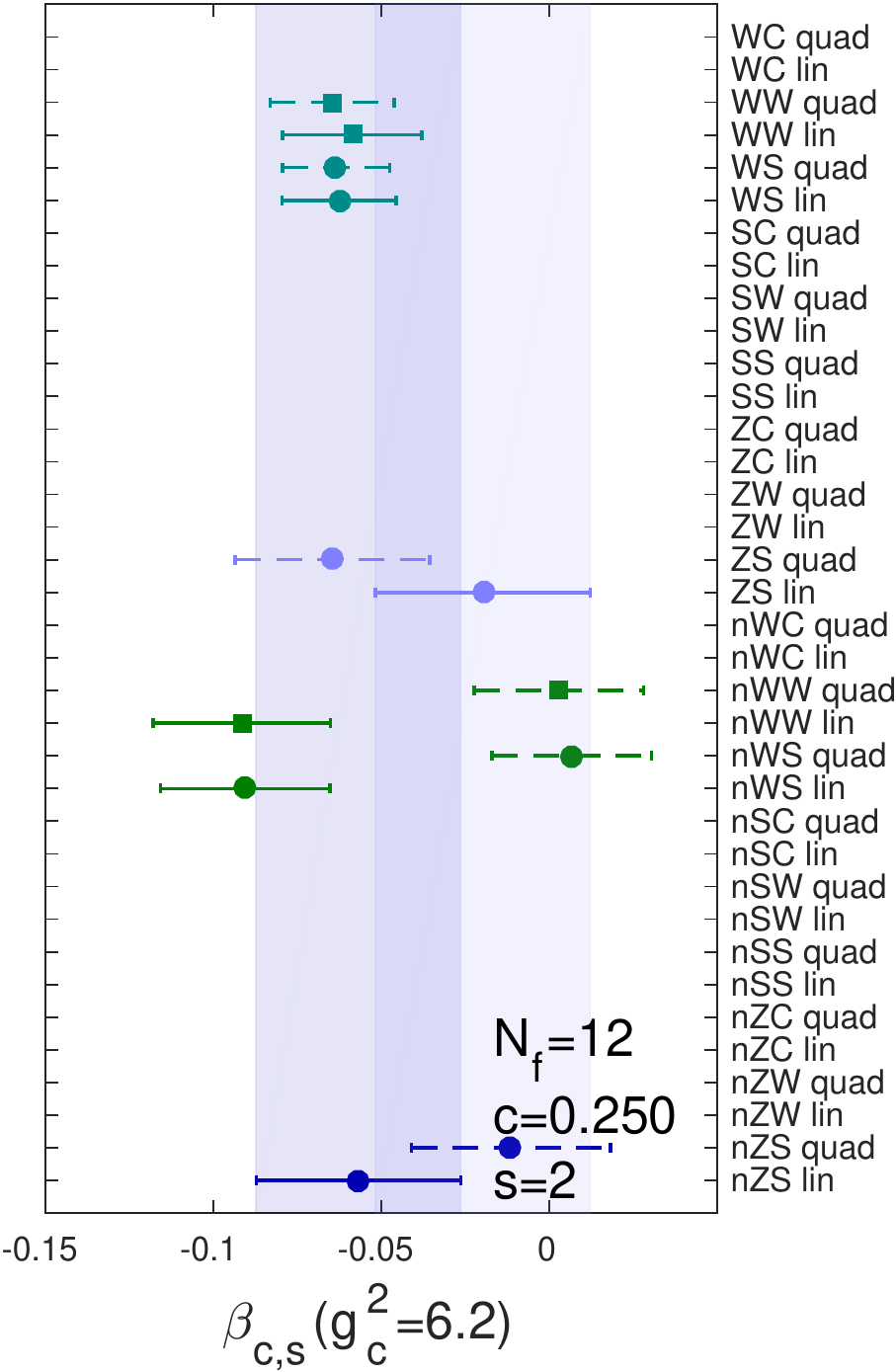}\\
  \includegraphics[height=0.304\textheight]{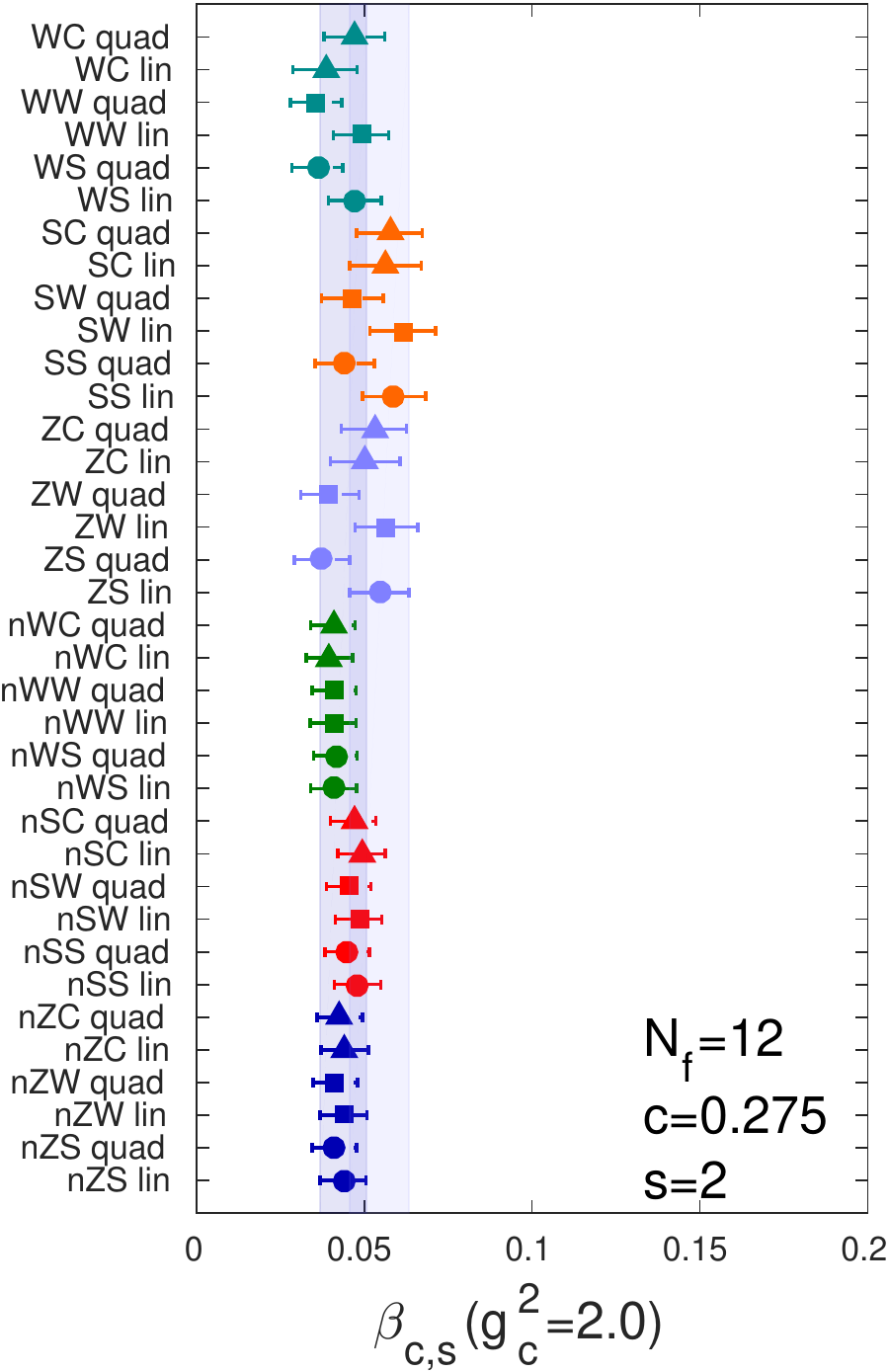}
  \includegraphics[height=0.304\textheight]{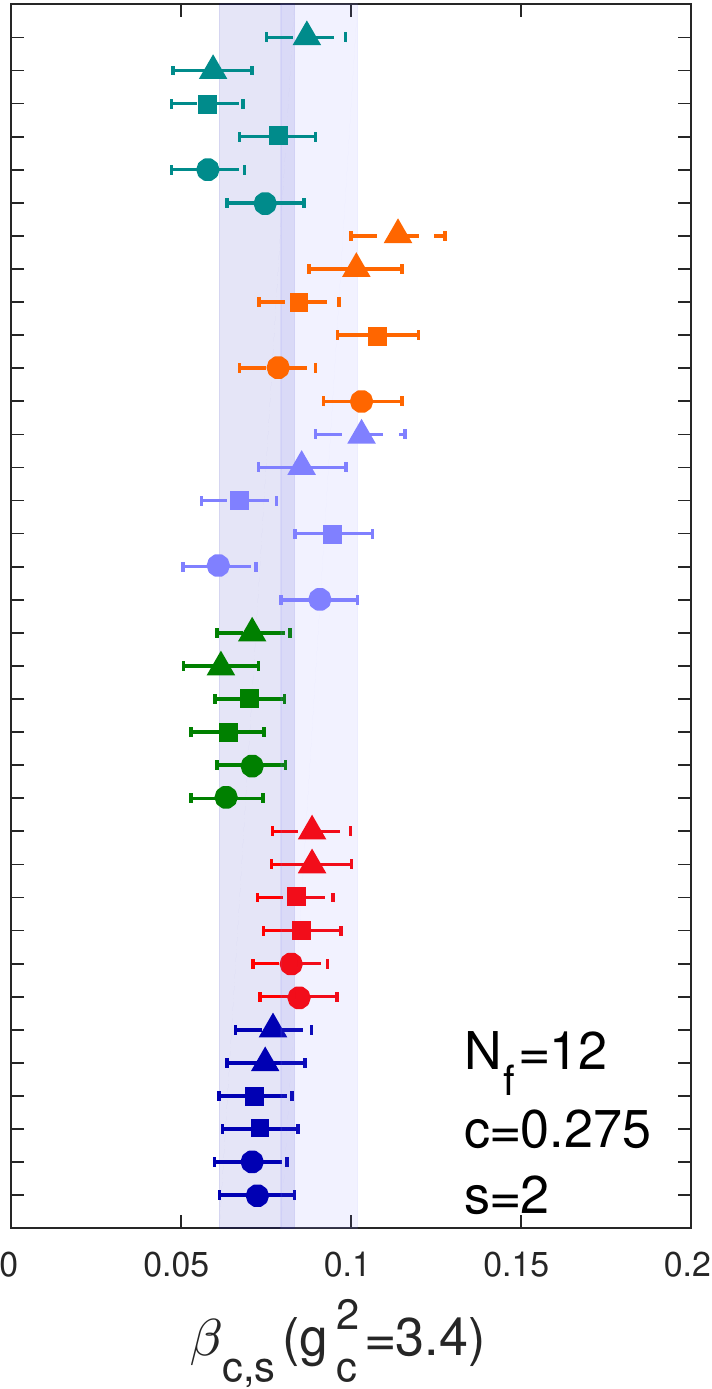}
  \includegraphics[height=0.304\textheight]{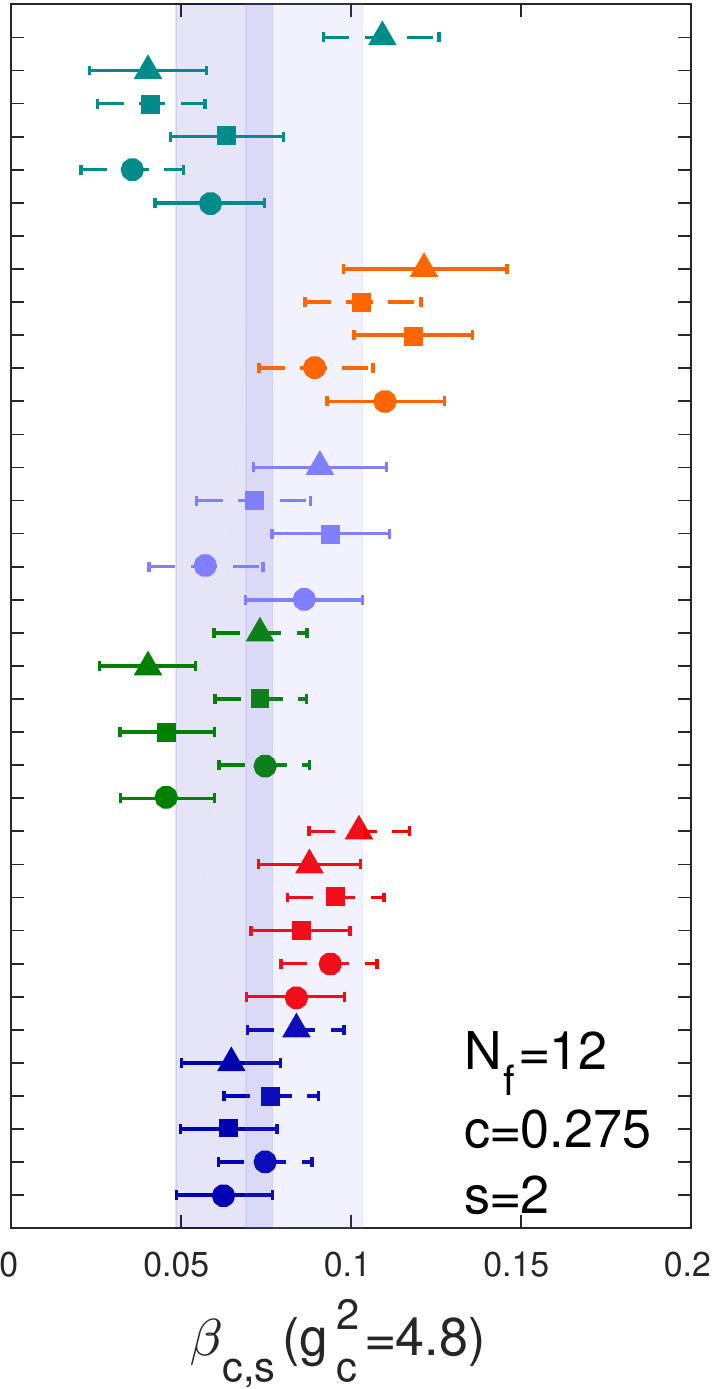}
  \includegraphics[height=0.304\textheight]{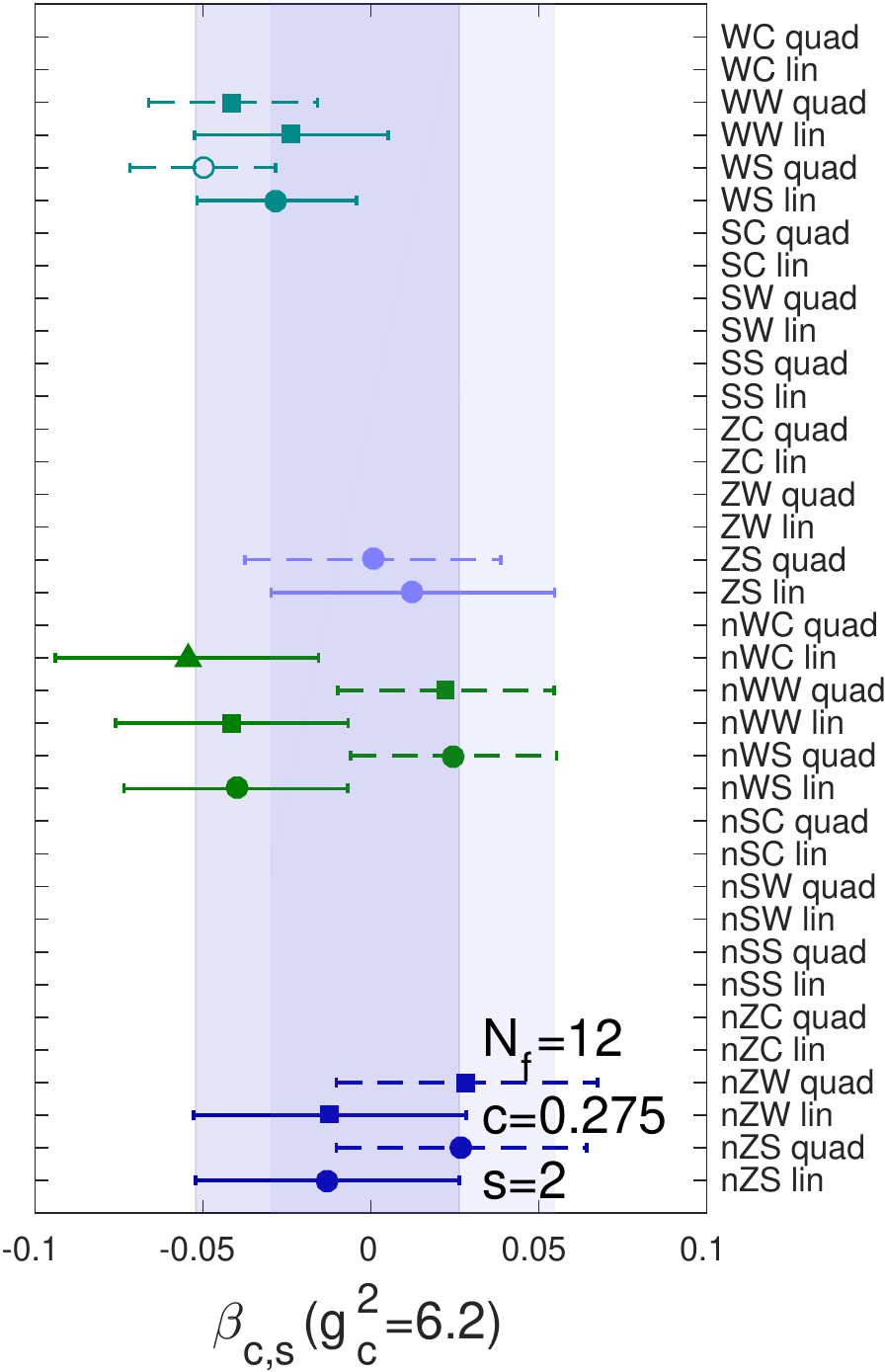}\\
  \includegraphics[height=0.304\textheight]{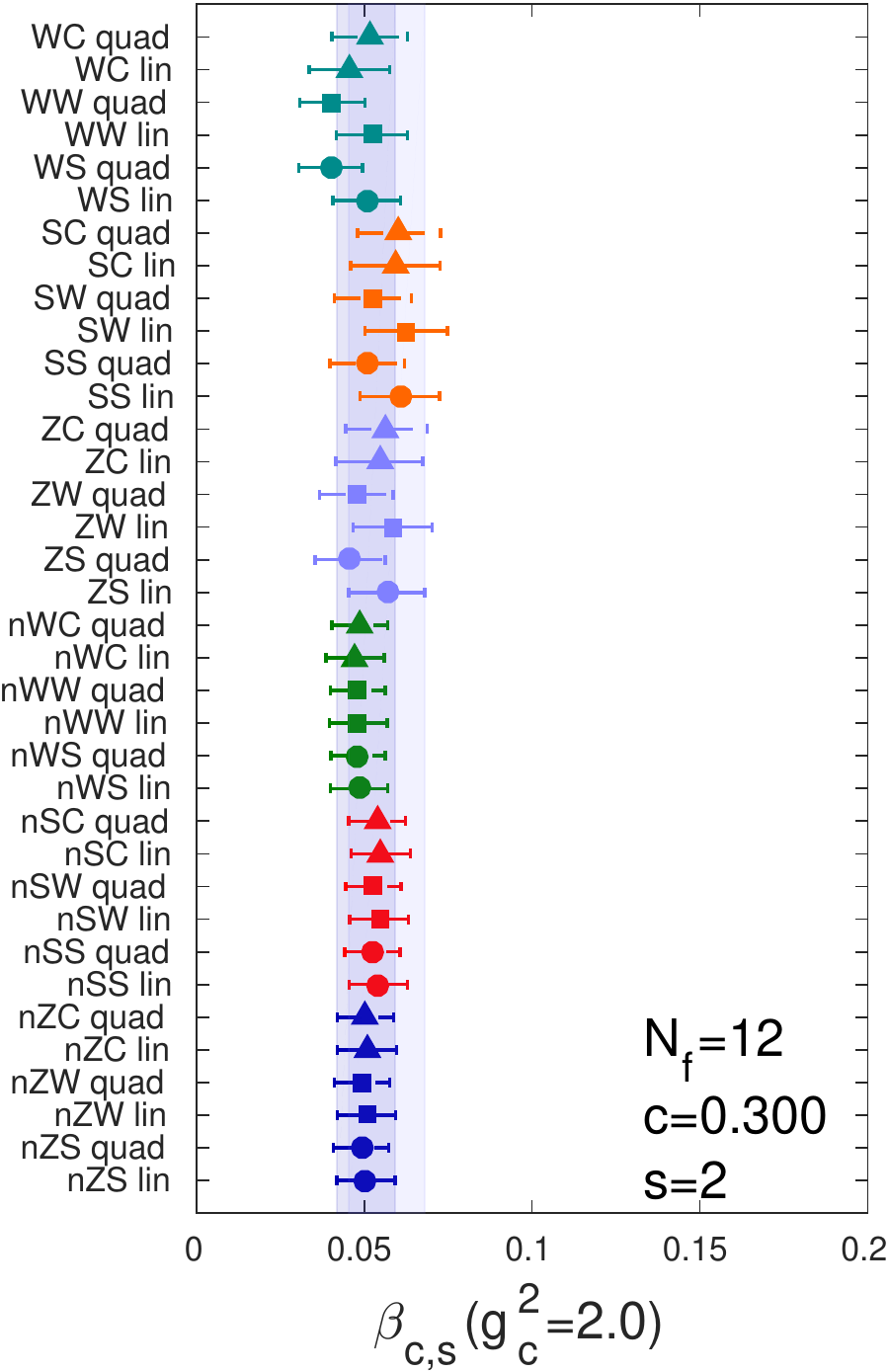}
  \includegraphics[height=0.304\textheight]{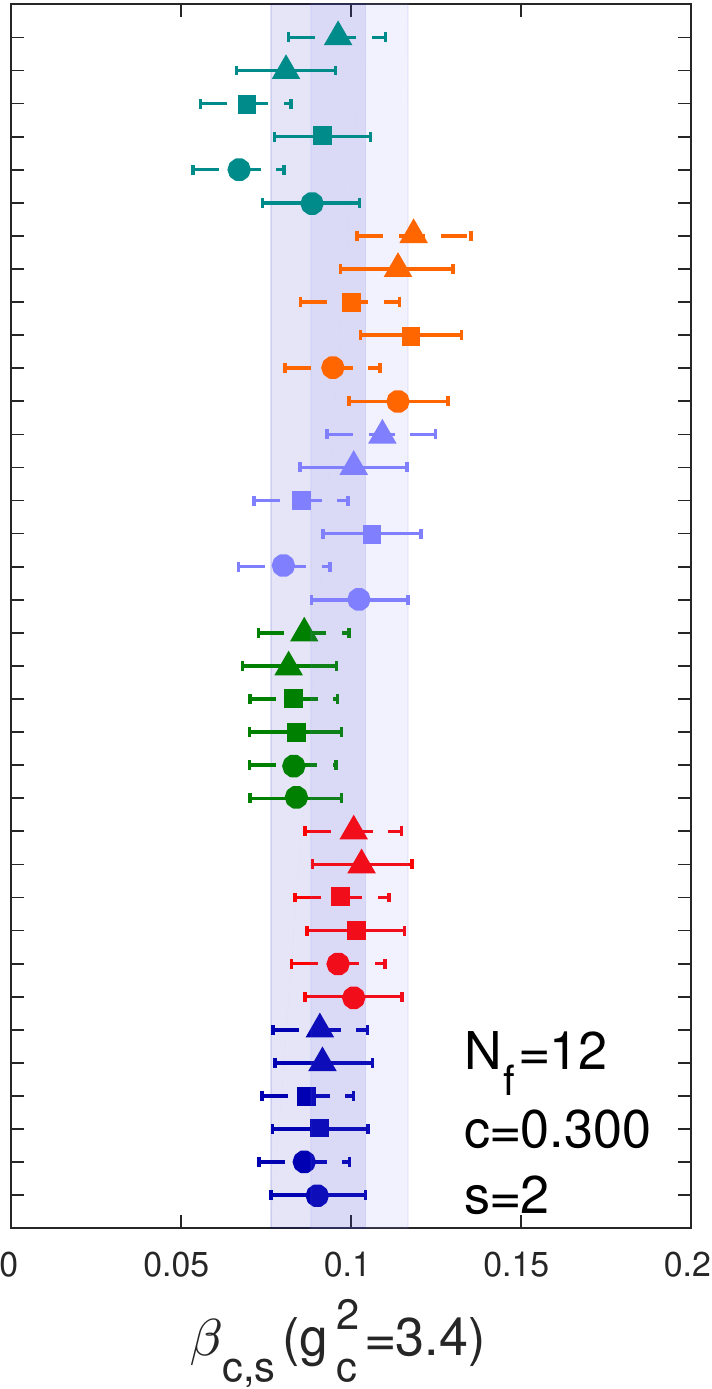}
  \includegraphics[height=0.304\textheight]{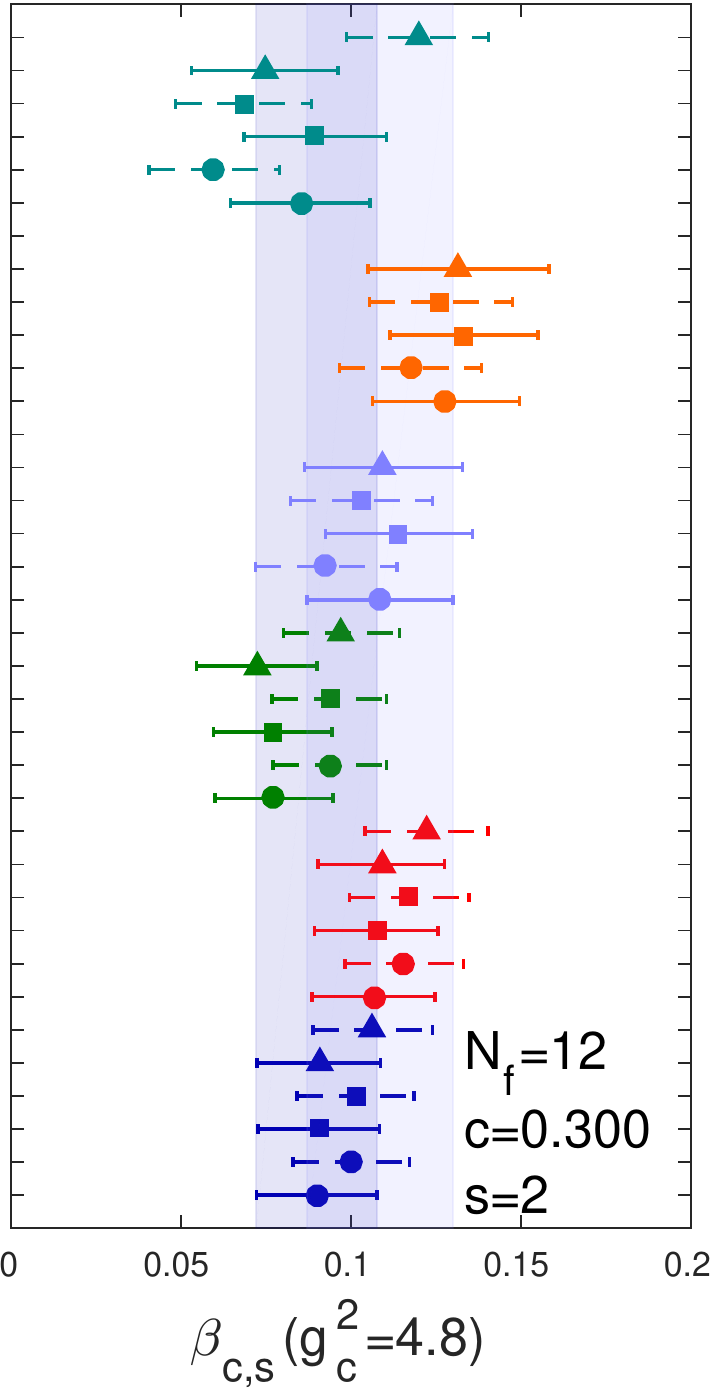}
  \includegraphics[height=0.304\textheight]{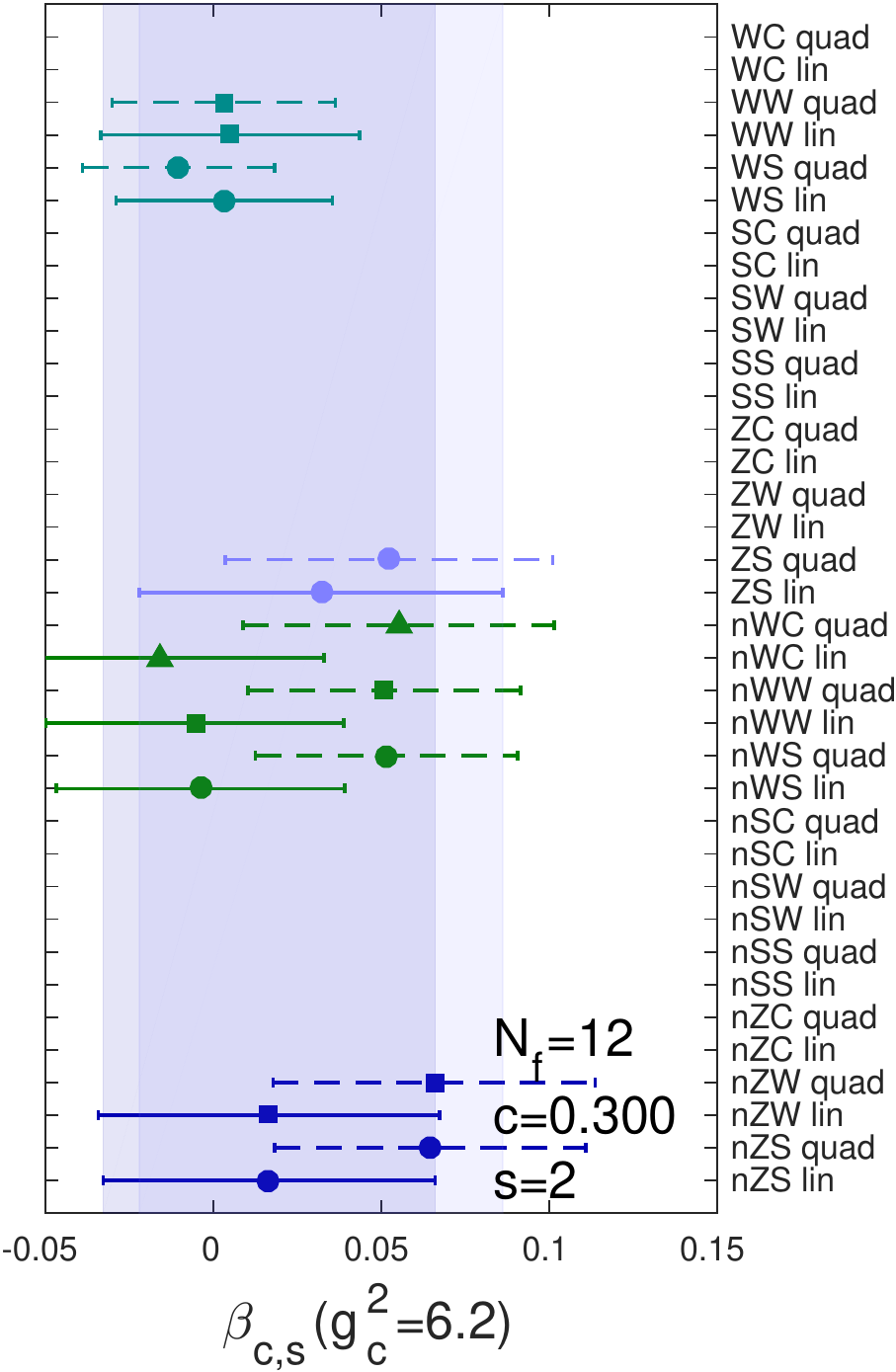}\\
  \caption{Systematic effects due to tree-level improvement, different flows and operators as well as linear extrapolations of the three largest volume pairs vs.~quadratic extrapolation of all volume pairs. The columns show our continuum limit results at selective $g_c^2$ values of 2.0, 3.4, 4.8, and 6.2; the rows correspond to renormalization schemes $c=0.250$, 0.275, 0.300. Open symbols indicate extrapolations with a $p$-value below 5\% and the vertical shaded bands highlight our preferred (n)ZS analysis.}
  \label{Fig.beta_sys}
\end{figure*}

In total we obtain 36 different continuum limit predictions  for the step-scaling function based on nine different flow/operator combinations, analyzing the data with and without tree-level normalization, and performing two different extrapolations to the 18 data sets. Choosing four representative couplings for the range of $g_c^2$ simulated, we present in Fig.~\ref{Fig.beta_sys} an overview of our results. The plots in one column correspond to $g_c^2=2.0$, 3.4, 4.8 and 6.2, while the plots in a row have the same $c$ value, from top to bottom $c=0.250$, 0.275, and 0.300. All plots show a fixed range of 0.2 in units of $\beta_{c,s}$ on the  abscissa. The upper half of each plot shows the results without tree-level improvement, the lower half with improvement. The different colors denote the three different flows: blue Zeuthen, red Symanzik, and green Wilson, whereas a circle marks the Symanzik operator, a square the Wilson plaquette operator, and the triangle the clover operator. Open symbols indicate continuum extrapolations with a $p$-value below 5\%. The two strongest couplings ($g_c^2=4.8,$ 6.2) are not reached by all flow-operator combinations, and therefore, e.g.~no red Symanzik flow symbols are shown in the rightmost plots. The different continuum extrapolations are indicated by a solid (linear fit) or dashed (quadratic fit) line denoting the statistical uncertainty. Highlighted by the vertical, blue-shaded bands are the results of our preferred (n)ZS analysis.

%=================================================

At the weakest coupling shown ($g_c^2=2.0$), the different analysis show very little scatter and in particular the tree-level improved results (bottom half) are well ``aligned.'' Without improvement, differences can be mostly observed between linear and quadratic extrapolations. These are likely due to discretization effects on the smallest $8\to 16$ volume pair. Looking at stronger coupling ($g_c^2=3.4$ and 4.8), we observe that the ``scatter'' of the results grows. Zeuthen and Symanzik flow with tree-level improvement as well as Wilson flow without improvement are still very consistent. Only extrapolations of unimproved SC or WC combinations at $c=0.250$ differ by more than $1\sigma$ from either of our preferred (n)ZS determinations. In both cases, the continuum extrapolation is less reliable because it covers a very large range (more than an order of magnitude) as can be seen e.g.~in the upper right plot of Fig.~\ref{Fig.beta_alt_SC}. The larger scatter for the intermediate couplings could be related to the observation that most of our extrapolations around $g_c^2\sim 5$ show a lower $p$-value, hence continuum predictions may be less reliable. At the strongest coupling, $g_c^2=6.2$, statistical errors are significantly larger and within these all (available) results agree at the $1\sigma$ level. 

Comparing the different step-scaling results for different $c$ values, the overall consistency improves significantly when $c$ increases. While at $c=0.250$ our preferred determinations based on nZS and ZS exhibit a small tension at weak and intermediate couplings, both determinations are consistent at the $1\sigma$ level for $c=0.275$ and 0.300. Being aware that $c=0.250$ might be affected by discretization effects on our smaller lattice volume pairs and perturbative improvement may not be fully justified at the stronger couplings, we therefore choose to quote the envelope covering both, nZS and ZS, determinations as our final result which also accounts for systematic effects. Systematic effects due to different flow/operator combinations are not resolved within the combined nZS+ZS uncertainty. In fact only a few extrapolations for alternative flow/operators at $c=0.250$ lead to results differing by more than $1\sigma$. Such extrapolations cover a large range and are less reliable because extrapolating all five points using the quadratic Ansatz tends to differ by several sigmas from extrapolating the three largest volume pairs using a linear Ansatz. While the analyzed choice of flow/operator combinations is by no means complete, in fact an arbitrary number of operators and flows could be considered, we do expect to have studied a representative set. Finally, it is noteworthy that the agreement at weakest and strongest couplings is better than at the intermediate range.

%%%%%%%%%%% FIGURE %%%%%%%%%%%%%%%%%%%%%%%%%%%
\begin{figure}[tb] % no figure before 1st section
  \centering
  \includegraphics[width=0.48\textwidth]{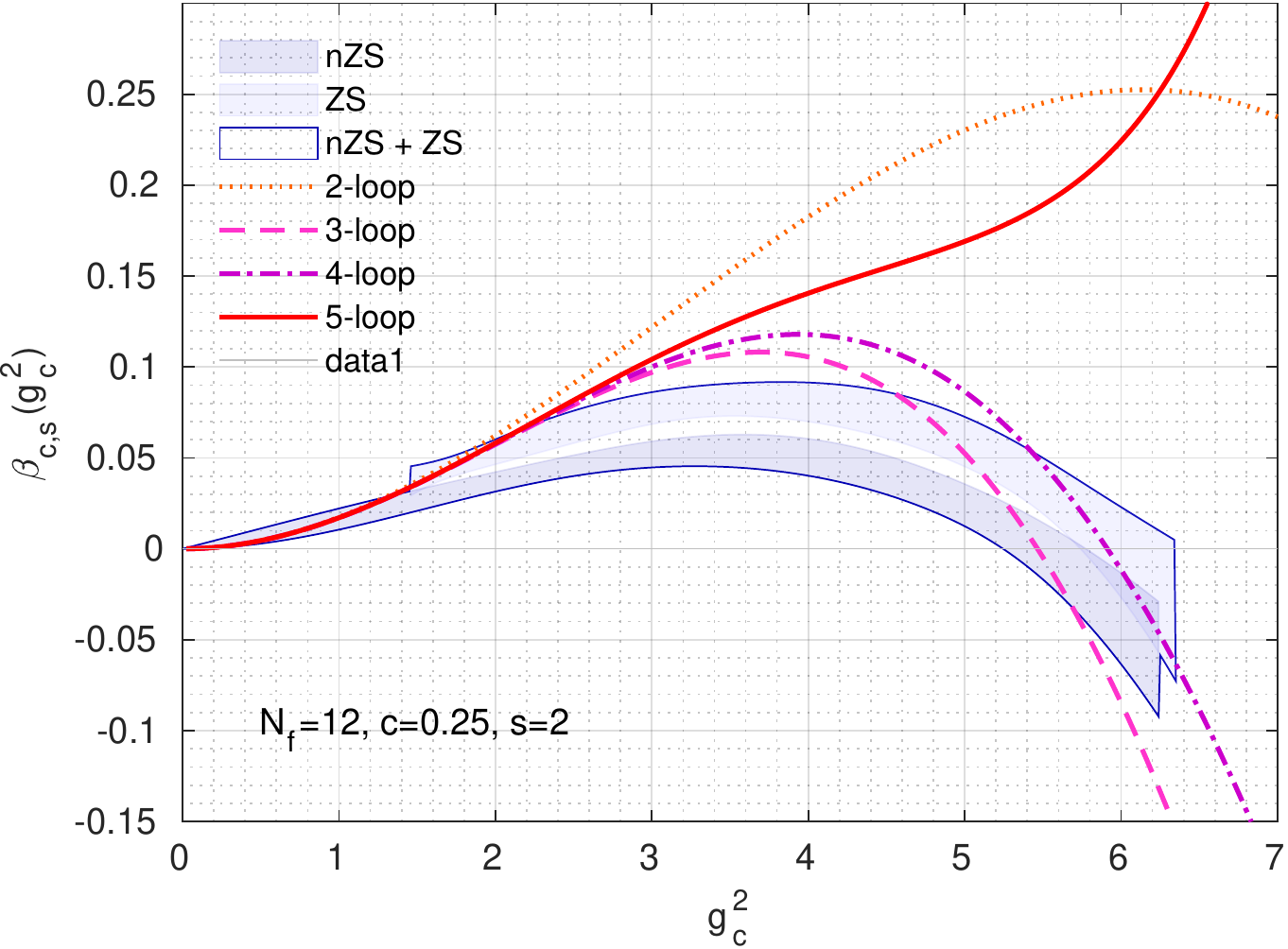}\\
  \includegraphics[width=0.48\textwidth]{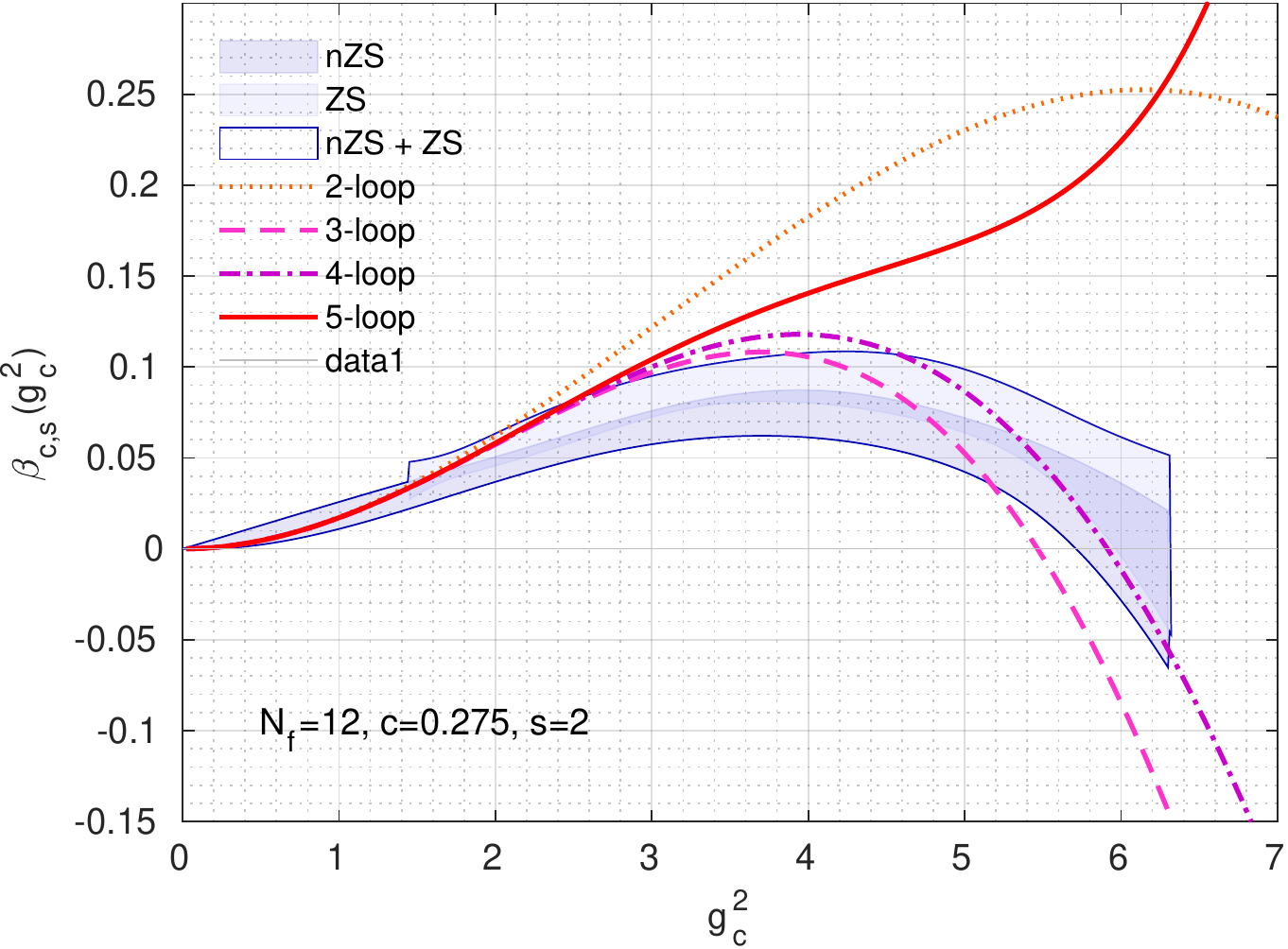}\\
  \includegraphics[width=0.48\textwidth]{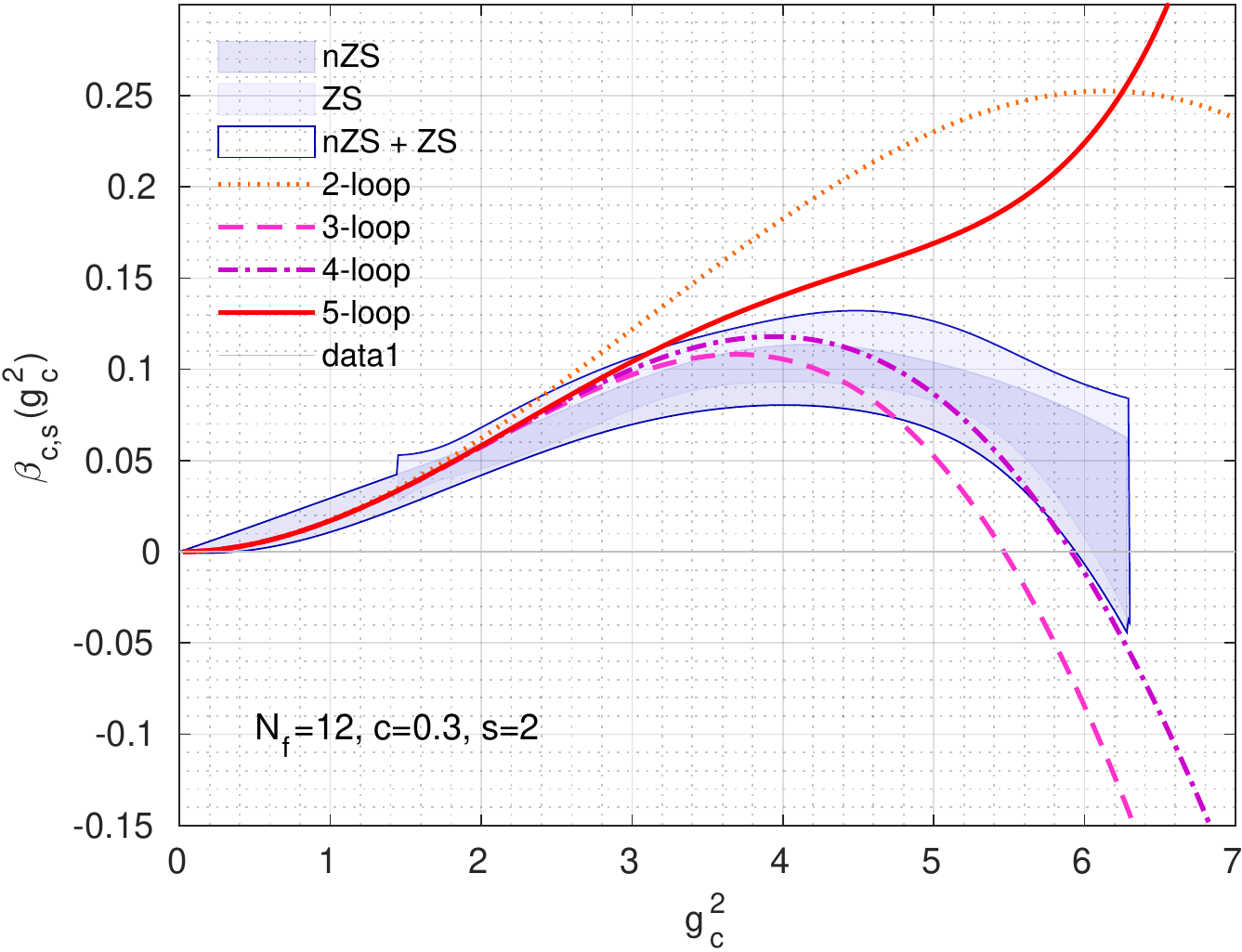}\\  
  \caption{Continuum extrapolations of our preferred (n)ZS data set for $c=0.250$, 0.275, and 0.300 in comparison to perturbative 2-, 3-, 4-, and 5-loop $\overline{\text{MS}}$ results.}
  \label{Fig.continuum}% Give a unique label
\end{figure}
%%%%%%%%%%%% FIGURE %%%%%%%%%%%%%%%%%%%%%%%%%%%

Our final continuum limit results for the gradient flow step-scaling function obtained in renormalization schemes $c=0.250$, 0.275, and 0.300 are shown in Fig.~\ref{Fig.continuum} where we also show  perturbative predictions~\cite{Baikov:2016tgj}. Our results are consistent with the perturbative predictions obtained at 3- or 4-loop in the $\overline{\text{MS}}$ scheme but differ significantly from the 5-loop, which grows rapidly for $g^2_c>4.0$. Our prediction in the $c=0.250$ scheme indicates a fixed point near the  4-loop $\overline{\text{MS}}$ value, whereas for $c=0.300$  the 4-loop $\overline{\text{MS}}$ value coincides with the lower bound of our prediction of a possible  fixed point  Unfortunately, our uncertainties are too large to establish or rule out a dependence on the renormalization scheme $c$. Our predicted range of couplings for an IRFP of an SU(3) gauge theory with twelve flavors is consistent with predictions by Ryttov and Shrock who improve perturbative convergence using Pad\'e approximants \cite{Ryttov:2016ner,Ryttov:2017kmx}.

%=================================================
\section{Conclusion}
\label{Sec.Conclusion}
%=================================================
We have presented details of our gradient flow step-scaling calculation for SU(3) with twelve dynamical flavors. Our calculations are based on  gauge field ensembles generated with Symanzik gauge action and three times stout-smeared M\"obius domain wall fermions. Using Zeuthen flow combined with the Symanzik operator, we determine renormalized couplings and predict the step-scaling $\beta$-function. We assign a systematic uncertainty to our numerical predictions that covers the difference between the continuum limit extrapolated results of various flows and operators, different extrapolation forms, and also the effect of the tree-level normalization.  For the $c=0.250$ scheme we can identify a sign change of the $\beta$-function (infrared fixed point) in the range of $5.2 \le g_c^2 \le 6.4$, while for schemes $c=0.275$ ($c=0.300$) we can only name a lower limit for a possible fixed point $5.7 \le g_c^2$ ($5.9\le g_c^2$). It appears that the step scaling function and the value of the fixed point exhibit a mild dependence on the renormalization scheme $c$, however, further data at even stronger couplings with  greater precision are required to resolve such a dependence. The set of available gauge field ensembles restricts the maximum $g_c^2$ value a given  flow/operator combination can reach. Less than half of the 36 combinations we consider reach $g_c^2=6.2$,  which unfortunately limits the control of our systematic error estimate around $g_c^2\sim 6.0$ where we observe the IRFP for $c=0.250$. Since the reach in $g_c^2$ depends on the flow/operator combination, the converse argument implies that on a given set of gauge field ensembles, an IRFP can be identified for some flow/operator combinations but missed for others.

\begin{figure}[tb] % no figure before 1st section
  \centering
  \includegraphics[width=0.48\textwidth]{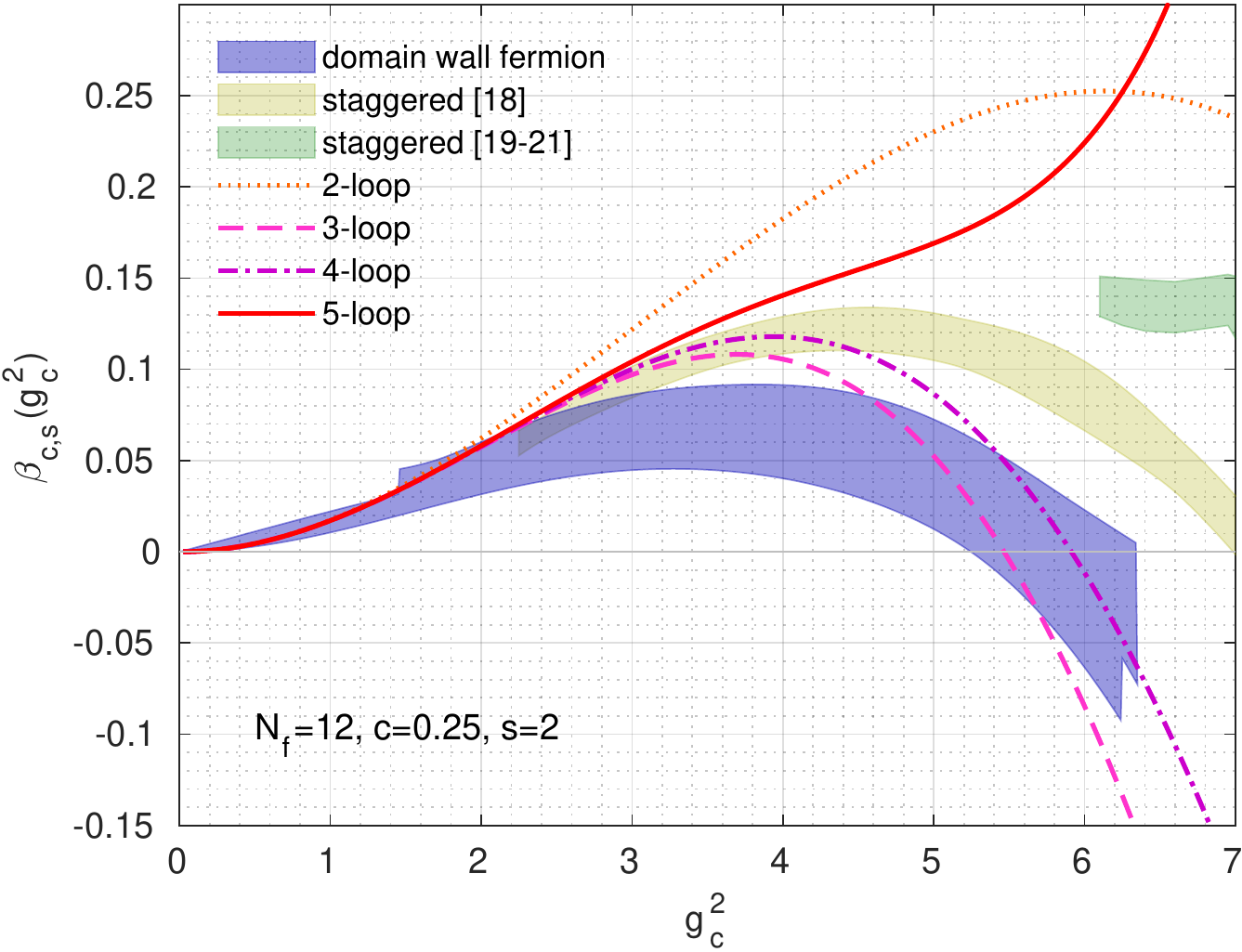}\\
  \includegraphics[width=0.48\textwidth]{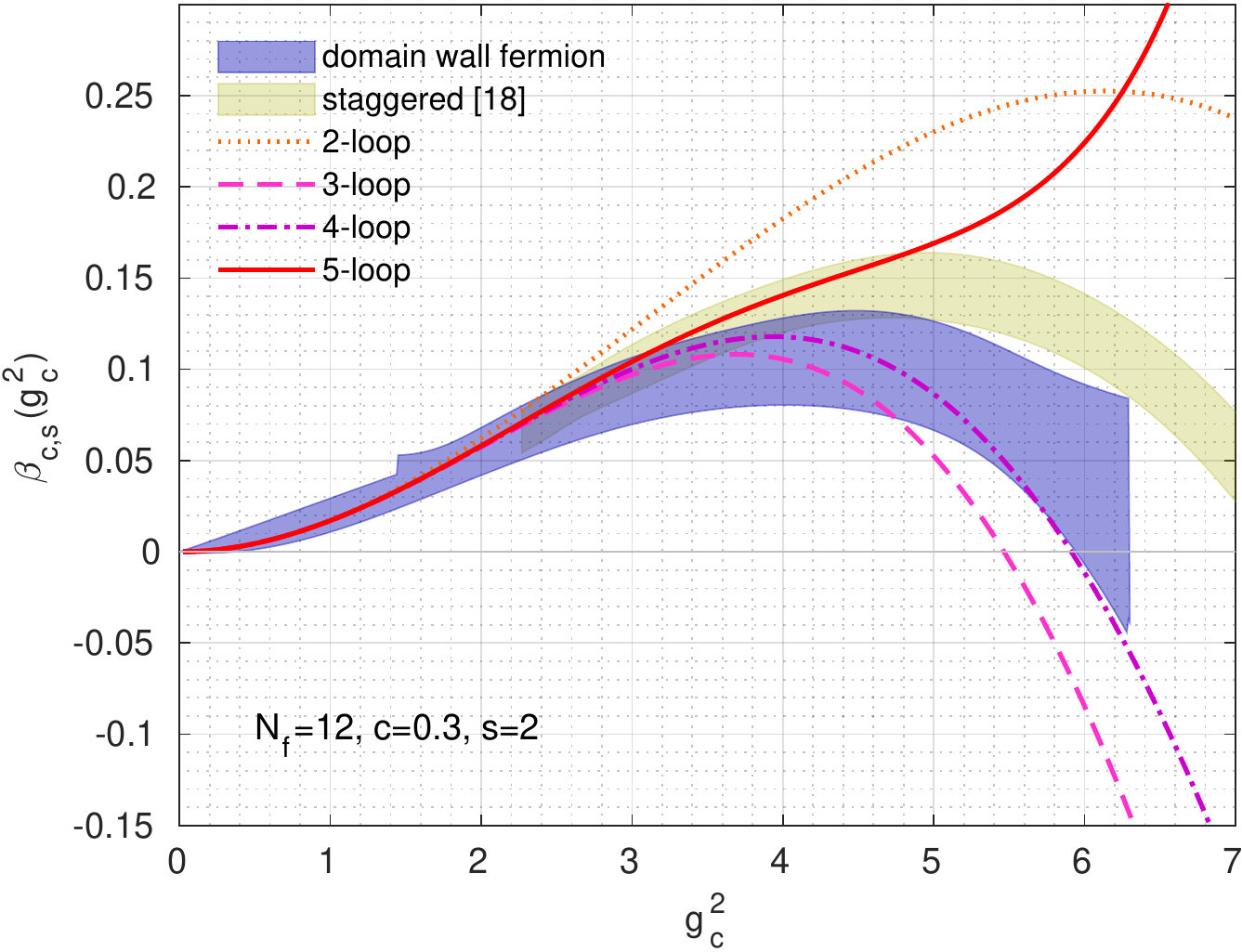}  
  \caption{Comparison of our continuum extrapolations including an estimate for systematic effects for renormalization schemes $c=0.250$ and 0.300 to nonperturbative determinations based on staggered fermions.  Reference \cite{Hasenfratz:2016dou} identifies an IRFP at somewhat stronger coupling than our determination and presents results for the $c=0.250$ and 0.300 schemes. Using only $c=0.250$ to determine $\beta_{c,s}$, Refs.~\cite{Fodor:2016zil,Fodor:2017gtj,Fodor:2017nlp} report a basically flat $\beta$ function for the range $6.0 \lesssim g_c^2 \lesssim 7.2$. All predictions use $s=2$.}
  \label{Fig.compare}% Give a unique label
\end{figure}

Our presented result are consistent with perturbative predictions obtained at 3- or 4-loop in the $\overline{\text{MS}}$ scheme. Finally, we compare our continuum limit results to other, nonperturbative determinations published in the literature. So far only calculations based on staggered fermions have been performed \cite{Lin:2015zpa,Hasenfratz:2016dou,Fodor:2016zil,Fodor:2017gtj,Fodor:2017nlp}. The work by Hasenfratz and Schaich \cite{Hasenfratz:2016dou} extends from the weak couplings at $g_c^2\sim 2.3$ to strong couplings $g_c^2 \sim 8.2$ identifying an IRFP around $g_c^2\sim 7.3$ for both $c=0.25$ and $c=0.3$. In the weak coupling their prediction is consistent with the findings of Ref.~\cite{Lin:2015zpa}  which however uses larger $c$ values; for $g_c^2\gtrsim 6.0$ it is in tension with the determination by the LatHC collaboration \cite{Fodor:2016zil,Fodor:2017gtj,Fodor:2017nlp}. In Fig.~\ref{Fig.compare} we compare our DWF result\footnote{ASCII files containing our final results (blue shaded bands) are uploaded as Supplemental Material.} to the prediction of the step-scaling function by Hasenfratz/Schaich \cite{Hasenfratz:2016dou} and LatHC \cite{Fodor:2016zil,Fodor:2017gtj,Fodor:2017nlp}.  Reference \cite{Hasenfratz:2016dou} presents results for the renormalization schemes $c=0.250$ and $0.300$ and accounts for systematic effects due to the extra- and interpolations. References \cite{Fodor:2016zil,Fodor:2017gtj,Fodor:2017nlp} only use $c=0.250$ without quantifying systematic effects. Extending what has so far been done in the literature, we  emphasize that our results account for the systematic effect estimated by considering tree-level normalization, different flow/operator combinations, and continuum limit extrapolations.

Our results exhibit a similar shape as the $\beta$-function predicted by Hasenfratz and Schaich in Ref.~\cite{Hasenfratz:2016dou}. While for $c=0.250$, the two  predictions differ by $\sim 2\sigma$ for intermediate and strong couplings, the differences almost vanish for $c=0.300$. This might suggest that uncertainties in particular for $c=0.250$ are underestimated.  In contrast, the results of Refs.~\cite{Fodor:2016zil,Fodor:2017gtj,Fodor:2017nlp} predict a qualitatively different step scaling function that is nearly constant in a wide $g_c^2$ coupling range, without a sign of an IRFP.  Further investigations are needed to track down the origin of this disagreement.

%%%%%%%%%%% FIGURE %%%%%%%%%%%%%%%%%%%%%%%%%%%
%%%%%%%%%%%% FIGURE %%%%%%%%%%%%%%%%%%%%%%%%%%%

%=================================================
\begin{acknowledgments}
  We are very grateful to Peter Boyle, Guido Cossu, Anontin Portelli, and Azusa Yamaguchi who develop the \texttt{GRID} software library providing the basis of this work and who assisted us in installing and running \texttt{Grid} on different architectures and computing centers. A.H.~and O.W.~acknowledge support by DOE Grant No.~DE-SC0010005 and C.R. by DOE Grant No.~DE-SC0015845. A.H.~would like to acknowledge the Mainz Institute for Theoretical Physics (MITP) of the Cluster of Excellence PRISMA+ (Project ID 39083149).  O.W.~acknowledges partial support by the Munich Institute for Astro- and Particle Physics (MIAPP) of the DFG cluster of excellence ``Origin and Structure of the Universe''. 

Computations for this work were carried out in part on facilities of the USQCD Collaboration, which are funded by the Office of Science of the U.S.~Department of Energy and the RMACC Summit supercomputer \cite{UCsummit}, which is supported by the National Science Foundation (awards ACI-1532235 and ACI-1532236), the University of Colorado Boulder, and Colorado State University. This work used the Extreme Science and Engineering Discovery Environment (XSEDE), which is supported by National Science Foundation grant number ACI-1548562 \cite{xsede} through allocation TG-PHY180005 on the XSEDE resource \texttt{stampede2}.  This research also used resources of the National Energy Research Scientific Computing Center (NERSC), a U.S. Department of Energy Office of Science User Facility operated under Contract No. DE-AC02-05CH11231.  We thank  Fermilab,  Jefferson Lab, NERSC, the University of Colorado Boulder, Texas Advanced Computing Center, the NSF, and the U.S.~DOE for providing the facilities essential for the completion of this work. 
\end{acknowledgments}
%=================================================
\clearpage
\appendix

\section{Tree-level normalization factors}
\label{Sec.tree-level}

\begin{longtable}{ccccc}
  \caption{Tree-level normalization coefficients $C(c,L/a)$ for renormalization schemes $c=0.250$, 0.275, and 0.300. Values are quotes for simulations with Symanzik gauge action and Zeuthen, Symanzik, or Wilson flow combined with Symanzik, Wilson plaquette, or clover operator.}  \label{Tab.tln}\\
  
  \hline \hline
  flow/operator& $L/a$ & $C(0.250,L/a)$& $C(0.275,L/a)$& $C(0.300,L/a)$\\
  \hline
  \endfirsthead
  \hline
  flow/operator& $L/a$ & $C(0.250,L/a)$& $C(0.275,L/a)$& $C(0.300,L/a)$\\
  \hline
  \endhead

  \hline
  \endfoot

  \hline \hline
  \endlastfoot

  \input{tln}

\end{longtable}

\newpage
\setlength{\LTcapwidth}{\textwidth}
\section{\texorpdfstring{Renormalized couplings $g_c^2$}{Renormalized couplings gc2}}
\label{Sec.RenCouplings}

\begin{longtable*}{cccccccccccc}
  \caption{Details of our preferred analysis based on Zeuthen flow and Symanzik operator. For each ensemble specified by the spatial extent $L/a$ and bare gauge coupling $\beta$ we list $N$, the number of measurements, as well as the renormalized couplings $g_c^2$ for the analysis with (nZS) and without tree-level improvement (ZS) for the three renormalization schemes $c=0.250$, 0.275 and 0.300. In addition the integrated autocorrelation times determined using the $\Gamma$-method \cite{Wolff:2003sm} are listed in units of 10 MDTU.} \label{Tab.nZS_ZS}\\
  
  \hline\hline
      &         &     & \multicolumn{3}{c}{$c=0.25$}&\multicolumn{3}{c}{$c=0.275$}&\multicolumn{3}{c}{$c=0.30$}\\
  $L/a$ & $\beta$ & $N$ & $g_c^2$(nZS) & $g_c^2$(ZS) & $\tau_\text{int}$& $g_c^2$(nZS) &  $g_c^2$(ZS)  & $\tau_\text{int}$ &$g_c^2$(nZS)  &$g_c^2$(ZS)  & $\tau_\text{int}$\\
  \hline
  \endfirsthead

  \hline
      &         &     & \multicolumn{3}{c}{$c=0.25$}&\multicolumn{3}{c}{$c=0.275$}&\multicolumn{3}{c}{$c=0.30$}\\
  $L/a$ & $\beta$ & $N$ & $g_c^2$(nZS) &  $g_c^2$(ZS) & $\tau_\text{int}$& $g_c^2$(nZS) &$g_c^2$(ZS) & $\tau_\text{int}$ & $g_c^2$(nZS) & $g_c^2$(ZS) & $\tau_\text{int}$\\
  \hline
  \endhead

  \hline
  \endfoot

  \hline \hline
  \endlastfoot

  \input{gcSq_Nf12_nZS_ZS}
\end{longtable*}

\begin{table*}[tb]
  \caption{Renormalized couplings determined on additional ensembles generated with an alternative choice of $L_s$ to test effects of residual chiral symmetry breaking. For easier comparison we also list and highlight in bold face the ensembles used in our main analysis.  Ensembles are specified by the spatial extent $L/a$, the bare gauge coupling $\beta$, and the value of the fifth dimension $L_s$. As above we list $N$, the number of measurements, as well as the renormalized couplings $g_c^2$ for the analysis with (nZS) and without tree-level improvement (ZS) for the three renormalization schemes $c=0.250$, 0.275 and 0.300. In addition the integrated autocorrelation times determined using the $\Gamma$-method \cite{Wolff:2003sm} are listed in units of 10 MDTU.} \label{Tab.nZS_ZS_Ls}
  \begin{tabular}{>{\rowmac}c>{\rowmac}c>{\rowmac}c>{\rowmac}c>{\rowmac}c>{\rowmac}c>{\rowmac}c>{\rowmac}c>{\rowmac}c>{\rowmac}c>{\rowmac}c>{\rowmac}c>{\rowmac}c<{\clearrow}}    
  \hline\hline
      &         & &     & \multicolumn{3}{c}{$c=0.25$}&\multicolumn{3}{c}{$c=0.275$}&\multicolumn{3}{c}{$c=0.30$}\\
  $L/a$ & $\beta$ & $L_s$ & $N$ & $g_c^2$(nZS) & $g_c^2$(ZS) & $\tau_\text{int}$& $g_c^2$(nZS) &  $g_c^2$(ZS)  & $\tau_\text{int}$ &$g_c^2$(nZS)  &$g_c^2$(ZS)  & $\tau_\text{int}$\\
  \hline
  \input{gcSq_Nf12_nZS_ZS_Ls}

  \hline \hline
\end{tabular}
\end{table*}

\clearpage
\section{Continuum limit extrapolations}
\label{Sec.ContLimitExtra}
\begin{figure}[hb]
  \includegraphics[width=0.98\columnwidth]{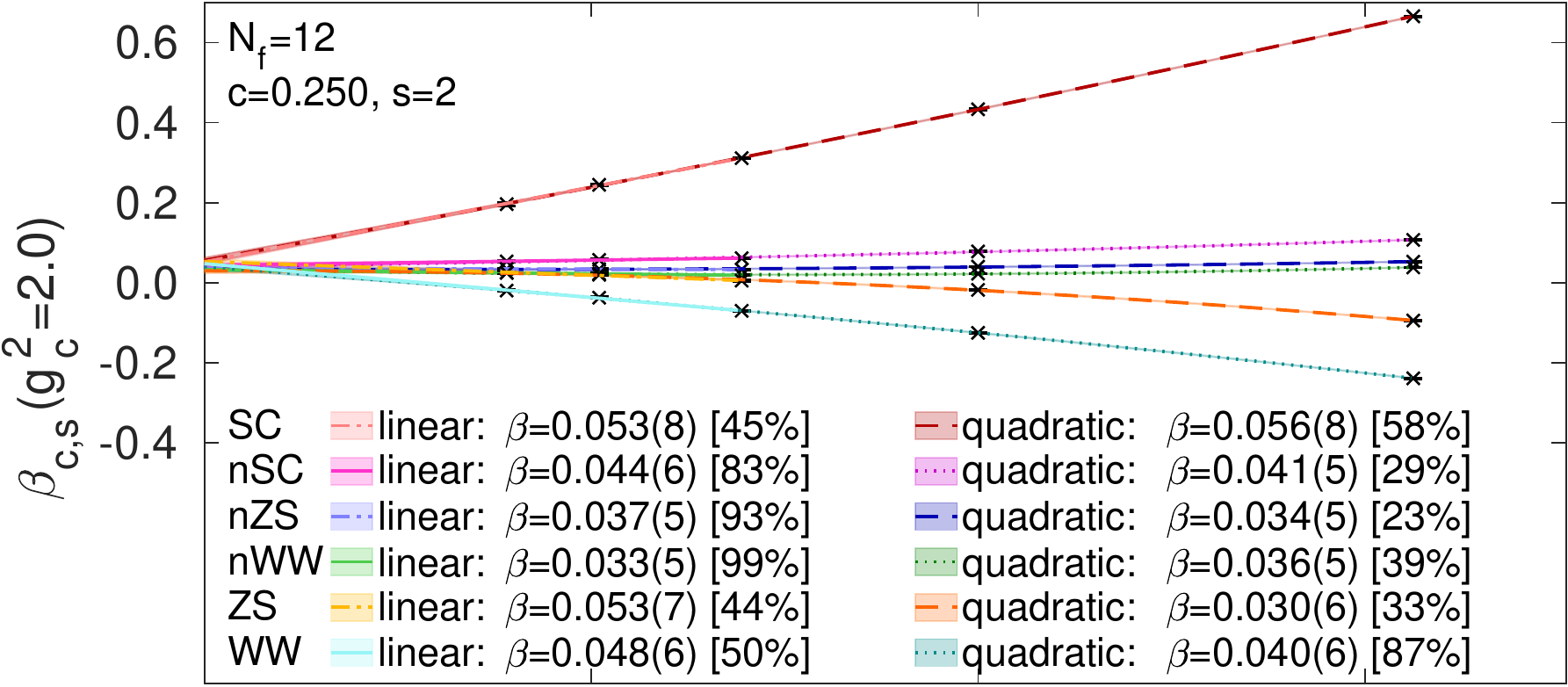}
  \includegraphics[width=0.98\columnwidth]{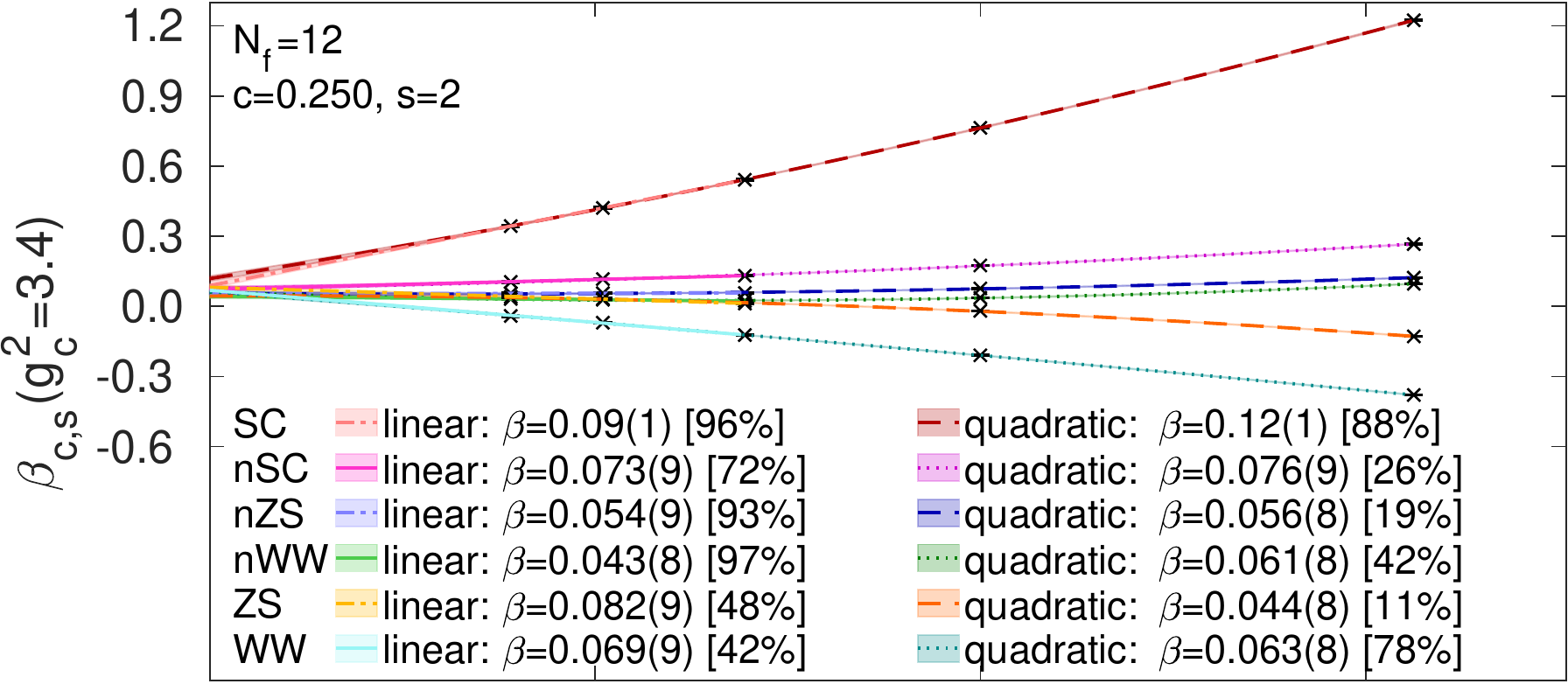}
  \includegraphics[width=0.98\columnwidth]{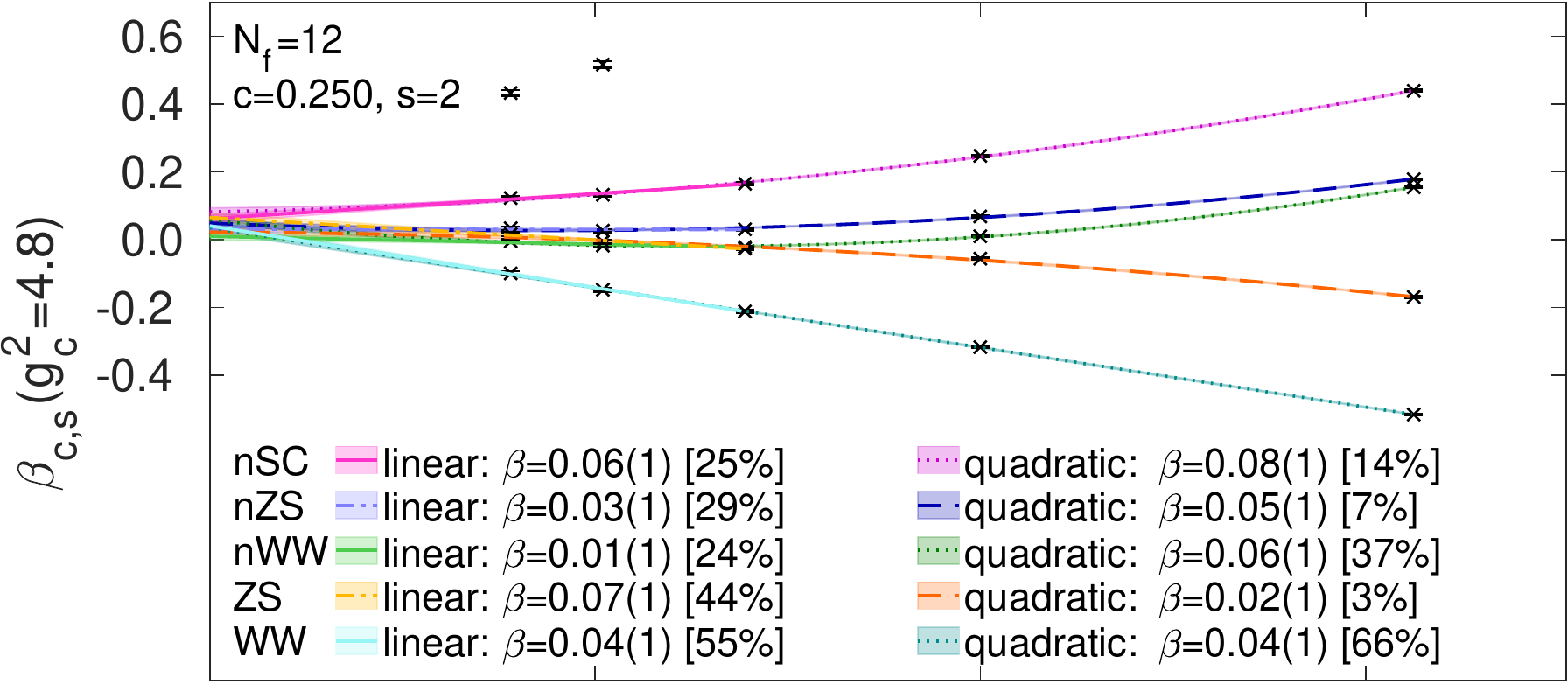}
  \includegraphics[width=0.98\columnwidth]{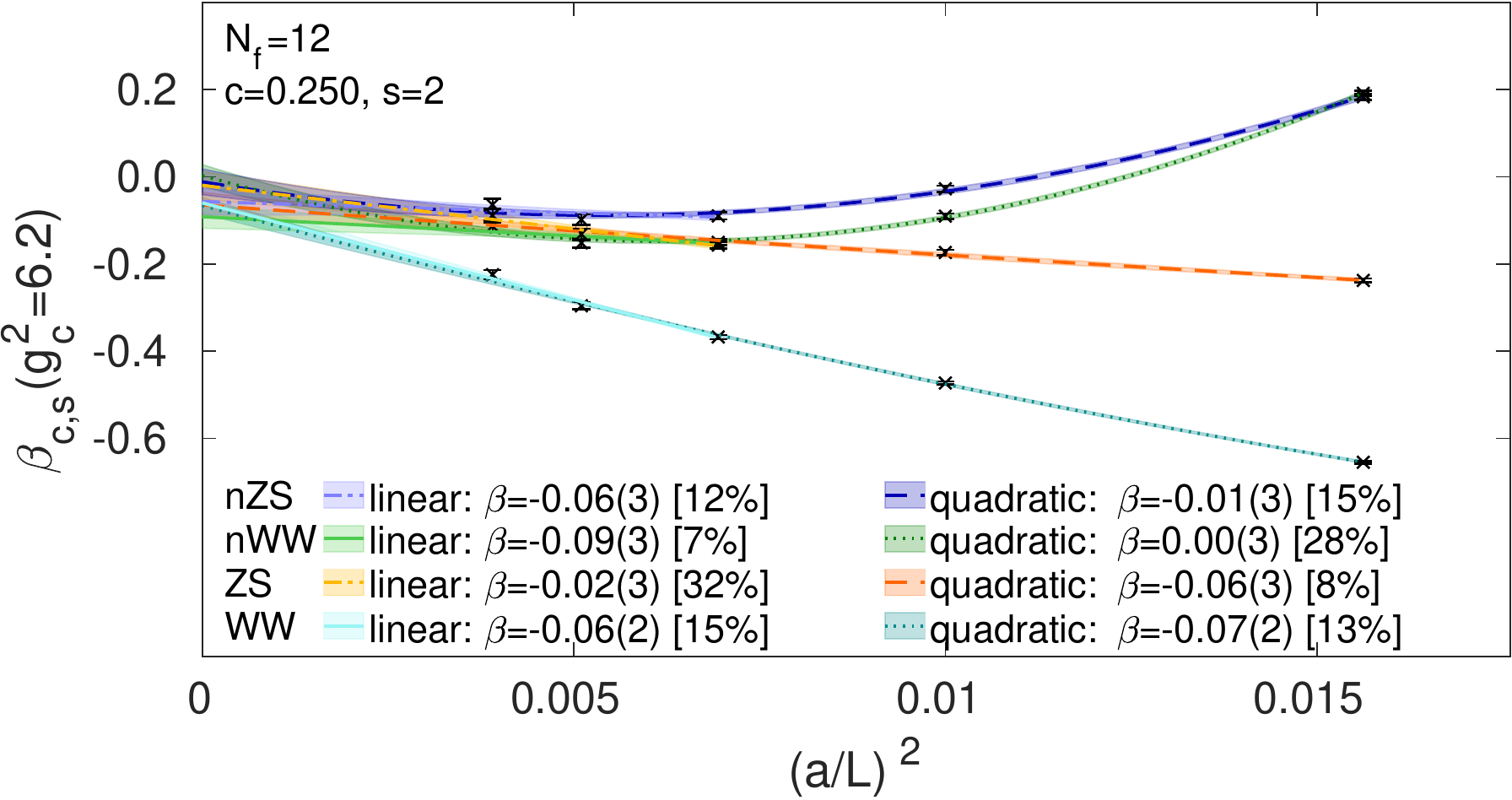}
  \caption{Details of the continuum extrapolation for our preferred (n)ZS data in comparison to alternative determinations based on (n)SC and (n)WW for $c=0.250$. Continuum limit values of $\beta_{c,s}$ and corresponding $p$-values of the extrapolation are quoted in the legend.}
  \label{Fig.cont_extra_c0250}
\end{figure}

\begin{figure}[hb]
  \vspace{13mm}
  \includegraphics[width=0.98\columnwidth]{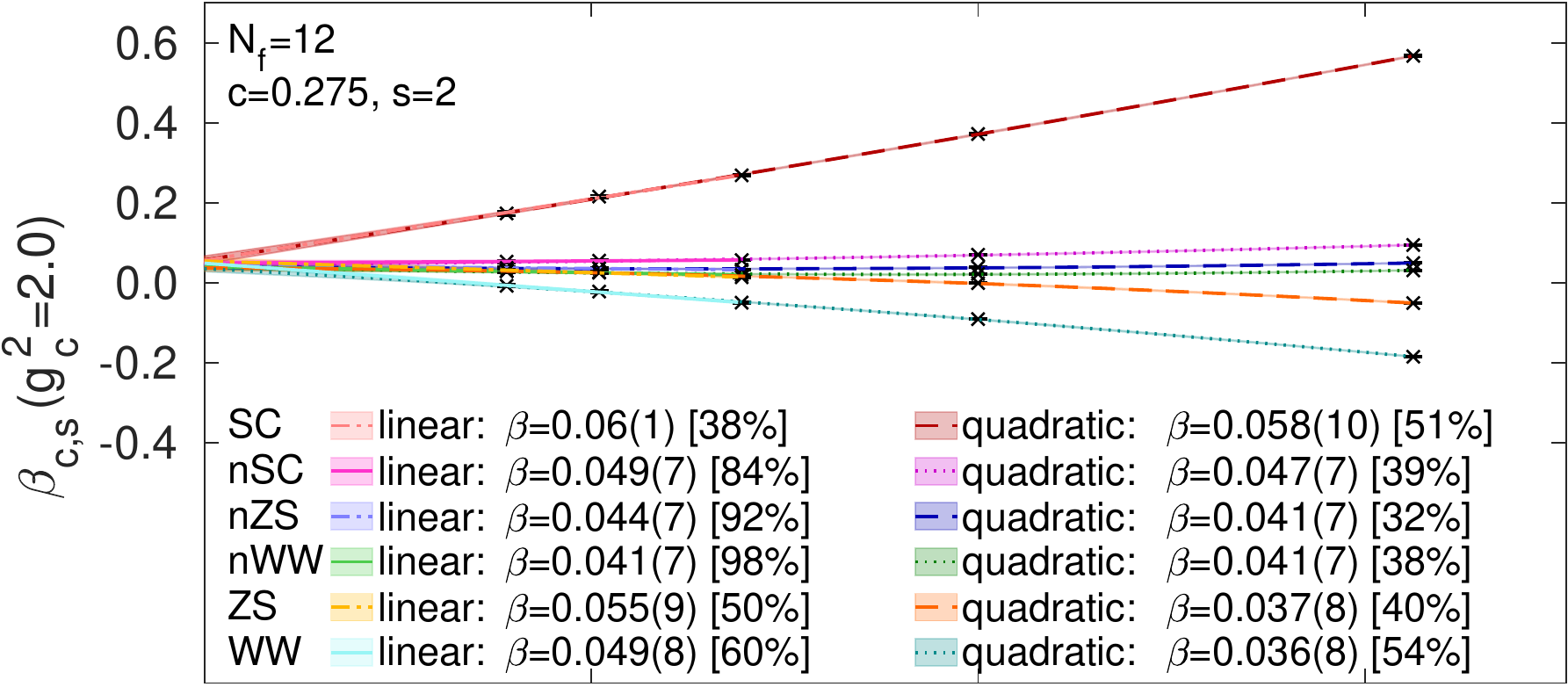}
  \includegraphics[width=0.98\columnwidth]{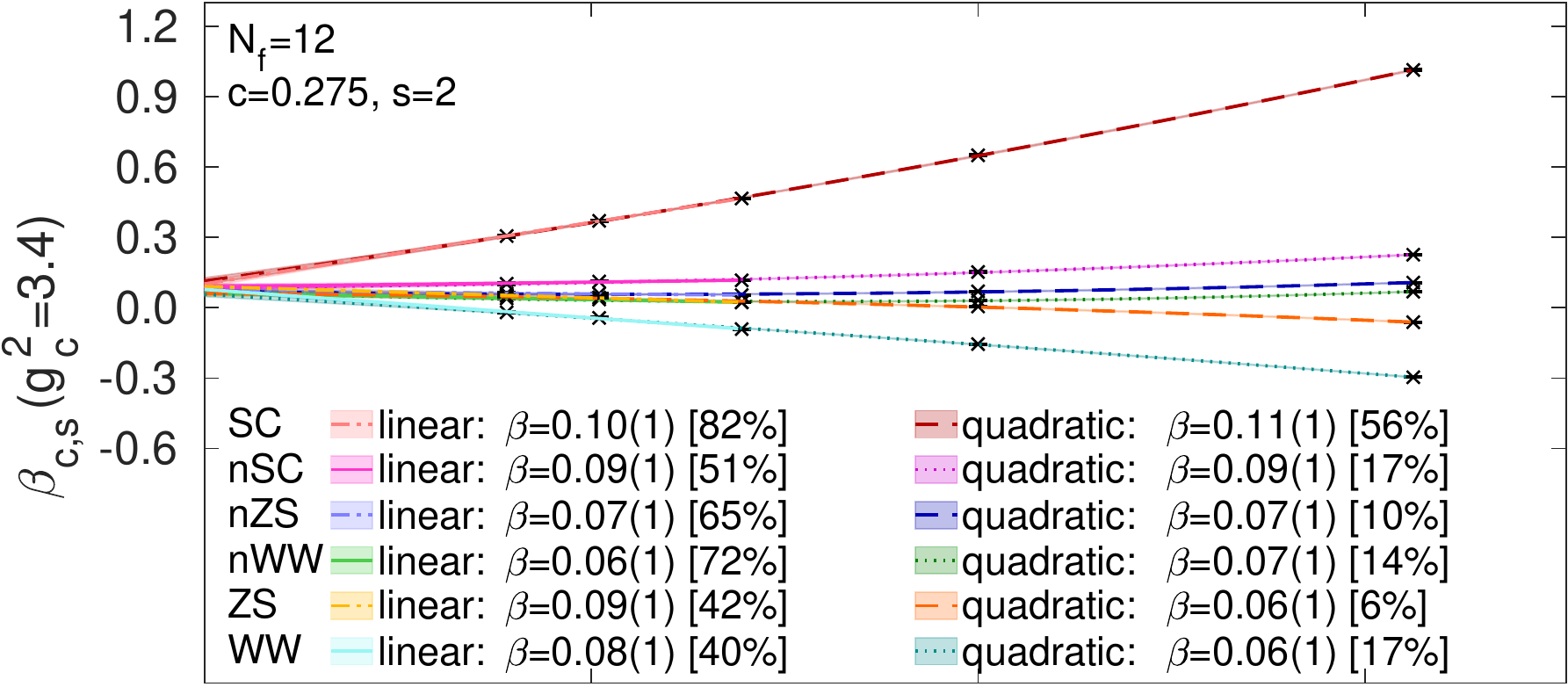}
  \includegraphics[width=0.98\columnwidth]{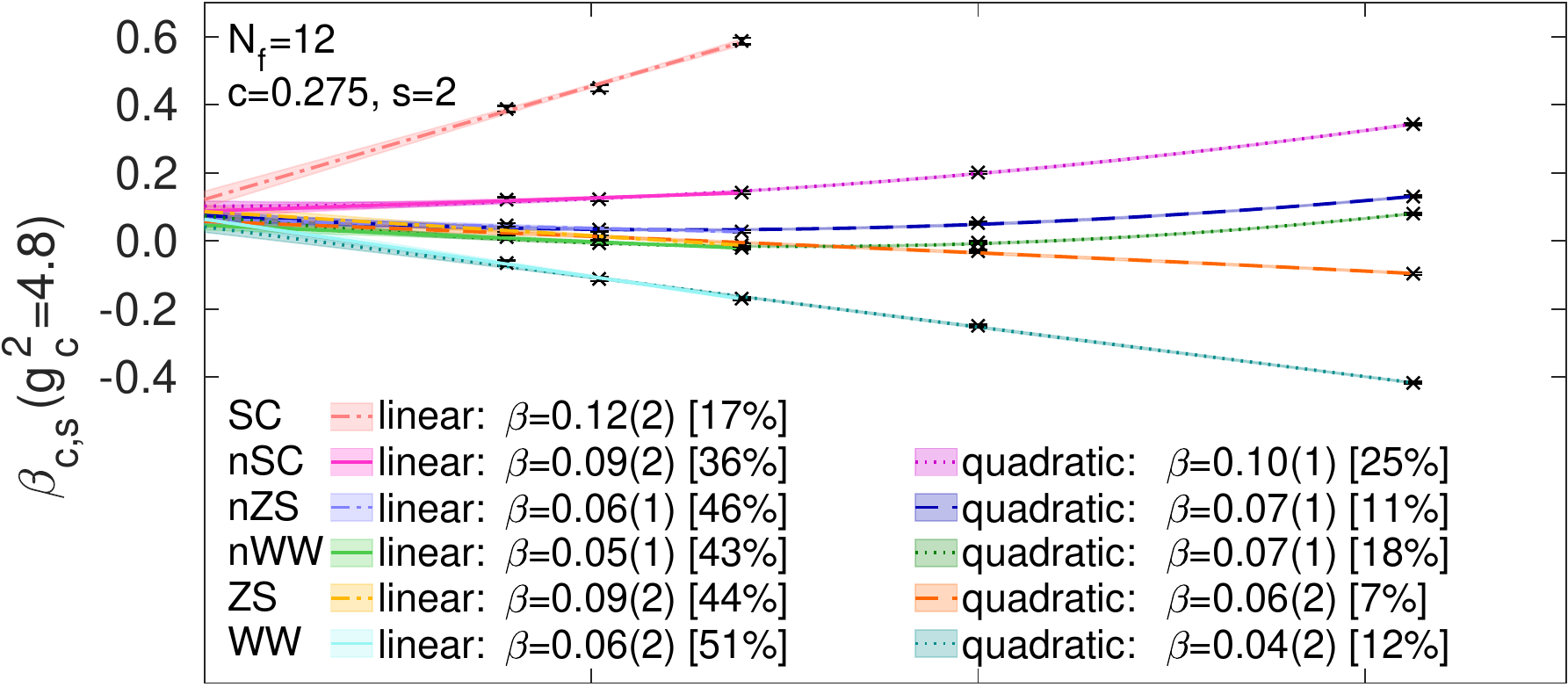}
  \includegraphics[width=0.98\columnwidth]{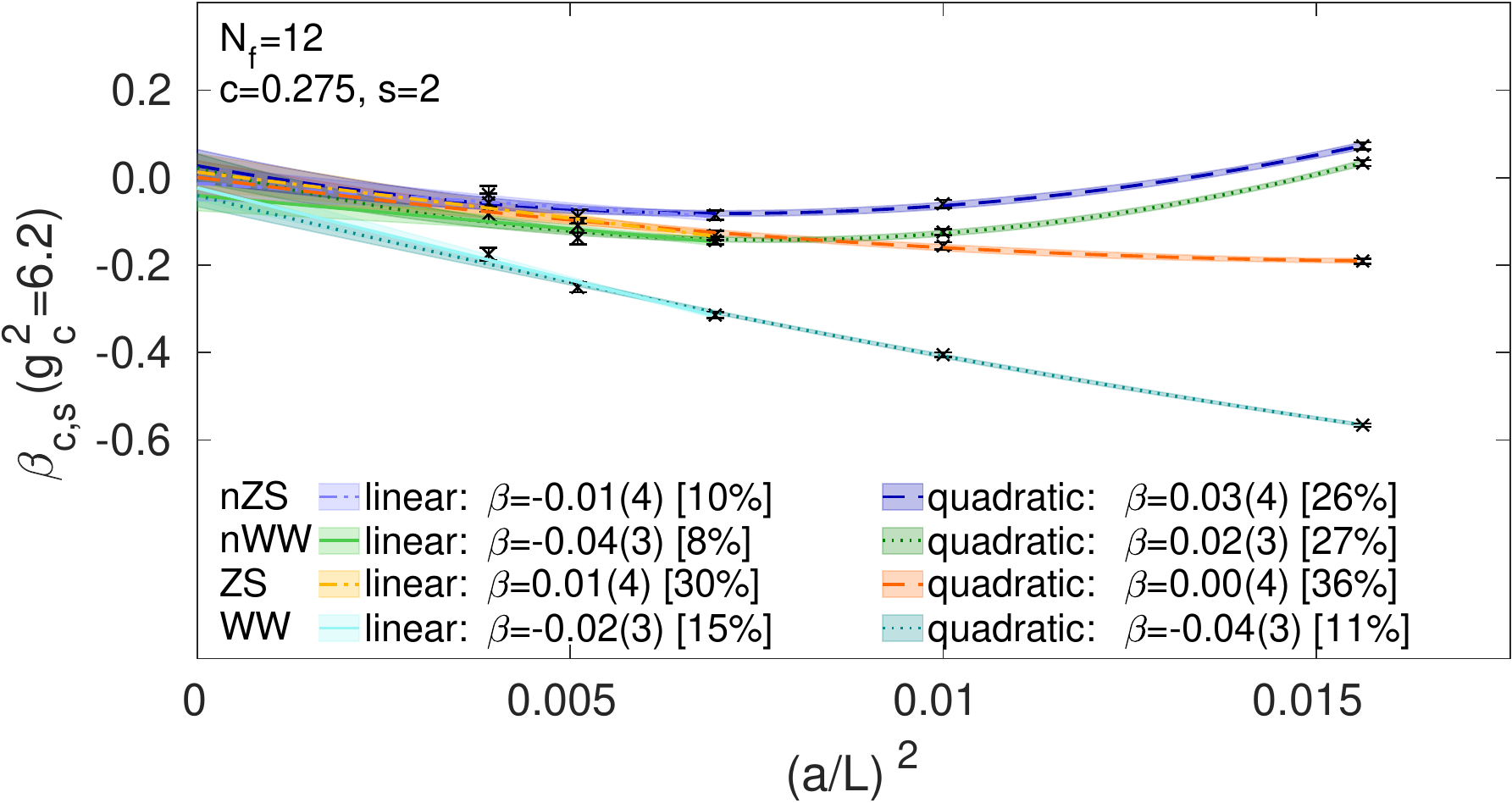}
  \caption{Details of the continuum extrapolation for our preferred (n)ZS data in comparison to alternative determinations based on (n)SC and (n)WW for $c=0.275$. Continuum limit values of $\beta_{c,s}$ and corresponding $p$-values of the extrapolation are quoted in the legend.}
  \label{Fig.cont_extra_c0275}
\end{figure}
\clearpage
\begin{figure}[htb]
  \includegraphics[width=0.98\columnwidth]{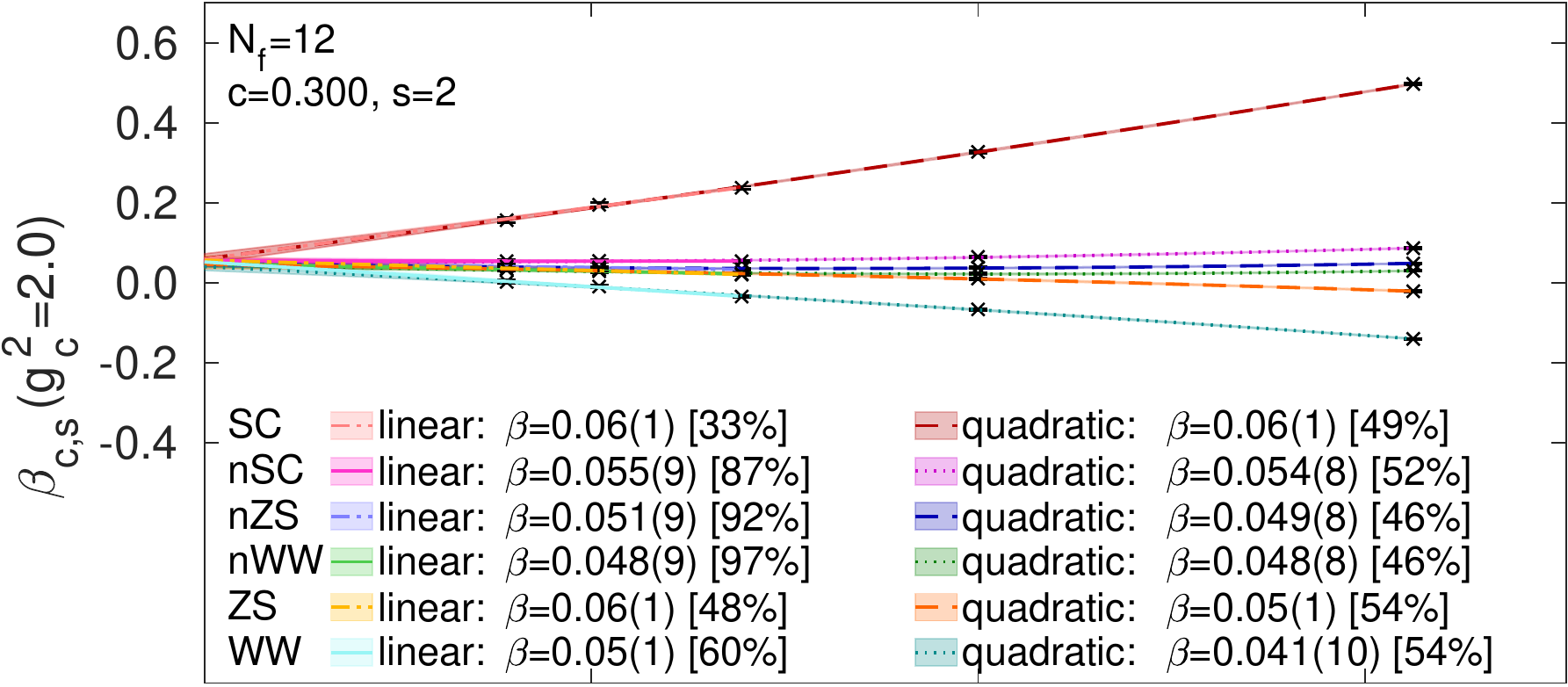}
  \includegraphics[width=0.98\columnwidth]{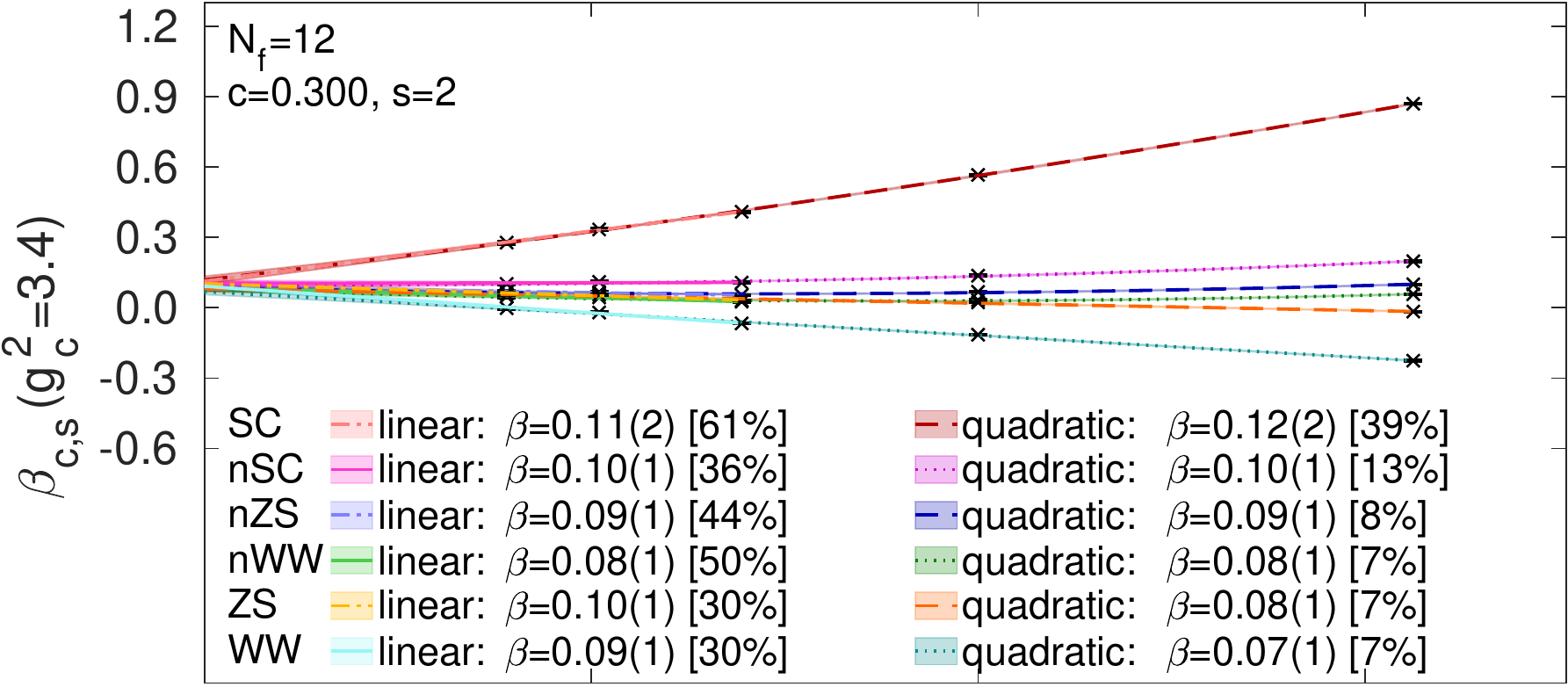}
  \includegraphics[width=0.98\columnwidth]{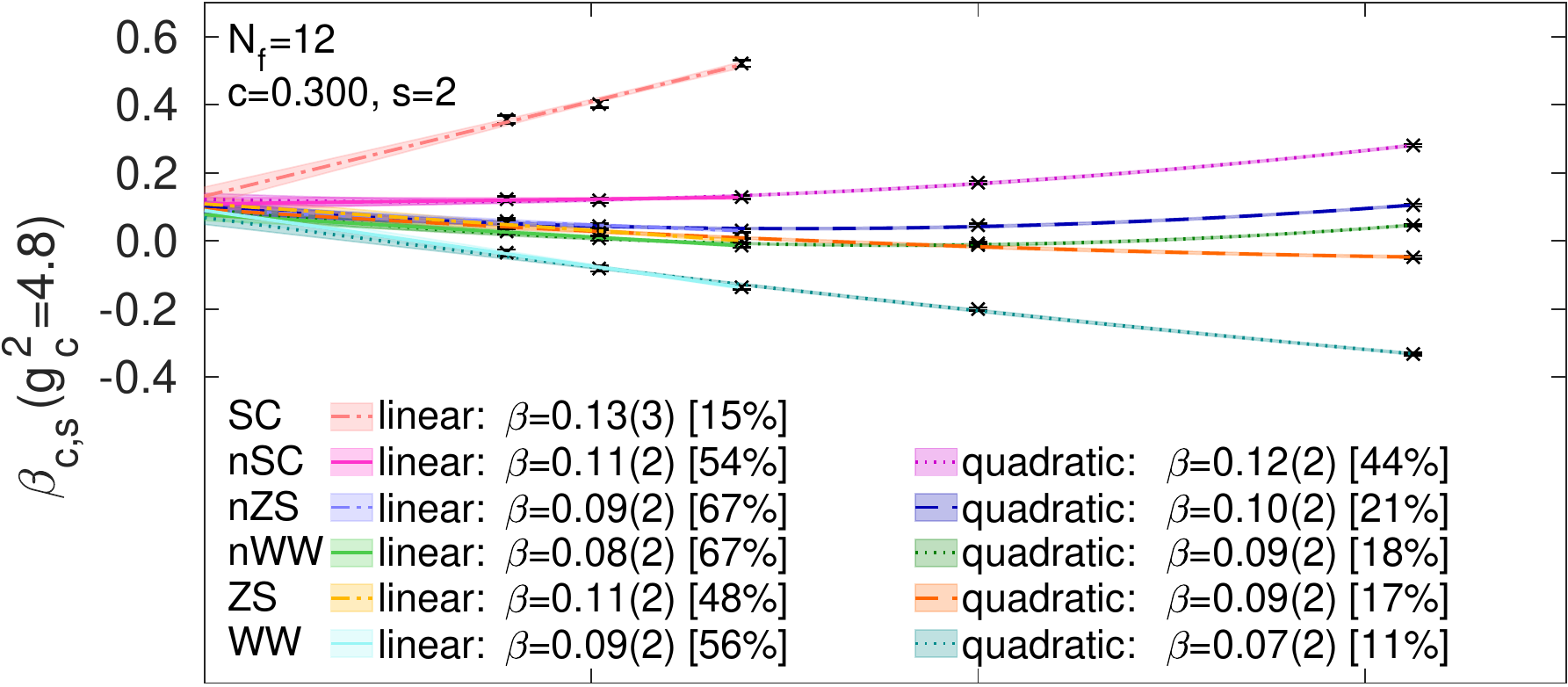}
  \includegraphics[width=0.98\columnwidth]{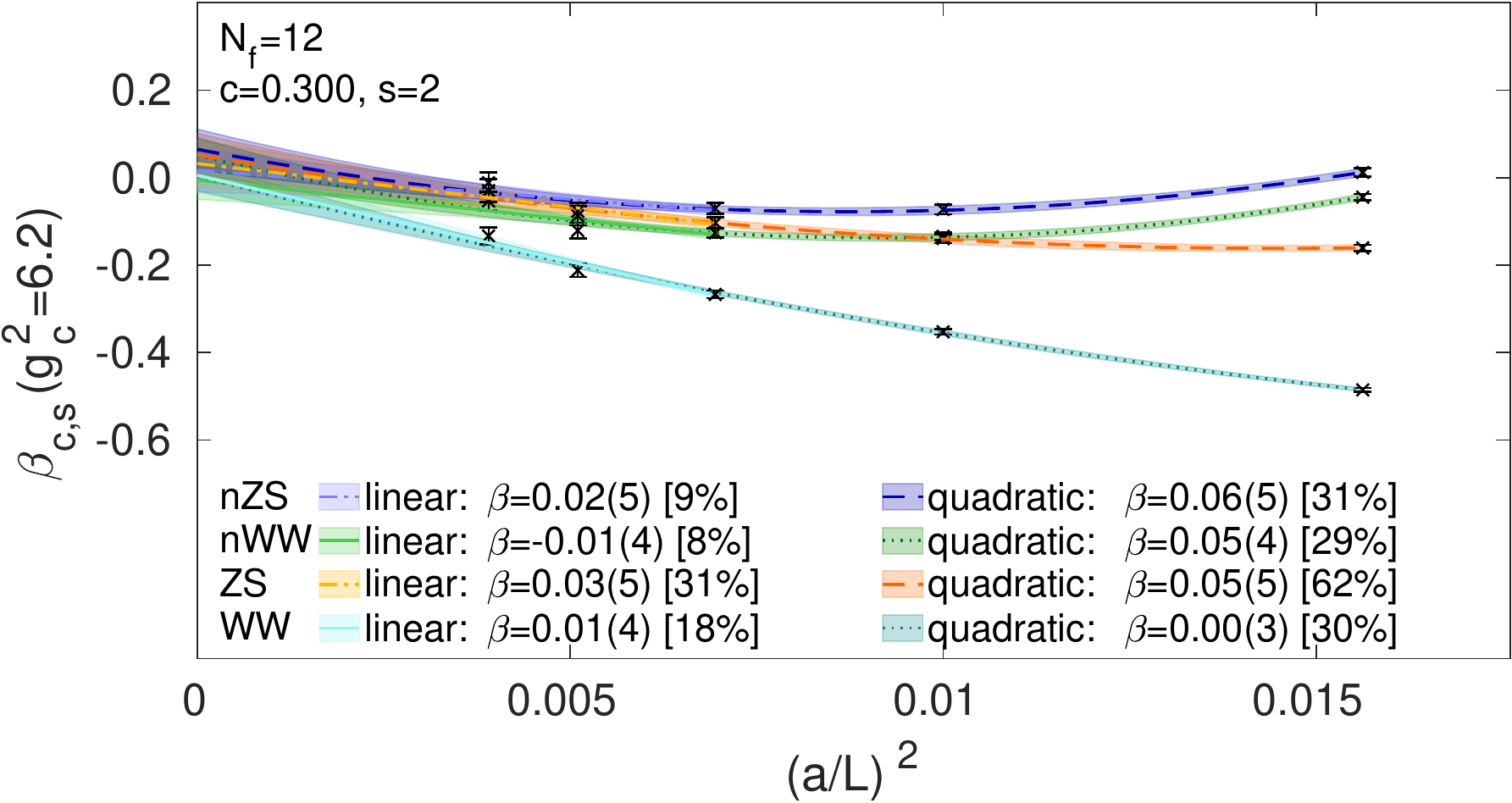}
  \caption{Details of the continuum extrapolation for our preferred (n)ZS data in comparison to alternative determinations based on (n)SC and (n)WW for $c=0.300$. Continuum limit values of $\beta_{c,s}$ and corresponding $p$-values of the extrapolation are quoted in the legend.}
  \label{Fig.cont_extra_c0300}  
\end{figure}
\clearpage

\bibliography{../General/BSM}
\bibliographystyle{apsrev4-1} % bst file
\end{document}

%% file: tln.tex
ZS & 8 & 1.098653 & 1.054568& 1.023078\\[1pt] 
ZS & 10 & 1.028937 & 1.008921& 0.992696\\[1pt] 
ZS & 12 & 1.006523 & 0.994355& 0.982724\\[1pt] 
ZS & 14 & 0.997574 & 0.988333& 0.978506\\[1pt] 
ZS & 16 & 0.993296 & 0.985414& 0.976448\\[1pt] 
ZS & 20 & 0.989702 & 0.982950& 0.974707\\[1pt] 
ZS & 24 & 0.988394 & 0.982052& 0.974073\\[1pt] 
ZS & 28 & 0.987826 & 0.981662& 0.973798\\[1pt] 
ZS & 32 & 0.987549 & 0.981472& 0.973664\\[1pt] 
\hline
ZW & 8 & 0.991436 & 0.967277& 0.949748\\[1pt] 
ZW & 10 & 0.963060 & 0.954041& 0.945790\\[1pt] 
ZW & 12 & 0.960664 & 0.955833& 0.949679\\[1pt] 
ZW & 14 & 0.963470 & 0.959650& 0.953905\\[1pt] 
ZW & 16 & 0.966875 & 0.963204& 0.957414\\[1pt] 
ZW & 20 & 0.972488 & 0.968506& 0.962350\\[1pt] 
ZW & 24 & 0.976300 & 0.971920& 0.965417\\[1pt] 
ZW & 28 & 0.978873 & 0.974170& 0.967403\\[1pt] 
ZW & 32 & 0.980657 & 0.975710& 0.968749\\[1pt] 
\hline
ZC & 8 & 0.692642 & 0.720334& 0.740038\\[1pt] 
ZC & 10 & 0.773982 & 0.795212& 0.809212\\[1pt] 
ZC & 12 & 0.827175 & 0.843095& 0.852580\\[1pt] 
ZC & 14 & 0.863400 & 0.875159& 0.881220\\[1pt] 
ZC & 16 & 0.888948 & 0.897503& 0.900980\\[1pt] 
ZC & 20 & 0.921408 & 0.925563& 0.925559\\[1pt] 
ZC & 24 & 0.940295 & 0.941715& 0.939585\\[1pt] 
ZC & 28 & 0.952163 & 0.951796& 0.948291\\[1pt] 
ZC & 32 & 0.960073 & 0.958485& 0.954048\\[1pt] 
\hline
SS & 8 & 1.004010 & 0.981303& 0.964323\\[1pt] 
SS & 10 & 0.975429 & 0.965750& 0.956887\\[1pt] 
SS & 12 & 0.970635 & 0.964851& 0.957992\\[1pt] 
SS & 14 & 0.971361 & 0.966657& 0.960279\\[1pt] 
SS & 16 & 0.973194 & 0.968759& 0.962429\\[1pt] 
SS & 20 & 0.976744 & 0.972206& 0.965663\\[1pt] 
SS & 24 & 0.979337 & 0.974545& 0.967756\\[1pt] 
SS & 28 & 0.981140 & 0.976124& 0.969139\\[1pt] 
SS & 32 & 0.982411 & 0.977218& 0.970087\\[1pt] 
\hline
SW & 8 & 0.913774 & 0.905341& 0.898897\\[1pt] 
SW & 10 & 0.916176 & 0.915408& 0.913243\\[1pt] 
SW & 12 & 0.928013 & 0.928589& 0.926585\\[1pt] 
SW & 14 & 0.939067 & 0.939241& 0.936594\\[1pt] 
SW & 16 & 0.947869 & 0.947315& 0.943949\\[1pt] 
SW & 20 & 0.960001 & 0.958090& 0.953542\\[1pt] 
SW & 24 & 0.967478 & 0.964576& 0.959217\\[1pt] 
SW & 28 & 0.972316 & 0.968720& 0.962808\\[1pt] 
SW & 32 & 0.975597 & 0.971509& 0.965210\\[1pt] 
\hline
SC & 8 & 0.657535 & 0.687715& 0.710135\\[1pt] 
SC & 10 & 0.744876 & 0.769007& 0.785720\\[1pt] 
SC & 12 & 0.803525 & 0.822215& 0.834139\\[1pt] 
SC & 14 & 0.844127 & 0.858375& 0.866552\\[1pt] 
SC & 16 & 0.873090 & 0.883830& 0.889124\\[1pt] 
SC & 20 & 0.910297 & 0.916108& 0.917444\\[1pt] 
SC & 24 & 0.932164 & 0.934852& 0.933730\\[1pt] 
SC & 28 & 0.945990 & 0.946612& 0.943886\\[1pt] 
SC & 32 & 0.955241 & 0.954442& 0.950622\\[1pt] 
\hline
WS & 8 & 1.417786 & 1.309105& 1.228189\\[1pt] 
WS & 10 & 1.212382 & 1.153760& 1.111117\\[1pt] 
WS & 12 & 1.124191 & 1.089496& 1.061970\\[1pt] 
WS & 14 & 1.081241 & 1.056847& 1.035972\\[1pt] 
WS & 16 & 1.056513 & 1.037441& 1.020202\\[1pt] 
WS & 20 & 1.029792 & 1.016056& 1.002600\\[1pt] 
WS & 24 & 1.016164 & 1.005008& 0.993423\\[1pt] 
WS & 28 & 1.008214 & 0.998522& 0.988011\\[1pt] 
WS & 32 & 1.003155 & 0.994380& 0.984546\\[1pt] 
\hline
WW & 8 & 1.255986 & 1.183080& 1.127496\\[1pt] 
WW & 10 & 1.123465 & 1.083497& 1.053546\\[1pt] 
WW & 12 & 1.067677 & 1.043763& 1.023836\\[1pt] 
WW & 14 & 1.041450 & 1.024261& 1.008595\\[1pt] 
WW & 16 & 1.026736 & 1.012918& 0.999521\\[1pt] 
WW & 20 & 1.011175 & 1.000644& 0.989555\\[1pt] 
WW & 24 & 1.003383 & 0.994402& 0.984431\\[1pt] 
WW & 28 & 0.998885 & 0.990771& 0.981433\\[1pt] 
WW & 32 & 0.996043 & 0.988465& 0.979523\\[1pt] 
\hline
WC & 8 & 0.808940 & 0.829206& 0.841227\\[1pt] 
WC & 10 & 0.869590 & 0.880908& 0.886360\\[1pt] 
WC & 12 & 0.903618 & 0.910174& 0.911931\\[1pt] 
WC & 14 & 0.924867 & 0.928370& 0.927770\\[1pt] 
WC & 16 & 0.938990 & 0.940417& 0.938230\\[1pt] 
WC & 20 & 0.955954 & 0.954837& 0.950722\\[1pt] 
WC & 24 & 0.965340 & 0.962790& 0.957597\\[1pt] 
WC & 28 & 0.971059 & 0.967628& 0.961775\\[1pt] 
WC & 32 & 0.974797 & 0.970786& 0.964499\\[1pt] 
\hline

%% file: gcSq_Nf12_nZS_ZS.tex
8 & 4.13 & 968 & 6.210(22)  & 6.912(24) & 1.6(3) & 6.369(27)  & 6.845(29) & 1.6(3) & 6.469(32)  & 6.799(33) & 1.7(3)\\ 
 8 & 4.15 & 905 & 5.922(11)  & 6.590(12) & 0.56(7) & 6.040(13)  & 6.492(14) & 0.57(8) & 6.105(15)  & 6.417(16) & 0.59(8)\\ 
 8 & 4.17 & 954 & 5.7406(98)  & 6.389(11) & 0.61(8) & 5.853(12)  & 6.291(13) & 0.62(8) & 5.918(14)  & 6.220(14) & 0.62(8)\\ 
 8 & 4.20 & 901 & 5.456(10)  & 6.072(11) & 0.9(1) & 5.552(12)  & 5.967(13) & 0.8(1) & 5.607(14)  & 5.892(15) & 0.8(1)\\ 
 8 & 4.25 & 1707 & 5.0721(53)  & 5.6451(59) & 0.54(5) & 5.1485(63)  & 5.5335(68) & 0.55(5) & 5.1924(74)  & 5.4570(78) & 0.57(5)\\ 
 8 & 4.30 & 1702 & 4.7804(53)  & 5.3204(59) & 0.67(7) & 4.8452(63)  & 5.2075(68) & 0.66(7) & 4.8824(74)  & 5.1312(78) & 0.66(7)\\ 
 8 & 4.40 & 1707 & 4.3109(39)  & 4.7978(43) & 0.50(4) & 4.3629(47)  & 4.6892(51) & 0.52(4) & 4.3940(57)  & 4.6179(59) & 0.54(4)\\ 
 8 & 4.50 & 1704 & 3.9479(31)  & 4.3938(35) & 0.48(3) & 3.9906(39)  & 4.2890(42) & 0.50(3) & 4.0161(50)  & 4.2208(53) & 0.60(6)\\ 
 8 & 4.60 & 1187 & 3.6564(36)  & 4.0694(40) & 0.49(4) & 3.6909(45)  & 3.9669(48) & 0.51(4) & 3.7106(53)  & 3.8997(56) & 0.52(5)\\ 
 8 & 4.70 & 759 & 3.4121(40)  & 3.7975(44) & 0.49(6) & 3.4417(48)  & 3.6990(52) & 0.48(6) & 3.4579(57)  & 3.6341(60) & 0.47(6)\\ 
 8 & 4.80 & 565 & 3.1971(53)  & 3.5583(59) & 0.7(1) & 3.2236(66)  & 3.4646(71) & 0.7(1) & 3.2390(81)  & 3.4040(85) & 0.7(1)\\ 
 8 & 5.00 & 741 & 2.8541(32)  & 3.1765(35) & 0.49(4) & 2.8735(39)  & 3.0884(42) & 0.48(4) & 2.8841(47)  & 3.0310(49) & 0.48(4)\\ 
 8 & 5.20 & 566 & 2.5805(37)  & 2.8719(41) & 0.6(1) & 2.5960(45)  & 2.7901(48) & 0.6(1) & 2.6043(52)  & 2.7370(54) & 0.57(8)\\ 
 8 & 5.50 & 1051 & 2.2618(21)  & 2.5173(23) & 0.48(3) & 2.2735(26)  & 2.4435(28) & 0.48(3) & 2.2797(31)  & 2.3958(33) & 0.49(3)\\ 
 8 & 6.00 & 821 & 1.8826(20)  & 2.0952(23) & 0.51(5) & 1.8900(25)  & 2.0313(27) & 0.51(5) & 1.8936(30)  & 1.9900(31) & 0.51(5)\\ 
 8 & 6.50 & 1051 & 1.6121(16)  & 1.7942(18) & 0.60(7) & 1.6152(20)  & 1.7360(21) & 0.58(7) & 1.6152(23)  & 1.6975(24) & 0.55(6)\\ 
 8 & 7.00 & 1081 & 1.4172(12)  & 1.5773(14) & 0.49(3) & 1.4200(16)  & 1.5261(17) & 0.50(3) & 1.4202(19)  & 1.4925(20) & 0.50(4)\\ 
 \hline 
10 & 4.15 & 901 & 6.190(10)  & 6.452(11) & 0.60(8) & 6.228(12)  & 6.404(12) & 0.60(8) & 6.237(13)  & 6.360(14) & 0.63(8)\\ 
 10 & 4.17 & 962 & 5.9430(97)  & 6.195(10) & 0.67(9) & 5.975(11)  & 6.144(12) & 0.69(10) & 5.985(13)  & 6.103(13) & 0.7(1)\\ 
 10 & 4.20 & 902 & 5.6467(84)  & 5.8858(88) & 0.61(8) & 5.6778(98)  & 5.838(10) & 0.61(8) & 5.690(11)  & 5.802(12) & 0.63(8)\\ 
 10 & 4.25 & 1111 & 5.2430(61)  & 5.4650(63) & 0.54(6) & 5.2737(74)  & 5.4227(76) & 0.59(7) & 5.2901(93)  & 5.3946(95) & 0.69(9)\\ 
 10 & 4.30 & 1595 & 4.9319(51)  & 5.1407(53) & 0.61(6) & 4.9620(64)  & 5.1022(66) & 0.71(8) & 4.9794(76)  & 5.0778(78) & 0.75(9)\\ 
 10 & 4.40 & 1623 & 4.4451(40)  & 4.6332(41) & 0.53(5) & 4.4740(48)  & 4.6004(49) & 0.55(5) & 4.4929(56)  & 4.5816(57) & 0.55(5)\\ 
 10 & 4.50 & 1679 & 4.0646(40)  & 4.2367(41) & 0.67(7) & 4.0891(48)  & 4.2046(50) & 0.70(7) & 4.1051(59)  & 4.1862(60) & 0.74(9)\\ 
 10 & 4.60 & 1050 & 3.7524(41)  & 3.9113(43) & 0.53(7) & 3.7735(51)  & 3.8802(52) & 0.56(7) & 3.7872(62)  & 3.8620(64) & 0.59(7)\\ 
 10 & 4.70 & 741 & 3.4948(49)  & 3.6427(51) & 0.62(9) & 3.5136(59)  & 3.6129(61) & 0.63(9) & 3.5256(71)  & 3.5953(72) & 0.64(9)\\ 
 10 & 4.80 & 851 & 3.2774(42)  & 3.4161(44) & 0.62(8) & 3.2953(50)  & 3.3884(51) & 0.60(8) & 3.3072(58)  & 3.3726(59) & 0.59(8)\\ 
 10 & 5.00 & 741 & 2.9110(39)  & 3.0343(40) & 0.63(9) & 2.9245(48)  & 3.0071(49) & 0.64(9) & 2.9332(61)  & 2.9912(62) & 0.7(1)\\ 
 10 & 5.20 & 821 & 2.6244(28)  & 2.7355(29) & 0.47(5) & 2.6333(35)  & 2.7077(36) & 0.50(5) & 2.6378(40)  & 2.6899(40) & 0.45(5)\\ 
 10 & 5.50 & 741 & 2.3000(29)  & 2.3973(30) & 0.57(7) & 2.3077(37)  & 2.3728(38) & 0.61(9) & 2.3119(46)  & 2.3576(47) & 0.7(1)\\ 
 10 & 6.00 & 681 & 1.9123(23)  & 1.9932(24) & 0.48(4) & 1.9179(28)  & 1.9721(29) & 0.49(4) & 1.9211(33)  & 1.9591(34) & 0.48(5)\\ 
 10 & 6.50 & 673 & 1.6349(21)  & 1.7041(22) & 0.56(9) & 1.6383(26)  & 1.6846(27) & 0.59(9) & 1.6397(32)  & 1.6720(32) & 0.61(9)\\ 
 10 & 7.00 & 660 & 1.4306(19)  & 1.4911(20) & 0.65(10) & 1.4324(23)  & 1.4728(24) & 0.65(10) & 1.4324(28)  & 1.4607(29) & 0.7(1)\\ 
 \hline 
12 & 4.15 & 1012 & 6.2402(96)  & 6.3627(98) & 0.72(10) & 6.230(11)  & 6.314(11) & 0.71(10) & 6.212(12)  & 6.271(12) & 0.7(1)\\ 
 12 & 4.17 & 962 & 6.008(10)  & 6.126(10) & 0.9(1) & 6.004(12)  & 6.084(12) & 1.0(2) & 5.993(14)  & 6.050(14) & 1.1(2)\\ 
 12 & 4.20 & 1385 & 5.7112(75)  & 5.8233(77) & 0.9(1) & 5.7173(92)  & 5.7940(93) & 1.0(1) & 5.716(11)  & 5.770(11) & 1.0(2)\\ 
 12 & 4.25 & 1094 & 5.3090(65)  & 5.4132(66) & 0.63(8) & 5.3204(80)  & 5.3917(81) & 0.68(9) & 5.323(10)  & 5.374(10) & 0.8(1)\\ 
 12 & 4.30 & 1050 & 4.9847(57)  & 5.0826(58) & 0.62(8) & 4.9982(72)  & 5.0652(72) & 0.69(9) & 5.0048(90)  & 5.0524(91) & 0.8(1)\\ 
 12 & 4.40 & 1062 & 4.5043(51)  & 4.5927(52) & 0.62(8) & 4.5218(65)  & 4.5824(66) & 0.69(9) & 4.5325(82)  & 4.5755(83) & 0.8(1)\\ 
 12 & 4.50 & 1032 & 4.1135(46)  & 4.1942(47) & 0.58(7) & 4.1281(57)  & 4.1834(58) & 0.62(8) & 4.1361(72)  & 4.1754(73) & 0.70(10)\\ 
 12 & 4.60 & 900 & 3.8003(49)  & 3.8749(50) & 0.65(8) & 3.8157(60)  & 3.8668(61) & 0.67(10) & 3.8253(73)  & 3.8616(74) & 0.7(1)\\ 
 12 & 4.70 & 995 & 3.5377(38)  & 3.6071(39) & 0.53(6) & 3.5499(50)  & 3.5975(51) & 0.62(8) & 3.5566(62)  & 3.5904(62) & 0.68(9)\\ 
 12 & 4.80 & 790 & 3.3122(37)  & 3.3772(38) & 0.49(5) & 3.3231(48)  & 3.3676(48) & 0.56(8) & 3.3290(59)  & 3.3606(59) & 0.61(9)\\ 
 12 & 5.00 & 796 & 2.9477(37)  & 3.0055(38) & 0.61(8) & 2.9580(46)  & 2.9977(46) & 0.63(9) & 2.9645(56)  & 2.9927(56) & 0.66(9)\\ 
 12 & 5.20 & 628 & 2.6628(37)  & 2.7151(37) & 0.55(8) & 2.6716(44)  & 2.7075(45) & 0.57(8) & 2.6771(56)  & 2.7025(57) & 0.7(1)\\ 
 12 & 5.50 & 685 & 2.3221(33)  & 2.3676(34) & 0.65(10) & 2.3278(39)  & 2.3590(40) & 0.6(1) & 2.3307(48)  & 2.3528(48) & 0.7(1)\\ 
 12 & 6.00 & 614 & 1.9256(29)  & 1.9634(29) & 0.7(1) & 1.9290(36)  & 1.9549(37) & 0.8(1) & 1.9299(44)  & 1.9483(45) & 0.8(2)\\ 
 12 & 6.50 & 617 & 1.6526(23)  & 1.6850(24) & 0.7(1) & 1.6559(29)  & 1.6781(30) & 0.7(1) & 1.6571(36)  & 1.6728(36) & 0.7(1)\\ 
 12 & 7.00 & 673 & 1.4446(17)  & 1.4729(18) & 0.49(7) & 1.4467(21)  & 1.4661(22) & 0.51(7) & 1.4473(26)  & 1.4611(26) & 0.54(7)\\ 
 \hline 
14 & 4.15 & 1400 & 6.2397(79)  & 6.3056(79) & 0.9(1) & 6.2190(93)  & 6.2642(93) & 0.9(1) & 6.203(11)  & 6.235(11) & 1.0(1)\\ 
 14 & 4.17 & 1012 & 6.0033(86)  & 6.0667(87) & 0.8(1) & 5.995(11)  & 6.038(11) & 0.9(1) & 5.987(13)  & 6.018(13) & 1.0(2)\\ 
 14 & 4.20 & 1002 & 5.7077(81)  & 5.7679(82) & 0.8(1) & 5.7053(98)  & 5.7468(99) & 0.8(1) & 5.702(12)  & 5.732(12) & 0.8(1)\\ 
 14 & 4.25 & 1001 & 5.3289(60)  & 5.3852(60) & 0.53(7) & 5.3348(74)  & 5.3736(75) & 0.59(7) & 5.3385(92)  & 5.3661(92) & 0.65(9)\\ 
 14 & 4.30 & 1001 & 5.0298(69)  & 5.0829(70) & 0.9(1) & 5.0386(84)  & 5.0752(85) & 0.9(1) & 5.044(10)  & 5.070(10) & 0.9(1)\\ 
 14 & 4.40 & 1002 & 4.5157(54)  & 4.5634(55) & 0.66(9) & 4.5201(67)  & 4.5529(68) & 0.7(1) & 4.5198(83)  & 4.5431(84) & 0.8(1)\\ 
 14 & 4.50 & 1002 & 4.1451(63)  & 4.1889(63) & 1.0(2) & 4.1559(79)  & 4.1862(79) & 1.1(2) & 4.1633(96)  & 4.1848(97) & 1.1(2)\\ 
 14 & 4.60 & 1001 & 3.8283(50)  & 3.8688(50) & 0.8(1) & 3.8374(61)  & 3.8653(62) & 0.9(1) & 3.8426(76)  & 3.8624(76) & 0.9(2)\\ 
 14 & 4.70 & 1004 & 3.5659(54)  & 3.6036(55) & 1.0(2) & 3.5746(69)  & 3.6006(70) & 1.1(2) & 3.5796(86)  & 3.5982(86) & 1.2(2)\\ 
 14 & 4.80 & 1003 & 3.3349(39)  & 3.3701(40) & 0.69(10) & 3.3423(50)  & 3.3666(50) & 0.7(1) & 3.3467(63)  & 3.3640(63) & 0.8(1)\\ 
 14 & 5.00 & 1002 & 2.9738(38)  & 3.0052(38) & 0.70(10) & 2.9825(48)  & 3.0041(48) & 0.8(1) & 2.9885(58)  & 3.0039(58) & 0.8(1)\\ 
 14 & 5.20 & 1001 & 2.6755(30)  & 2.7038(30) & 0.59(8) & 2.6810(37)  & 2.7005(37) & 0.63(9) & 2.6838(45)  & 2.6977(45) & 0.69(10)\\ 
 14 & 5.50 & 1002 & 2.3444(26)  & 2.3691(26) & 0.60(8) & 2.3509(32)  & 2.3680(33) & 0.64(8) & 2.3554(39)  & 2.3675(39) & 0.67(9)\\ 
 14 & 6.00 & 1001 & 1.9407(23)  & 1.9612(23) & 0.7(1) & 1.9449(29)  & 1.9591(29) & 0.8(1) & 1.9476(37)  & 1.9577(38) & 0.9(1)\\ 
 14 & 6.50 & 1002 & 1.6569(17)  & 1.6744(18) & 0.58(7) & 1.6586(22)  & 1.6707(22) & 0.61(8) & 1.6586(26)  & 1.6672(26) & 0.64(8)\\ 
 14 & 7.00 & 1003 & 1.4516(16)  & 1.4669(16) & 0.63(8) & 1.4536(21)  & 1.4642(21) & 0.71(10) & 1.4547(26)  & 1.4622(26) & 0.8(1)\\ 
 \hline 
16 & 4.13 & 608 & 6.491(13)  & 6.532(13) & 0.9(2) & 6.454(16)  & 6.482(16) & 1.0(2) & 6.427(19)  & 6.446(19) & 1.1(3)\\ 
 16 & 4.15 & 657 & 6.219(12)  & 6.257(12) & 1.0(2) & 6.200(15)  & 6.227(15) & 1.2(2) & 6.187(19)  & 6.206(19) & 1.3(3)\\ 
 16 & 4.17 & 961 & 5.9924(83)  & 6.0297(84) & 0.8(1) & 5.976(10)  & 6.002(10) & 0.9(1) & 5.965(13)  & 5.983(13) & 1.0(2)\\ 
 16 & 4.20 & 922 & 5.679(10)  & 5.715(10) & 1.3(2) & 5.667(12)  & 5.691(12) & 1.3(2) & 5.654(15)  & 5.671(15) & 1.4(3)\\ 
 16 & 4.25 & 886 & 5.3213(77)  & 5.3545(77) & 0.8(1) & 5.3192(97)  & 5.3421(98) & 0.9(1) & 5.317(12)  & 5.333(12) & 0.9(2)\\ 
 16 & 4.30 & 889 & 5.0244(77)  & 5.0557(77) & 0.9(1) & 5.0273(100)  & 5.049(10) & 1.0(2) & 5.030(12)  & 5.045(12) & 1.1(2)\\ 
 16 & 4.40 & 896 & 4.5468(63)  & 4.5751(64) & 0.9(2) & 4.5554(80)  & 4.5749(80) & 1.0(2) & 4.5634(98)  & 4.5773(99) & 1.0(2)\\ 
 16 & 4.50 & 877 & 4.1536(68)  & 4.1795(68) & 1.1(2) & 4.1592(84)  & 4.1771(84) & 1.2(2) & 4.163(10)  & 4.176(10) & 1.2(2)\\ 
 16 & 4.60 & 783 & 3.8457(53)  & 3.8697(54) & 0.66(9) & 3.8529(66)  & 3.8694(66) & 0.7(1) & 3.8578(82)  & 3.8696(83) & 0.8(1)\\ 
 16 & 4.70 & 675 & 3.5801(73)  & 3.6023(73) & 1.4(3) & 3.589(10)  & 3.604(10) & 1.9(5) & 3.597(14)  & 3.608(14) & 2.3(6)\\ 
 16 & 4.80 & 602 & 3.3545(74)  & 3.3754(75) & 1.4(3) & 3.3619(97)  & 3.3764(98) & 1.6(4) & 3.368(12)  & 3.378(12) & 1.9(5)\\ 
 16 & 5.00 & 621 & 2.9900(49)  & 3.0086(49) & 0.9(2) & 2.9983(62)  & 3.0112(62) & 0.9(2) & 3.0052(76)  & 3.0144(76) & 1.0(2)\\ 
 16 & 5.20 & 514 & 2.6931(46)  & 2.7099(47) & 0.7(1) & 2.6998(60)  & 2.7114(61) & 0.8(2) & 2.7052(78)  & 2.7135(79) & 1.0(2)\\ 
 16 & 5.50 & 561 & 2.3487(34)  & 2.3633(35) & 0.6(1) & 2.3529(43)  & 2.3630(43) & 0.7(1) & 2.3556(50)  & 2.3628(51) & 0.7(1)\\ 
 16 & 6.00 & 537 & 1.9470(37)  & 1.9591(37) & 0.9(2) & 1.9503(46)  & 1.9587(47) & 1.1(2) & 1.9522(57)  & 1.9582(57) & 1.1(2)\\ 
 16 & 6.50 & 543 & 1.6695(31)  & 1.6799(32) & 1.0(2) & 1.6737(39)  & 1.6809(39) & 1.0(2) & 1.6771(47)  & 1.6822(48) & 1.1(2)\\ 
 16 & 7.00 & 471 & 1.4536(20)  & 1.4626(20) & 0.43(6) & 1.4537(24)  & 1.4599(25) & 0.47(6) & 1.4525(29)  & 1.4569(29) & 0.50(8)\\ 
 \hline 
20 & 4.15 & 701 & 6.134(12)  & 6.150(12) & 1.2(2) & 6.118(15)  & 6.129(15) & 1.3(3) & 6.108(18)  & 6.116(18) & 1.4(3)\\ 
 20 & 4.17 & 642 & 5.941(14)  & 5.956(14) & 1.6(4) & 5.931(19)  & 5.942(19) & 2.0(5) & 5.928(24)  & 5.935(24) & 2.3(6)\\ 
 20 & 4.20 & 499 & 5.678(14)  & 5.692(14) & 1.3(3) & 5.669(17)  & 5.679(17) & 1.5(4) & 5.662(22)  & 5.669(22) & 1.7(4)\\ 
 20 & 4.25 & 902 & 5.3314(93)  & 5.3452(94) & 1.3(2) & 5.334(12)  & 5.344(12) & 1.4(3) & 5.341(15)  & 5.348(15) & 1.6(3)\\ 
 20 & 4.30 & 923 & 5.0325(89)  & 5.0455(89) & 1.3(3) & 5.036(11)  & 5.045(11) & 1.4(3) & 5.042(13)  & 5.048(13) & 1.4(3)\\ 
 20 & 4.40 & 851 & 4.562(12)  & 4.573(12) & 2.6(7) & 4.570(15)  & 4.578(15) & 2.9(7) & 4.580(18)  & 4.586(18) & 3.1(8)\\ 
 20 & 4.50 & 985 & 4.1592(70)  & 4.1699(70) & 1.2(2) & 4.1640(90)  & 4.1714(90) & 1.4(3) & 4.169(11)  & 4.174(11) & 1.5(3)\\ 
 20 & 4.60 & 863 & 3.8684(97)  & 3.8784(97) & 2.2(5) & 3.877(13)  & 3.884(13) & 2.5(6) & 3.885(16)  & 3.890(16) & 2.9(7)\\ 
 20 & 4.70 & 759 & 3.6016(65)  & 3.6109(65) & 1.2(2) & 3.6112(83)  & 3.6177(83) & 1.3(3) & 3.621(10)  & 3.626(10) & 1.4(3)\\ 
 20 & 4.80 & 751 & 3.3777(75)  & 3.3864(75) & 1.6(4) & 3.3865(94)  & 3.3926(94) & 1.8(4) & 3.395(11)  & 3.399(11) & 1.8(4)\\ 
 20 & 5.00 & 801 & 2.9961(44)  & 3.0038(44) & 0.9(1) & 3.0017(57)  & 3.0070(57) & 1.0(2) & 3.0066(75)  & 3.0104(75) & 1.3(3)\\ 
 20 & 5.20 & 691 & 2.7139(40)  & 2.7209(40) & 0.9(2) & 2.7219(53)  & 2.7268(53) & 1.0(2) & 2.7295(70)  & 2.7329(70) & 1.3(3)\\ 
 20 & 5.50 & 621 & 2.3736(51)  & 2.3797(51) & 1.4(3) & 2.3796(65)  & 2.3839(65) & 1.6(4) & 2.3849(81)  & 2.3879(81) & 1.7(4)\\ 
 20 & 6.00 & 606 & 1.9601(40)  & 1.9652(41) & 1.2(3) & 1.9632(51)  & 1.9667(51) & 1.3(3) & 1.9653(65)  & 1.9678(65) & 1.5(4)\\ 
 20 & 6.50 & 610 & 1.6738(30)  & 1.6782(30) & 1.1(2) & 1.6743(37)  & 1.6773(37) & 1.1(2) & 1.6735(44)  & 1.6757(44) & 1.1(2)\\ 
 20 & 7.00 & 505 & 1.4654(27)  & 1.4692(27) & 0.9(2) & 1.4669(35)  & 1.4695(35) & 1.0(2) & 1.4673(44)  & 1.4692(44) & 1.1(3)\\ 
 \hline 
24 & 4.15 & 435 & 6.094(18)  & 6.102(18) & 1.8(5) & 6.091(22)  & 6.096(22) & 2.0(6) & 6.094(28)  & 6.098(28) & 2.2(7)\\ 
 24 & 4.17 & 615 & 5.893(14)  & 5.900(14) & 1.6(4) & 5.893(18)  & 5.899(18) & 1.8(5) & 5.901(23)  & 5.905(23) & 2.0(5)\\ 
 24 & 4.20 & 552 & 5.675(17)  & 5.683(17) & 2.3(6) & 5.679(22)  & 5.684(22) & 2.5(7) & 5.687(27)  & 5.691(27) & 2.7(8)\\ 
 24 & 4.25 & 663 & 5.324(13)  & 5.330(13) & 1.8(5) & 5.334(17)  & 5.339(17) & 2.0(5) & 5.350(21)  & 5.353(21) & 2.2(6)\\ 
 24 & 4.30 & 840 & 5.041(12)  & 5.047(12) & 2.0(5) & 5.055(15)  & 5.059(15) & 2.3(5) & 5.072(19)  & 5.075(19) & 2.6(7)\\ 
 24 & 4.40 & 1056 & 4.5645(91)  & 4.5702(91) & 2.0(4) & 4.574(12)  & 4.578(12) & 2.2(5) & 4.584(15)  & 4.587(15) & 2.5(6)\\ 
 24 & 4.50 & 932 & 4.1981(86)  & 4.2034(86) & 1.8(4) & 4.213(11)  & 4.217(11) & 2.0(4) & 4.230(14)  & 4.233(14) & 2.2(5)\\ 
 24 & 4.60 & 910 & 3.8613(78)  & 3.8662(79) & 1.7(4) & 3.8653(98)  & 3.8687(98) & 1.8(4) & 3.868(12)  & 3.871(12) & 2.0(5)\\ 
 24 & 4.70 & 786 & 3.6073(56)  & 3.6118(56) & 1.0(2) & 3.6139(71)  & 3.6171(71) & 1.2(2) & 3.6200(91)  & 3.6222(91) & 1.4(3)\\ 
 24 & 4.80 & 642 & 3.3858(74)  & 3.3900(74) & 1.5(4) & 3.3888(93)  & 3.3917(93) & 1.6(4) & 3.389(11)  & 3.391(11) & 1.7(4)\\ 
 24 & 5.00 & 556 & 3.0222(58)  & 3.0260(58) & 1.1(2) & 3.0307(75)  & 3.0334(75) & 1.2(3) & 3.039(10)  & 3.040(10) & 1.6(4)\\ 
 24 & 5.20 & 610 & 2.7263(66)  & 2.7297(66) & 1.8(5) & 2.7335(88)  & 2.7359(88) & 2.2(6) & 2.740(12)  & 2.742(12) & 2.6(8)\\ 
 24 & 5.50 & 502 & 2.3910(53)  & 2.3940(53) & 1.3(3) & 2.3988(68)  & 2.4008(68) & 1.4(4) & 2.4058(87)  & 2.4073(87) & 1.6(4)\\ 
 24 & 6.00 & 483 & 1.9684(68)  & 1.9708(68) & 2.9(9) & 1.9723(87)  & 1.9740(87) & 3(1) & 1.976(11)  & 1.977(11) & 3(1)\\ 
 24 & 6.50 & 502 & 1.6848(34)  & 1.6869(34) & 1.1(3) & 1.6892(43)  & 1.6906(43) & 1.2(3) & 1.6932(55)  & 1.6942(55) & 1.5(4)\\ 
 24 & 7.00 & 502 & 1.4721(33)  & 1.4740(33) & 1.3(3) & 1.4741(41)  & 1.4754(41) & 1.4(3) & 1.4752(51)  & 1.4761(51) & 1.5(4)\\ 
 \hline 
28 & 4.15 & 254 & 6.075(40)  & 6.079(40) & 4(2) & 6.077(49)  & 6.080(49) & 5(2) & 6.088(58)  & 6.090(58) & 4(2)\\ 
 28 & 4.17 & 372 & 5.889(23)  & 5.893(23) & 2.3(8) & 5.894(30)  & 5.897(30) & 2.6(9) & 5.907(38)  & 5.909(38) & 3(1)\\ 
 28 & 4.20 & 433 & 5.673(18)  & 5.677(18) & 2.2(7) & 5.685(24)  & 5.687(24) & 2.5(8) & 5.702(31)  & 5.703(31) & 2.8(10)\\ 
 28 & 4.25 & 472 & 5.327(21)  & 5.331(21) & 3(1) & 5.340(27)  & 5.342(27) & 3(1) & 5.356(33)  & 5.357(33) & 4(1)\\ 
 28 & 4.30 & 341 & 5.042(21)  & 5.045(21) & 3(1) & 5.060(26)  & 5.063(26) & 3(1) & 5.082(33)  & 5.083(33) & 3(1)\\ 
 28 & 4.40 & 322 & 4.579(14)  & 4.582(14) & 1.3(4) & 4.590(17)  & 4.592(17) & 1.4(5) & 4.600(21)  & 4.602(21) & 1.6(5)\\ 
 28 & 4.50 & 362 & 4.195(14)  & 4.198(14) & 2.1(7) & 4.205(18)  & 4.207(18) & 2.3(8) & 4.215(22)  & 4.217(22) & 2.6(9)\\ 
 28 & 4.60 & 361 & 3.890(12)  & 3.893(12) & 1.6(5) & 3.903(16)  & 3.905(16) & 2.1(7) & 3.917(23)  & 3.918(23) & 2.8(10)\\ 
 28 & 4.70 & 334 & 3.637(11)  & 3.639(11) & 1.7(6) & 3.650(14)  & 3.651(14) & 2.0(7) & 3.662(19)  & 3.663(19) & 2.6(9)\\ 
 28 & 4.80 & 361 & 3.419(10)  & 3.421(10) & 1.7(5) & 3.434(13)  & 3.436(13) & 1.9(6) & 3.450(17)  & 3.451(17) & 2.3(8)\\ 
 28 & 5.00 & 340 & 3.0554(84)  & 3.0575(84) & 1.6(5) & 3.072(11)  & 3.074(11) & 1.8(6) & 3.090(14)  & 3.091(14) & 2.0(7)\\ 
 28 & 5.20 & 323 & 2.7563(77)  & 2.7582(77) & 1.6(5) & 2.7685(98)  & 2.7698(98) & 1.8(6) & 2.780(13)  & 2.781(13) & 2.2(7)\\ 
 28 & 5.50 & 329 & 2.3881(99)  & 2.3898(99) & 2.7(10) & 2.395(13)  & 2.396(13) & 3(1) & 2.402(18)  & 2.403(18) & 4(2)\\ 
 28 & 6.00 & 360 & 1.9800(80)  & 1.9813(80) & 3(1) & 1.984(11)  & 1.985(11) & 3(1) & 1.986(14)  & 1.987(14) & 4(2)\\ 
 28 & 6.50 & 361 & 1.6940(70)  & 1.6951(70) & 3(1) & 1.6967(91)  & 1.6975(91) & 3(1) & 1.698(11)  & 1.699(11) & 4(1)\\ 
 28 & 7.00 & 361 & 1.4809(42)  & 1.4819(42) & 1.7(5) & 1.4843(57)  & 1.4850(57) & 2.1(7) & 1.4872(72)  & 1.4877(72) & 2.3(8)\\ 
 \hline 
32 & 4.15 & 263 & 6.069(44)  & 6.071(44) & 6(3) & 6.076(54)  & 6.077(54) & 6(3) & 6.086(64)  & 6.087(64) & 7(3)\\ 
 32 & 4.17 & 302 & 5.906(27)  & 5.908(27) & 3(1) & 5.920(36)  & 5.921(36) & 3(1) & 5.937(46)  & 5.939(46) & 4(2)\\ 
 32 & 4.20 & 446 & 5.686(20)  & 5.689(20) & 2.4(8) & 5.708(25)  & 5.710(25) & 2.6(9) & 5.735(32)  & 5.736(32) & 3(1)\\ 
 32 & 4.25 & 371 & 5.351(23)  & 5.353(23) & 3(1) & 5.372(31)  & 5.373(31) & 4(2) & 5.397(38)  & 5.398(38) & 4(2)\\ 
 32 & 4.30 & 347 & 5.063(18)  & 5.065(18) & 2.3(8) & 5.081(25)  & 5.083(25) & 3(1) & 5.101(32)  & 5.102(32) & 3(1)\\ 
 32 & 4.40 & 322 & 4.606(15)  & 4.608(15) & 2.1(7) & 4.631(19)  & 4.633(19) & 2.3(8) & 4.659(23)  & 4.660(23) & 2.4(9)\\ 
 32 & 4.50 & 321 & 4.236(17)  & 4.237(17) & 2.2(7) & 4.253(22)  & 4.254(22) & 2.4(9) & 4.272(28)  & 4.273(28) & 3(1)\\ 
 32 & 4.60 & 478 & 3.898(16)  & 3.899(16) & 4(1) & 3.910(21)  & 3.912(21) & 4(1) & 3.923(26)  & 3.924(26) & 5(2)\\ 
 32 & 4.70 & 318 & 3.656(20)  & 3.657(20) & 3(1) & 3.672(26)  & 3.673(26) & 4(2) & 3.687(34)  & 3.688(34) & 5(2)\\ 
 32 & 4.80 & 319 & 3.423(14)  & 3.424(14) & 3(1) & 3.433(17)  & 3.434(17) & 3(1) & 3.440(21)  & 3.441(21) & 3(1)\\ 
 32 & 5.00 & 339 & 3.062(17)  & 3.064(17) & 5(2) & 3.080(22)  & 3.081(22) & 5(2) & 3.099(28)  & 3.099(29) & 6(3)\\ 
 32 & 5.20 & 350 & 2.753(19)  & 2.754(19) & 7(3) & 2.763(24)  & 2.764(24) & 7(3) & 2.773(30)  & 2.773(30) & 8(4)\\ 
 32 & 5.50 & 319 & 2.407(12)  & 2.408(12) & 4(2) & 2.416(15)  & 2.417(15) & 4(2) & 2.424(18)  & 2.424(18) & 5(2)\\ 
 32 & 6.00 & 351 & 2.0029(58)  & 2.0038(58) & 1.8(6) & 2.0134(76)  & 2.0139(76) & 2.2(7) & 2.024(10)  & 2.024(10) & 3(1)\\ 
 32 & 6.50 & 351 & 1.6998(52)  & 1.7005(52) & 1.6(5) & 1.7041(68)  & 1.7046(68) & 1.9(6) & 1.7078(87)  & 1.7081(87) & 2.2(8)\\ 
 32 & 7.00 & 351 & 1.4875(61)  & 1.4881(61) & 3(1) & 1.4924(75)  & 1.4928(75) & 3(1) & 1.4972(89)  & 1.4975(89) & 3(1)\\ 
 

%% file: gcSq_Nf12_nZS_ZS_Ls.tex
8 & 4.15 & 12 & 461 & 5.911(19)  & 6.578(21) & 0.9(2) & 6.025(23)  & 6.475(25) & 1.0(2) & 6.084(26)  & 6.394(28) & 1.0(2)\\ 
\setrow{\bfseries} 8 & 4.15 & 16 & 905 & 5.922(11)  & 6.590(12) & 0.56(7) & 6.040(13)  & 6.492(14) & 0.57(8) & 6.105(15)  & 6.417(16) & 0.60(8)\\ 
 8 & 4.15 & 32 & 986 & 5.925(12)  & 6.594(14) & 0.8(1) & 6.050(15)  & 6.502(16) & 0.8(1) & 6.125(17)  & 6.437(18) & 0.8(1)\\ 
 8 & 4.20 & 12 & 986 & 5.4209(83)  & 6.0332(92) & 0.59(7) & 5.5108(98)  & 5.923(11) & 0.58(7) & 5.561(11)  & 5.845(12) & 0.58(7)\\ 
\setrow{\bfseries} 8 & 4.20 & 16 & 901 & 5.456(10)  & 6.072(11) & 0.8(1) & 5.552(12)  & 5.967(13) & 0.8(1) & 5.607(14)  & 5.892(15) & 0.8(1)\\ 
 \hline 
10 & 4.15 & 12 & 501 & 6.151(14)  & 6.411(15) & 0.7(1) & 6.180(16)  & 6.355(16) & 0.7(1) & 6.185(17)  & 6.307(18) & 0.7(1)\\ 
\setrow{\bfseries} 10 & 4.15 & 16 & 901 & 6.190(10)  & 6.452(11) & 0.60(8) & 6.228(12)  & 6.404(12) & 0.61(8) & 6.237(13)  & 6.360(14) & 0.64(8)\\ 
 10 & 4.20 & 12 & 941 & 5.6509(81)  & 5.8901(84) & 0.58(7) & 5.6845(93)  & 5.8451(96) & 0.58(8) & 5.700(11)  & 5.812(11) & 0.59(8)\\ 
\setrow{\bfseries} 10 & 4.20 & 16 & 902 & 5.6467(84)  & 5.8858(88) & 0.61(8) & 5.6778(98)  & 5.838(10) & 0.61(8) & 5.690(11)  & 5.802(12) & 0.63(8)\\ 
 \hline 
12 & 4.15 & 12 & 543 & 6.215(14)  & 6.337(14) & 0.8(2) & 6.207(17)  & 6.290(17) & 0.9(2) & 6.193(21)  & 6.252(21) & 1.1(2)\\ 
\setrow{\bfseries} 12 & 4.15 & 16 & 1012 & 6.2402(96)  & 6.3627(98) & 0.72(10) & 6.230(11)  & 6.314(11) & 0.71(10) & 6.212(12)  & 6.271(12) & 0.7(1)\\ 
 12 & 4.20 & 12 & 909 & 5.6942(72)  & 5.8060(74) & 0.58(8) & 5.6993(85)  & 5.7757(86) & 0.59(8) & 5.697(10)  & 5.751(10) & 0.61(8)\\ 
\setrow{\bfseries} 12 & 4.20 & 16 & 1385 & 5.7112(75)  & 5.8233(77) & 0.9(1) & 5.7173(92)  & 5.7940(93) & 1.0(1) & 5.716(11)  & 5.770(11) & 1.0(2)\\ 
 \hline 
16 & 4.15 & 12 & 521 & 6.166(12)  & 6.205(12) & 0.9(2) & 6.152(15)  & 6.178(15) & 1.0(2) & 6.142(19)  & 6.161(19) & 1.1(3)\\ 
\setrow{\bfseries} 16 & 4.15 & 24 & 657 & 6.219(12)  & 6.257(12) & 1.0(2) & 6.200(15)  & 6.227(15) & 1.2(3) & 6.187(19)  & 6.206(19) & 1.3(3)\\ 
 16 & 4.15 & 32 & 806 & 6.230(13)  & 6.269(13) & 1.5(3) & 6.208(16)  & 6.235(16) & 1.7(4) & 6.194(20)  & 6.213(20) & 1.9(4)\\ 
 16 & 4.20 & 12 & 535 & 5.694(10)  & 5.730(10) & 0.8(2) & 5.689(13)  & 5.714(13) & 0.9(2) & 5.686(17)  & 5.704(17) & 1.2(3)\\ 
\setrow{\bfseries} 16 & 4.20 & 16 & 922 & 5.679(10)  & 5.715(10) & 1.3(2) & 5.667(12)  & 5.691(12) & 1.3(2) & 5.654(15)  & 5.671(15) & 1.4(3)\\ 
 16 & 4.20 & 24 & 19 & 5.688(65)  & 5.724(66) & 2(1) & 5.683(81)  & 5.707(82) & 2(1) & 5.686(96)  & 5.703(96) & 2(1)\\ 
 \hline 
20 & 4.20 & 12 & 702 & 5.658(13)  & 5.673(13) & 1.6(4) & 5.649(17)  & 5.659(17) & 1.9(5) & 5.644(22)  & 5.651(22) & 2.1(5)\\ 
\setrow{\bfseries} 20 & 4.20 & 16 & 499 & 5.678(14)  & 5.692(14) & 1.3(3) & 5.669(17)  & 5.679(17) & 1.5(4) & 5.662(22)  & 5.669(22) & 1.7(4)\\ 
 \hline 
24 & 4.15 & 16 & 97 & 6.147(55)  & 6.155(55) & 3(2) & 6.154(66)  & 6.159(66) & 3(2) & 6.164(80)  & 6.168(80) & 3(2)\\ 
\setrow{\bfseries} 24 & 4.15 & 24 & 435 & 6.094(18)  & 6.102(18) & 1.8(5) & 6.091(22)  & 6.096(22) & 2.0(6) & 6.094(28)  & 6.098(28) & 2.2(7)\\ 
 24 & 4.15 & 32 & 417 & 6.134(26)  & 6.142(27) & 3(1) & 6.135(32)  & 6.140(32) & 3(1) & 6.142(39)  & 6.145(39) & 3(1)\\ 
 \hline 
32 & 4.30 & 12 & 132 & 5.092(31)  & 5.094(31) & 3(1) & 5.125(40)  & 5.126(40) & 3(1) & 5.164(51)  & 5.165(51) & 3(2)\\ 
\setrow{\bfseries} 32 & 4.30 & 16 & 347 & 5.063(18)  & 5.065(18) & 2.3(8) & 5.081(25)  & 5.083(25) & 3(1) & 5.101(32)  & 5.102(32) & 3(1)\\

%% file: Nf12StepScaling.bbl
\begin{thebibliography}{50}%
\makeatletter
\providecommand \@ifxundefined [1]{%
 \@ifx{#1\undefined}
}%
\providecommand \@ifnum [1]{%
 \ifnum #1\expandafter \@firstoftwo
 \else \expandafter \@secondoftwo
 \fi
}%
\providecommand \@ifx [1]{%
 \ifx #1\expandafter \@firstoftwo
 \else \expandafter \@secondoftwo
 \fi
}%
\providecommand \natexlab [1]{#1}%
\providecommand \enquote  [1]{``#1''}%
\providecommand \bibnamefont  [1]{#1}%
\providecommand \bibfnamefont [1]{#1}%
\providecommand \citenamefont [1]{#1}%
\providecommand \href@noop [0]{\@secondoftwo}%
\providecommand \href [0]{\begingroup \@sanitize@url \@href}%
\providecommand \@href[1]{\@@startlink{#1}\@@href}%
\providecommand \@@href[1]{\endgroup#1\@@endlink}%
\providecommand \@sanitize@url [0]{\catcode `\\12\catcode `\$12\catcode
  `\&12\catcode `\#12\catcode `\^12\catcode `\_12\catcode `\%12\relax}%
\providecommand \@@startlink[1]{}%
\providecommand \@@endlink[0]{}%
\providecommand \url  [0]{\begingroup\@sanitize@url \@url }%
\providecommand \@url [1]{\endgroup\@href {#1}{\urlprefix }}%
\providecommand \urlprefix  [0]{URL }%
\providecommand \Eprint [0]{\href }%
\providecommand \doibase [0]{http://dx.doi.org/}%
\providecommand \selectlanguage [0]{\@gobble}%
\providecommand \bibinfo  [0]{\@secondoftwo}%
\providecommand \bibfield  [0]{\@secondoftwo}%
\providecommand \translation [1]{[#1]}%
\providecommand \BibitemOpen [0]{}%
\providecommand \bibitemStop [0]{}%
\providecommand \bibitemNoStop [0]{.\EOS\space}%
\providecommand \EOS [0]{\spacefactor3000\relax}%
\providecommand \BibitemShut  [1]{\csname bibitem#1\endcsname}%
\let\auto@bib@innerbib\@empty
%</preamble>
\bibitem [{\citenamefont {Banks}\ and\ \citenamefont
  {Zaks}(1982)}]{Banks:1981nn}%
  \BibitemOpen
  \bibfield  {author} {\bibinfo {author} {\bibfnamefont {T.}~\bibnamefont
  {Banks}}\ and\ \bibinfo {author} {\bibfnamefont {A.}~\bibnamefont {Zaks}},\
  }\href {\doibase 10.1016/0550-3213(82)90035-9} {\bibfield  {journal}
  {\bibinfo  {journal} {Nucl. Phys.}\ }\textbf {\bibinfo {volume} {B196}},\
  \bibinfo {pages} {189} (\bibinfo {year} {1982})}\BibitemShut {NoStop}%
%%CITATION = NUPHA,B196,189;%%
\bibitem [{\citenamefont {DeGrand}(2016)}]{DeGrand:2015zxa}%
  \BibitemOpen
  \bibfield  {author} {\bibinfo {author} {\bibfnamefont {T.}~\bibnamefont
  {DeGrand}},\ }\href {\doibase 10.1103/RevModPhys.88.015001} {\bibfield
  {journal} {\bibinfo  {journal} {Rev. Mod. Phys.}\ }\textbf {\bibinfo {volume}
  {88}},\ \bibinfo {pages} {015001} (\bibinfo {year} {2016})},\ \Eprint
  {http://arxiv.org/abs/1510.05018} {arXiv:1510.05018 [hep-ph]} \BibitemShut
  {NoStop}%
%%CITATION = ARXIV:1510.05018;%%
\bibitem [{\citenamefont {Nogradi}\ and\ \citenamefont
  {Patella}(2016)}]{Nogradi:2016qek}%
  \BibitemOpen
  \bibfield  {author} {\bibinfo {author} {\bibfnamefont {D.}~\bibnamefont
  {Nogradi}}\ and\ \bibinfo {author} {\bibfnamefont {A.}~\bibnamefont
  {Patella}},\ }\href {\doibase 10.1142/S0217751X1643003X} {\bibfield
  {journal} {\bibinfo  {journal} {Int. J. Mod. Phys.}\ }\textbf {\bibinfo
  {volume} {A31}},\ \bibinfo {pages} {1643003} (\bibinfo {year} {2016})},\
  \Eprint {http://arxiv.org/abs/1607.07638} {arXiv:1607.07638 [hep-lat]}
  \BibitemShut {NoStop}%
%%CITATION = ARXIV:1607.07638;%%
\bibitem [{\citenamefont {Witzel}(2019)}]{Witzel:2019jbe}%
  \BibitemOpen
  \bibfield  {author} {\bibinfo {author} {\bibfnamefont {O.}~\bibnamefont
  {Witzel}},\ }\href {\doibase 10.22323/1.334.0006} {\bibfield  {journal}
  {\bibinfo  {journal} {PoS}\ }\textbf {\bibinfo {volume} {LATTICE2018}},\
  \bibinfo {pages} {006} (\bibinfo {year} {2019})},\ \Eprint
  {http://arxiv.org/abs/1901.08216} {arXiv:1901.08216 [hep-lat]} \BibitemShut
  {NoStop}%
%%CITATION = ARXIV:1901.08216;%%
\bibitem [{\citenamefont {Baikov}\ \emph {et~al.}(2017)\citenamefont {Baikov},
  \citenamefont {Chetyrkin},\ and\ \citenamefont {Kühn}}]{Baikov:2016tgj}%
  \BibitemOpen
  \bibfield  {author} {\bibinfo {author} {\bibfnamefont {P.~A.}\ \bibnamefont
  {Baikov}}, \bibinfo {author} {\bibfnamefont {K.~G.}\ \bibnamefont
  {Chetyrkin}}, \ and\ \bibinfo {author} {\bibfnamefont {J.~H.}\ \bibnamefont
  {Kühn}},\ }\href {\doibase 10.1103/PhysRevLett.118.082002} {\bibfield
  {journal} {\bibinfo  {journal} {Phys. Rev. Lett.}\ }\textbf {\bibinfo
  {volume} {118}},\ \bibinfo {pages} {082002} (\bibinfo {year} {2017})},\
  \Eprint {http://arxiv.org/abs/1606.08659} {arXiv:1606.08659 [hep-ph]}
  \BibitemShut {NoStop}%
%%CITATION = ARXIV:1606.08659;%%
\bibitem [{\citenamefont {Ryttov}\ and\ \citenamefont
  {Shrock}(2011)}]{Ryttov:2010iz}%
  \BibitemOpen
  \bibfield  {author} {\bibinfo {author} {\bibfnamefont {T.~A.}\ \bibnamefont
  {Ryttov}}\ and\ \bibinfo {author} {\bibfnamefont {R.}~\bibnamefont
  {Shrock}},\ }\href {\doibase 10.1103/PhysRevD.83.056011} {\bibfield
  {journal} {\bibinfo  {journal} {Phys. Rev.}\ }\textbf {\bibinfo {volume}
  {D83}},\ \bibinfo {pages} {056011} (\bibinfo {year} {2011})},\ \Eprint
  {http://arxiv.org/abs/1011.4542} {arXiv:1011.4542} \BibitemShut {NoStop}%
\bibitem [{\citenamefont {Ryttov}\ and\ \citenamefont
  {Shrock}(2016{\natexlab{a}})}]{Ryttov:2016ner}%
  \BibitemOpen
  \bibfield  {author} {\bibinfo {author} {\bibfnamefont {T.~A.}\ \bibnamefont
  {Ryttov}}\ and\ \bibinfo {author} {\bibfnamefont {R.}~\bibnamefont
  {Shrock}},\ }\href {\doibase 10.1103/PhysRevD.94.105015} {\bibfield
  {journal} {\bibinfo  {journal} {Phys. Rev.}\ }\textbf {\bibinfo {volume}
  {D94}},\ \bibinfo {pages} {105015} (\bibinfo {year} {2016}{\natexlab{a}})},\
  \Eprint {http://arxiv.org/abs/1607.06866} {arXiv:1607.06866 [hep-th]}
  \BibitemShut {NoStop}%
%%CITATION = ARXIV:1607.06866;%%
\bibitem [{\citenamefont {Ryttov}\ and\ \citenamefont
  {Shrock}(2016{\natexlab{b}})}]{Ryttov:2016hal}%
  \BibitemOpen
  \bibfield  {author} {\bibinfo {author} {\bibfnamefont {T.~A.}\ \bibnamefont
  {Ryttov}}\ and\ \bibinfo {author} {\bibfnamefont {R.}~\bibnamefont
  {Shrock}},\ }\href {\doibase 10.1103/PhysRevD.94.125005} {\bibfield
  {journal} {\bibinfo  {journal} {Phys. Rev.}\ }\textbf {\bibinfo {volume}
  {D94}},\ \bibinfo {pages} {125005} (\bibinfo {year} {2016}{\natexlab{b}})},\
  \Eprint {http://arxiv.org/abs/1610.00387} {arXiv:1610.00387 [hep-th]}
  \BibitemShut {NoStop}%
%%CITATION = ARXIV:1610.00387;%%
\bibitem [{\citenamefont {DeGrand}(2011)}]{DeGrand:2011cu}%
  \BibitemOpen
  \bibfield  {author} {\bibinfo {author} {\bibfnamefont {T.}~\bibnamefont
  {DeGrand}},\ }\href {\doibase 10.1103/PhysRevD.84.116901} {\bibfield
  {journal} {\bibinfo  {journal} {Phys.Rev.}\ }\textbf {\bibinfo {volume}
  {D84}},\ \bibinfo {pages} {116901} (\bibinfo {year} {2011})},\ \Eprint
  {http://arxiv.org/abs/1109.1237} {arXiv:1109.1237 [hep-lat]} \BibitemShut
  {NoStop}%
%%CITATION = ARXIV:1109.1237;%%
\bibitem [{\citenamefont {Fodor}\ \emph {et~al.}(2011)\citenamefont {Fodor},
  \citenamefont {Holland}, \citenamefont {Kuti}, \citenamefont {Nogradi},\ and\
  \citenamefont {Schroeder}}]{Fodor:2011tu}%
  \BibitemOpen
  \bibfield  {author} {\bibinfo {author} {\bibfnamefont {Z.}~\bibnamefont
  {Fodor}}, \bibinfo {author} {\bibfnamefont {K.}~\bibnamefont {Holland}},
  \bibinfo {author} {\bibfnamefont {J.}~\bibnamefont {Kuti}}, \bibinfo {author}
  {\bibfnamefont {D.}~\bibnamefont {Nogradi}}, \ and\ \bibinfo {author}
  {\bibfnamefont {C.}~\bibnamefont {Schroeder}},\ }\href {\doibase
  10.1016/j.physletb.2011.07.037} {\bibfield  {journal} {\bibinfo  {journal}
  {Phys. Lett.}\ }\textbf {\bibinfo {volume} {B703}},\ \bibinfo {pages} {348}
  (\bibinfo {year} {2011})},\ \Eprint {http://arxiv.org/abs/1104.3124}
  {arXiv:1104.3124 [hep-lat]} \BibitemShut {NoStop}%
%%CITATION = ARXIV:1104.3124;%%
\bibitem [{\citenamefont {Fodor}\ \emph
  {et~al.}(2012{\natexlab{a}})\citenamefont {Fodor}, \citenamefont {Holland},
  \citenamefont {Kuti}, \citenamefont {Nogradi}, \citenamefont {Schroeder},\
  and\ \citenamefont {Wong}}]{Fodor:2012et}%
  \BibitemOpen
  \bibfield  {author} {\bibinfo {author} {\bibfnamefont {Z.}~\bibnamefont
  {Fodor}}, \bibinfo {author} {\bibfnamefont {K.}~\bibnamefont {Holland}},
  \bibinfo {author} {\bibfnamefont {J.}~\bibnamefont {Kuti}}, \bibinfo {author}
  {\bibfnamefont {D.}~\bibnamefont {Nogradi}}, \bibinfo {author} {\bibfnamefont
  {C.}~\bibnamefont {Schroeder}}, \ and\ \bibinfo {author} {\bibfnamefont
  {C.~H.}\ \bibnamefont {Wong}},\ }\href
  {http://pos.sissa.it/archive/conferences/164/279/Lattice 2012_279.pdf}
  {\bibfield  {journal} {\bibinfo  {journal} {PoS}\ }\textbf {\bibinfo {volume}
  {Lattice 2012}},\ \bibinfo {pages} {279} (\bibinfo {year}
  {2012}{\natexlab{a}})},\ \Eprint {http://arxiv.org/abs/1211.4238}
  {arXiv:1211.4238} \BibitemShut {NoStop}%
\bibitem [{\citenamefont {Aoki}\ \emph {et~al.}(2012)\citenamefont {Aoki},
  \citenamefont {Aoyama}, \citenamefont {Kurachi}, \citenamefont {Maskawa},
  \citenamefont {Nagai}, \citenamefont {Ohki}, \citenamefont {Shibata},
  \citenamefont {Yamawaki},\ and\ \citenamefont {Yamazaki}}]{Aoki:2012eq}%
  \BibitemOpen
  \bibfield  {author} {\bibinfo {author} {\bibfnamefont {Y.}~\bibnamefont
  {Aoki}}, \bibinfo {author} {\bibfnamefont {T.}~\bibnamefont {Aoyama}},
  \bibinfo {author} {\bibfnamefont {M.}~\bibnamefont {Kurachi}}, \bibinfo
  {author} {\bibfnamefont {T.}~\bibnamefont {Maskawa}}, \bibinfo {author}
  {\bibfnamefont {K.-i.}\ \bibnamefont {Nagai}}, \bibinfo {author}
  {\bibfnamefont {H.}~\bibnamefont {Ohki}}, \bibinfo {author} {\bibfnamefont
  {A.}~\bibnamefont {Shibata}}, \bibinfo {author} {\bibfnamefont
  {K.}~\bibnamefont {Yamawaki}}, \ and\ \bibinfo {author} {\bibfnamefont
  {T.}~\bibnamefont {Yamazaki}} (\bibinfo {collaboration} {{LatKMI}}),\ }\href
  {\doibase 10.1103/PhysRevD.86.059903, 10.1103/PhysRevD.86.054506} {\bibfield
  {journal} {\bibinfo  {journal} {Phys. Rev.}\ }\textbf {\bibinfo {volume}
  {D86}},\ \bibinfo {pages} {054506} (\bibinfo {year} {2012})},\ \Eprint
  {http://arxiv.org/abs/1207.3060} {arXiv:1207.3060 [hep-lat]} \BibitemShut
  {NoStop}%
%%CITATION = ARXIV:1207.3060;%%
\bibitem [{\citenamefont {Itou}(2013)}]{Itou:2013ofa}%
  \BibitemOpen
  \bibfield  {author} {\bibinfo {author} {\bibfnamefont {E.}~\bibnamefont
  {Itou}},\ }\href@noop {} {\bibfield  {journal} {\bibinfo  {journal} {PoS}\
  }\textbf {\bibinfo {volume} {LATTICE2013}},\ \bibinfo {pages} {481} (\bibinfo
  {year} {2013})},\ \Eprint {http://arxiv.org/abs/1311.2998} {arXiv:1311.2998
  [hep-lat]} \BibitemShut {NoStop}%
%%CITATION = ARXIV:1311.2998;%%
\bibitem [{\citenamefont {Cheng}\ \emph {et~al.}(2013)\citenamefont {Cheng},
  \citenamefont {Hasenfratz}, \citenamefont {Petropoulos},\ and\ \citenamefont
  {Schaich}}]{Cheng:2013eu}%
  \BibitemOpen
  \bibfield  {author} {\bibinfo {author} {\bibfnamefont {A.}~\bibnamefont
  {Cheng}}, \bibinfo {author} {\bibfnamefont {A.}~\bibnamefont {Hasenfratz}},
  \bibinfo {author} {\bibfnamefont {G.}~\bibnamefont {Petropoulos}}, \ and\
  \bibinfo {author} {\bibfnamefont {D.}~\bibnamefont {Schaich}},\ }\href
  {\doibase 10.1007/JHEP07(2013)061} {\bibfield  {journal} {\bibinfo  {journal}
  {JHEP}\ }\textbf {\bibinfo {volume} {1307}},\ \bibinfo {pages} {061}
  (\bibinfo {year} {2013})},\ \Eprint {http://arxiv.org/abs/1301.1355}
  {arXiv:1301.1355 [hep-lat]} \BibitemShut {NoStop}%
%%CITATION = ARXIV:1301.1355;%%
\bibitem [{\citenamefont {Cheng}\ \emph
  {et~al.}(2014{\natexlab{a}})\citenamefont {Cheng}, \citenamefont
  {Hasenfratz}, \citenamefont {Liu}, \citenamefont {Petropoulos},\ and\
  \citenamefont {Schaich}}]{Cheng:2013xha}%
  \BibitemOpen
  \bibfield  {author} {\bibinfo {author} {\bibfnamefont {A.}~\bibnamefont
  {Cheng}}, \bibinfo {author} {\bibfnamefont {A.}~\bibnamefont {Hasenfratz}},
  \bibinfo {author} {\bibfnamefont {Y.}~\bibnamefont {Liu}}, \bibinfo {author}
  {\bibfnamefont {G.}~\bibnamefont {Petropoulos}}, \ and\ \bibinfo {author}
  {\bibfnamefont {D.}~\bibnamefont {Schaich}},\ }\href {\doibase
  10.1103/PhysRevD.90.014509} {\bibfield  {journal} {\bibinfo  {journal}
  {Phys.Rev.}\ }\textbf {\bibinfo {volume} {D90}},\ \bibinfo {pages} {014509}
  (\bibinfo {year} {2014}{\natexlab{a}})},\ \Eprint
  {http://arxiv.org/abs/1401.0195} {arXiv:1401.0195 [hep-lat]} \BibitemShut
  {NoStop}%
%%CITATION = ARXIV:1401.0195;%%
\bibitem [{\citenamefont {Cheng}\ \emph
  {et~al.}(2014{\natexlab{b}})\citenamefont {Cheng}, \citenamefont
  {Hasenfratz}, \citenamefont {Liu}, \citenamefont {Petropoulos},\ and\
  \citenamefont {Schaich}}]{Cheng:2014jba}%
  \BibitemOpen
  \bibfield  {author} {\bibinfo {author} {\bibfnamefont {A.}~\bibnamefont
  {Cheng}}, \bibinfo {author} {\bibfnamefont {A.}~\bibnamefont {Hasenfratz}},
  \bibinfo {author} {\bibfnamefont {Y.}~\bibnamefont {Liu}}, \bibinfo {author}
  {\bibfnamefont {G.}~\bibnamefont {Petropoulos}}, \ and\ \bibinfo {author}
  {\bibfnamefont {D.}~\bibnamefont {Schaich}},\ }\href {\doibase
  10.1007/JHEP05(2014)137} {\bibfield  {journal} {\bibinfo  {journal} {JHEP}\
  }\textbf {\bibinfo {volume} {1405}},\ \bibinfo {pages} {137} (\bibinfo {year}
  {2014}{\natexlab{b}})},\ \Eprint {http://arxiv.org/abs/1404.0984}
  {arXiv:1404.0984 [hep-lat]} \BibitemShut {NoStop}%
%%CITATION = ARXIV:1404.0984;%%
\bibitem [{\citenamefont {Lin}\ \emph {et~al.}(2015)\citenamefont {Lin},
  \citenamefont {Ogawa},\ and\ \citenamefont {Ramos}}]{Lin:2015zpa}%
  \BibitemOpen
  \bibfield  {author} {\bibinfo {author} {\bibfnamefont {C.~J.~D.}\
  \bibnamefont {Lin}}, \bibinfo {author} {\bibfnamefont {K.}~\bibnamefont
  {Ogawa}}, \ and\ \bibinfo {author} {\bibfnamefont {A.}~\bibnamefont
  {Ramos}},\ }\href {\doibase 10.1007/JHEP12(2015)103} {\bibfield  {journal}
  {\bibinfo  {journal} {JHEP}\ }\textbf {\bibinfo {volume} {12}},\ \bibinfo
  {pages} {103} (\bibinfo {year} {2015})},\ \Eprint
  {http://arxiv.org/abs/1510.05755} {arXiv:1510.05755 [hep-lat]} \BibitemShut
  {NoStop}%
%%CITATION = ARXIV:1510.05755;%%
\bibitem [{\citenamefont {Hasenfratz}\ and\ \citenamefont
  {Schaich}(2018)}]{Hasenfratz:2016dou}%
  \BibitemOpen
  \bibfield  {author} {\bibinfo {author} {\bibfnamefont {A.}~\bibnamefont
  {Hasenfratz}}\ and\ \bibinfo {author} {\bibfnamefont {D.}~\bibnamefont
  {Schaich}},\ }\href {\doibase 10.1007/JHEP02(2018)132} {\bibfield  {journal}
  {\bibinfo  {journal} {JHEP}\ }\textbf {\bibinfo {volume} {02}},\ \bibinfo
  {pages} {132} (\bibinfo {year} {2018})},\ \Eprint
  {http://arxiv.org/abs/1610.10004} {arXiv:1610.10004 [hep-lat]} \BibitemShut
  {NoStop}%
%%CITATION = ARXIV:1610.10004;%%
\bibitem [{\citenamefont {Fodor}\ \emph {et~al.}(2016)\citenamefont {Fodor},
  \citenamefont {Holland}, \citenamefont {Kuti}, \citenamefont {Mondal},
  \citenamefont {Nogradi},\ and\ \citenamefont {Wong}}]{Fodor:2016zil}%
  \BibitemOpen
  \bibfield  {author} {\bibinfo {author} {\bibfnamefont {Z.}~\bibnamefont
  {Fodor}}, \bibinfo {author} {\bibfnamefont {K.}~\bibnamefont {Holland}},
  \bibinfo {author} {\bibfnamefont {J.}~\bibnamefont {Kuti}}, \bibinfo {author}
  {\bibfnamefont {S.}~\bibnamefont {Mondal}}, \bibinfo {author} {\bibfnamefont
  {D.}~\bibnamefont {Nogradi}}, \ and\ \bibinfo {author} {\bibfnamefont
  {C.~H.}\ \bibnamefont {Wong}},\ }\href {\doibase 10.1103/PhysRevD.94.091501}
  {\bibfield  {journal} {\bibinfo  {journal} {Phys. Rev.}\ }\textbf {\bibinfo
  {volume} {D94}},\ \bibinfo {pages} {091501} (\bibinfo {year} {2016})},\
  \Eprint {http://arxiv.org/abs/1607.06121} {arXiv:1607.06121 [hep-lat]}
  \BibitemShut {NoStop}%
%%CITATION = ARXIV:1607.06121;%%
\bibitem [{\citenamefont {Fodor}\ \emph
  {et~al.}(2018{\natexlab{a}})\citenamefont {Fodor}, \citenamefont {Holland},
  \citenamefont {Kuti}, \citenamefont {Nogradi},\ and\ \citenamefont
  {Wong}}]{Fodor:2017gtj}%
  \BibitemOpen
  \bibfield  {author} {\bibinfo {author} {\bibfnamefont {Z.}~\bibnamefont
  {Fodor}}, \bibinfo {author} {\bibfnamefont {K.}~\bibnamefont {Holland}},
  \bibinfo {author} {\bibfnamefont {J.}~\bibnamefont {Kuti}}, \bibinfo {author}
  {\bibfnamefont {D.}~\bibnamefont {Nogradi}}, \ and\ \bibinfo {author}
  {\bibfnamefont {C.~H.}\ \bibnamefont {Wong}},\ }\href {\doibase
  10.1016/j.physletb.2018.02.008} {\bibfield  {journal} {\bibinfo  {journal}
  {Phys. Lett.}\ }\textbf {\bibinfo {volume} {B779}},\ \bibinfo {pages} {230}
  (\bibinfo {year} {2018}{\natexlab{a}})},\ \Eprint
  {http://arxiv.org/abs/1710.09262} {arXiv:1710.09262 [hep-lat]} \BibitemShut
  {NoStop}%
%%CITATION = ARXIV:1710.09262;%%
\bibitem [{\citenamefont {Fodor}\ \emph
  {et~al.}(2018{\natexlab{b}})\citenamefont {Fodor}, \citenamefont {Holland},
  \citenamefont {Kuti}, \citenamefont {Nogradi},\ and\ \citenamefont
  {Wong}}]{Fodor:2017nlp}%
  \BibitemOpen
  \bibfield  {author} {\bibinfo {author} {\bibfnamefont {Z.}~\bibnamefont
  {Fodor}}, \bibinfo {author} {\bibfnamefont {K.}~\bibnamefont {Holland}},
  \bibinfo {author} {\bibfnamefont {J.}~\bibnamefont {Kuti}}, \bibinfo {author}
  {\bibfnamefont {D.}~\bibnamefont {Nogradi}}, \ and\ \bibinfo {author}
  {\bibfnamefont {C.~H.}\ \bibnamefont {Wong}},\ }\href {\doibase
  10.1051/epjconf/201817508015} {\bibfield  {journal} {\bibinfo  {journal} {EPJ
  Web Conf.}\ }\textbf {\bibinfo {volume} {175}},\ \bibinfo {pages} {08015}
  (\bibinfo {year} {2018}{\natexlab{b}})},\ \Eprint
  {http://arxiv.org/abs/1712.08594} {arXiv:1712.08594 [hep-lat]} \BibitemShut
  {NoStop}%
%%CITATION = ARXIV:1712.08594;%%
\bibitem [{\citenamefont {Kaplan}(1992)}]{Kaplan:1992bt}%
  \BibitemOpen
  \bibfield  {author} {\bibinfo {author} {\bibfnamefont {D.~B.}\ \bibnamefont
  {Kaplan}},\ }\href {\doibase 10.1016/0370-2693(92)91112-M} {\bibfield
  {journal} {\bibinfo  {journal} {Phys. Lett.}\ }\textbf {\bibinfo {volume}
  {B288}},\ \bibinfo {pages} {342} (\bibinfo {year} {1992})},\ \Eprint
  {http://arxiv.org/abs/hep-lat/9206013} {arXiv:hep-lat/9206013} \BibitemShut
  {NoStop}%
%%CITATION = HEP-LAT/9206013;%%
\bibitem [{\citenamefont {Shamir}(1993)}]{Shamir:1993zy}%
  \BibitemOpen
  \bibfield  {author} {\bibinfo {author} {\bibfnamefont {Y.}~\bibnamefont
  {Shamir}},\ }\href {\doibase 10.1016/0550-3213(93)90162-I} {\bibfield
  {journal} {\bibinfo  {journal} {Nucl. Phys.}\ }\textbf {\bibinfo {volume}
  {B406}},\ \bibinfo {pages} {90} (\bibinfo {year} {1993})},\ \Eprint
  {http://arxiv.org/abs/hep-lat/9303005} {arXiv:hep-lat/9303005} \BibitemShut
  {NoStop}%
%%CITATION = HEP-LAT/9303005;%%
\bibitem [{\citenamefont {Furman}\ and\ \citenamefont
  {Shamir}(1995)}]{Furman:1994ky}%
  \BibitemOpen
  \bibfield  {author} {\bibinfo {author} {\bibfnamefont {V.}~\bibnamefont
  {Furman}}\ and\ \bibinfo {author} {\bibfnamefont {Y.}~\bibnamefont
  {Shamir}},\ }\href {\doibase 10.1016/0550-3213(95)00031-M} {\bibfield
  {journal} {\bibinfo  {journal} {Nucl. Phys.}\ }\textbf {\bibinfo {volume}
  {B439}},\ \bibinfo {pages} {54} (\bibinfo {year} {1995})},\ \Eprint
  {http://arxiv.org/abs/hep-lat/9405004} {arXiv:hep-lat/9405004} \BibitemShut
  {NoStop}%
%%CITATION = HEP-LAT/9405004;%%
\bibitem [{\citenamefont {Brower}\ \emph {et~al.}(2017)\citenamefont {Brower},
  \citenamefont {Neff},\ and\ \citenamefont {Orginos}}]{Brower:2012vk}%
  \BibitemOpen
  \bibfield  {author} {\bibinfo {author} {\bibfnamefont {R.~C.}\ \bibnamefont
  {Brower}}, \bibinfo {author} {\bibfnamefont {H.}~\bibnamefont {Neff}}, \ and\
  \bibinfo {author} {\bibfnamefont {K.}~\bibnamefont {Orginos}},\ }\href
  {\doibase 10.1016/j.cpc.2017.01.024} {\bibfield  {journal} {\bibinfo
  {journal} {Comput. Phys. Commun.}\ }\textbf {\bibinfo {volume} {220}},\
  \bibinfo {pages} {1} (\bibinfo {year} {2017})},\ \Eprint
  {http://arxiv.org/abs/1206.5214} {arXiv:1206.5214 [hep-lat]} \BibitemShut
  {NoStop}%
%%CITATION = ARXIV:1206.5214;%%
\bibitem [{\citenamefont {Morningstar}\ and\ \citenamefont
  {Peardon}(2004)}]{Morningstar:2003gk}%
  \BibitemOpen
  \bibfield  {author} {\bibinfo {author} {\bibfnamefont {C.}~\bibnamefont
  {Morningstar}}\ and\ \bibinfo {author} {\bibfnamefont {M.~J.}\ \bibnamefont
  {Peardon}},\ }\href {\doibase 10.1103/PhysRevD.69.054501} {\bibfield
  {journal} {\bibinfo  {journal} {Phys. Rev.}\ }\textbf {\bibinfo {volume}
  {D69}},\ \bibinfo {pages} {054501} (\bibinfo {year} {2004})},\ \Eprint
  {http://arxiv.org/abs/hep-lat/0311018} {arXiv:hep-lat/0311018 [hep-lat]}
  \BibitemShut {NoStop}%
%%CITATION = HEP-LAT/0311018;%%
\bibitem [{\citenamefont {L{\"u}scher}\ and\ \citenamefont
  {Weisz}(1985{\natexlab{a}})}]{Luscher:1984xn}%
  \BibitemOpen
  \bibfield  {author} {\bibinfo {author} {\bibfnamefont {M.}~\bibnamefont
  {L{\"u}scher}}\ and\ \bibinfo {author} {\bibfnamefont {P.}~\bibnamefont
  {Weisz}},\ }\href {\doibase 10.1007/BF01206178} {\bibfield  {journal}
  {\bibinfo  {journal} {Commun. Math. Phys.}\ }\textbf {\bibinfo {volume}
  {97}},\ \bibinfo {pages} {59} (\bibinfo {year} {1985}{\natexlab{a}})},\
  \bibinfo {note} {[Erratum: Commun. Math. Phys.98,433(1985)]}\BibitemShut
  {NoStop}%
%%CITATION = CMPHA,97,59;%%
\bibitem [{\citenamefont {L{\"u}scher}\ and\ \citenamefont
  {Weisz}(1985{\natexlab{b}})}]{Luscher:1985zq}%
  \BibitemOpen
  \bibfield  {author} {\bibinfo {author} {\bibfnamefont {M.}~\bibnamefont
  {L{\"u}scher}}\ and\ \bibinfo {author} {\bibfnamefont {P.}~\bibnamefont
  {Weisz}},\ }\href {\doibase 10.1016/0370-2693(85)90966-9} {\bibfield
  {journal} {\bibinfo  {journal} {Phys. Lett.}\ }\textbf {\bibinfo {volume}
  {158B}},\ \bibinfo {pages} {250} (\bibinfo {year}
  {1985}{\natexlab{b}})}\BibitemShut {NoStop}%
%%CITATION = PHLTA,158B,250;%%
\bibitem [{\citenamefont {Sint}\ and\ \citenamefont
  {Ramos}(2015)}]{Ramos:2014kka}%
  \BibitemOpen
  \bibfield  {author} {\bibinfo {author} {\bibfnamefont {S.}~\bibnamefont
  {Sint}}\ and\ \bibinfo {author} {\bibfnamefont {A.}~\bibnamefont {Ramos}},\
  }\href {\doibase 10.22323/1.214.0329} {\bibfield  {journal} {\bibinfo
  {journal} {PoS}\ }\textbf {\bibinfo {volume} {LATTICE2014}},\ \bibinfo
  {pages} {329} (\bibinfo {year} {2015})},\ \Eprint
  {http://arxiv.org/abs/1411.6706} {arXiv:1411.6706 [hep-lat]} \BibitemShut
  {NoStop}%
%%CITATION = ARXIV:1411.6706;%%
\bibitem [{\citenamefont {Ramos}\ and\ \citenamefont
  {Sint}(2016)}]{Ramos:2015baa}%
  \BibitemOpen
  \bibfield  {author} {\bibinfo {author} {\bibfnamefont {A.}~\bibnamefont
  {Ramos}}\ and\ \bibinfo {author} {\bibfnamefont {S.}~\bibnamefont {Sint}},\
  }\href {\doibase 10.1140/epjc/s10052-015-3831-9} {\bibfield  {journal}
  {\bibinfo  {journal} {Eur. Phys. J.}\ }\textbf {\bibinfo {volume} {C76}},\
  \bibinfo {pages} {15} (\bibinfo {year} {2016})},\ \Eprint
  {http://arxiv.org/abs/1508.05552} {arXiv:1508.05552 [hep-lat]} \BibitemShut
  {NoStop}%
%%CITATION = ARXIV:1508.05552;%%
\bibitem [{\citenamefont {Fodor}\ \emph {et~al.}(2014)\citenamefont {Fodor},
  \citenamefont {Holland}, \citenamefont {Kuti}, \citenamefont {Mondal},
  \citenamefont {Nogradi},\ and\ \citenamefont {Wong}}]{Fodor:2014cpa}%
  \BibitemOpen
  \bibfield  {author} {\bibinfo {author} {\bibfnamefont {Z.}~\bibnamefont
  {Fodor}}, \bibinfo {author} {\bibfnamefont {K.}~\bibnamefont {Holland}},
  \bibinfo {author} {\bibfnamefont {J.}~\bibnamefont {Kuti}}, \bibinfo {author}
  {\bibfnamefont {S.}~\bibnamefont {Mondal}}, \bibinfo {author} {\bibfnamefont
  {D.}~\bibnamefont {Nogradi}}, \ and\ \bibinfo {author} {\bibfnamefont
  {C.~H.}\ \bibnamefont {Wong}},\ }\href {\doibase 10.1007/JHEP09(2014)018}
  {\bibfield  {journal} {\bibinfo  {journal} {JHEP}\ }\textbf {\bibinfo
  {volume} {09}},\ \bibinfo {pages} {018} (\bibinfo {year} {2014})},\ \Eprint
  {http://arxiv.org/abs/1406.0827} {arXiv:1406.0827 [hep-lat]} \BibitemShut
  {NoStop}%
%%CITATION = ARXIV:1406.0827;%%
\bibitem [{\citenamefont {Hasenfratz}\ \emph {et~al.}(2018)\citenamefont
  {Hasenfratz}, \citenamefont {Rebbi},\ and\ \citenamefont
  {Witzel}}]{Hasenfratz:2017mdh}%
  \BibitemOpen
  \bibfield  {author} {\bibinfo {author} {\bibfnamefont {A.}~\bibnamefont
  {Hasenfratz}}, \bibinfo {author} {\bibfnamefont {C.}~\bibnamefont {Rebbi}}, \
  and\ \bibinfo {author} {\bibfnamefont {O.}~\bibnamefont {Witzel}},\ }\href
  {\doibase 10.1051/epjconf/201817503006} {\bibfield  {journal} {\bibinfo
  {journal} {EPJ Web Conf.}\ }\textbf {\bibinfo {volume} {175}},\ \bibinfo
  {pages} {03006} (\bibinfo {year} {2018})},\ \Eprint
  {http://arxiv.org/abs/1708.03385} {arXiv:1708.03385 [hep-lat]} \BibitemShut
  {NoStop}%
%%CITATION = ARXIV:1708.03385;%%
\bibitem [{\citenamefont {Hasenfratz}\ \emph
  {et~al.}(2019{\natexlab{a}})\citenamefont {Hasenfratz}, \citenamefont
  {Rebbi},\ and\ \citenamefont {Witzel}}]{Hasenfratz:2017qyr}%
  \BibitemOpen
  \bibfield  {author} {\bibinfo {author} {\bibfnamefont {A.}~\bibnamefont
  {Hasenfratz}}, \bibinfo {author} {\bibfnamefont {C.}~\bibnamefont {Rebbi}}, \
  and\ \bibinfo {author} {\bibfnamefont {O.}~\bibnamefont {Witzel}},\ }\href
  {\doibase 10.1016/j.physletb.2019.134937} {\bibfield  {journal} {\bibinfo
  {journal} {Phys. Lett.}\ }\textbf {\bibinfo {volume} {B798}},\ \bibinfo
  {pages} {134937} (\bibinfo {year} {2019}{\natexlab{a}})},\ \Eprint
  {http://arxiv.org/abs/1710.11578} {arXiv:1710.11578 [hep-lat]} \BibitemShut
  {NoStop}%
%%CITATION = ARXIV:1710.11578;%%
\bibitem [{\citenamefont {Hasenfratz}\ \emph
  {et~al.}(2019{\natexlab{b}})\citenamefont {Hasenfratz}, \citenamefont
  {Rebbi},\ and\ \citenamefont {Witzel}}]{Hasenfratz:2018wpq}%
  \BibitemOpen
  \bibfield  {author} {\bibinfo {author} {\bibfnamefont {A.}~\bibnamefont
  {Hasenfratz}}, \bibinfo {author} {\bibfnamefont {C.}~\bibnamefont {Rebbi}}, \
  and\ \bibinfo {author} {\bibfnamefont {O.}~\bibnamefont {Witzel}},\ }\href
  {\doibase 10.22323/1.334.0306} {\bibfield  {journal} {\bibinfo  {journal}
  {PoS}\ }\textbf {\bibinfo {volume} {LATTICE2018}},\ \bibinfo {pages} {306}
  (\bibinfo {year} {2019}{\natexlab{b}})},\ \Eprint
  {http://arxiv.org/abs/1810.05176} {arXiv:1810.05176 [hep-lat]} \BibitemShut
  {NoStop}%
%%CITATION = ARXIV:1810.05176;%%
\bibitem [{\citenamefont {Hasenfratz}\ and\ \citenamefont
  {Witzel}(2019{\natexlab{a}})}]{Hasenfratz:2019puu}%
  \BibitemOpen
  \bibfield  {author} {\bibinfo {author} {\bibfnamefont {A.}~\bibnamefont
  {Hasenfratz}}\ and\ \bibinfo {author} {\bibfnamefont {O.}~\bibnamefont
  {Witzel}},\ }\href@noop {} {\  (\bibinfo {year} {2019}{\natexlab{a}})},\
  \Eprint {http://arxiv.org/abs/1911.11531} {arXiv:1911.11531 [hep-lat]}
  \BibitemShut {NoStop}%
%%CITATION = ARXIV:1911.11531;%%
\bibitem [{\citenamefont {Fodor}\ \emph
  {et~al.}(2012{\natexlab{b}})\citenamefont {Fodor}, \citenamefont {Holland},
  \citenamefont {Kuti}, \citenamefont {Nogradi},\ and\ \citenamefont
  {Wong}}]{Fodor:2012td}%
  \BibitemOpen
  \bibfield  {author} {\bibinfo {author} {\bibfnamefont {Z.}~\bibnamefont
  {Fodor}}, \bibinfo {author} {\bibfnamefont {K.}~\bibnamefont {Holland}},
  \bibinfo {author} {\bibfnamefont {J.}~\bibnamefont {Kuti}}, \bibinfo {author}
  {\bibfnamefont {D.}~\bibnamefont {Nogradi}}, \ and\ \bibinfo {author}
  {\bibfnamefont {C.~H.}\ \bibnamefont {Wong}},\ }\href {\doibase
  10.1007/JHEP11(2012)007} {\bibfield  {journal} {\bibinfo  {journal} {JHEP}\
  }\textbf {\bibinfo {volume} {1211}},\ \bibinfo {pages} {007} (\bibinfo {year}
  {2012}{\natexlab{b}})},\ \Eprint {http://arxiv.org/abs/1208.1051}
  {arXiv:1208.1051 [hep-lat]} \BibitemShut {NoStop}%
%%CITATION = ARXIV:1208.1051;%%
\bibitem [{\citenamefont {Kaneko}\ \emph {et~al.}(2014)\citenamefont {Kaneko},
  \citenamefont {Aoki}, \citenamefont {Cossu}, \citenamefont {Fukaya},
  \citenamefont {Hashimoto},\ and\ \citenamefont {Noaki}}]{Kaneko:2013jla}%
  \BibitemOpen
  \bibfield  {author} {\bibinfo {author} {\bibfnamefont {T.}~\bibnamefont
  {Kaneko}}, \bibinfo {author} {\bibfnamefont {S.}~\bibnamefont {Aoki}},
  \bibinfo {author} {\bibfnamefont {G.}~\bibnamefont {Cossu}}, \bibinfo
  {author} {\bibfnamefont {H.}~\bibnamefont {Fukaya}}, \bibinfo {author}
  {\bibfnamefont {S.}~\bibnamefont {Hashimoto}}, \ and\ \bibinfo {author}
  {\bibfnamefont {J.}~\bibnamefont {Noaki}} (\bibinfo {collaboration}
  {JLQCD}),\ }\href {\doibase 10.22323/1.187.0125} {\bibfield  {journal}
  {\bibinfo  {journal} {PoS}\ }\textbf {\bibinfo {volume} {LATTICE2013}},\
  \bibinfo {pages} {125} (\bibinfo {year} {2014})},\ \Eprint
  {http://arxiv.org/abs/1311.6941} {arXiv:1311.6941 [hep-lat]} \BibitemShut
  {NoStop}%
%%CITATION = ARXIV:1311.6941;%%
\bibitem [{\citenamefont {Noaki}\ \emph {et~al.}(2016)\citenamefont {Noaki},
  \citenamefont {Cossu}, \citenamefont {Ishikawa}, \citenamefont {Iwasaki},\
  and\ \citenamefont {Yoshie}}]{Noaki:2015xpx}%
  \BibitemOpen
  \bibfield  {author} {\bibinfo {author} {\bibfnamefont {J.}~\bibnamefont
  {Noaki}}, \bibinfo {author} {\bibfnamefont {G.}~\bibnamefont {Cossu}},
  \bibinfo {author} {\bibfnamefont {K.-I.}\ \bibnamefont {Ishikawa}}, \bibinfo
  {author} {\bibfnamefont {Y.}~\bibnamefont {Iwasaki}}, \ and\ \bibinfo
  {author} {\bibfnamefont {T.}~\bibnamefont {Yoshie}},\ }\href {\doibase
  10.22323/1.251.0312} {\bibfield  {journal} {\bibinfo  {journal} {PoS}\
  }\textbf {\bibinfo {volume} {LATTICE2015}},\ \bibinfo {pages} {312} (\bibinfo
  {year} {2016})},\ \Eprint {http://arxiv.org/abs/1511.06474} {arXiv:1511.06474
  [hep-lat]} \BibitemShut {NoStop}%
%%CITATION = ARXIV:1511.06474;%%
\bibitem [{\citenamefont {Duane}\ \emph {et~al.}(1987)\citenamefont {Duane},
  \citenamefont {Kennedy}, \citenamefont {Pendleton},\ and\ \citenamefont
  {Roweth}}]{Duane:1987de}%
  \BibitemOpen
  \bibfield  {author} {\bibinfo {author} {\bibfnamefont {S.}~\bibnamefont
  {Duane}}, \bibinfo {author} {\bibfnamefont {A.}~\bibnamefont {Kennedy}},
  \bibinfo {author} {\bibfnamefont {B.}~\bibnamefont {Pendleton}}, \ and\
  \bibinfo {author} {\bibfnamefont {D.}~\bibnamefont {Roweth}},\ }\href
  {\doibase 10.1016/0370-2693(87)91197-X} {\bibfield  {journal} {\bibinfo
  {journal} {Phys.Lett.}\ }\textbf {\bibinfo {volume} {B195}},\ \bibinfo
  {pages} {216} (\bibinfo {year} {1987})}\BibitemShut {NoStop}%
%%CITATION = PHLTA,B195,216;%%
\bibitem [{\citenamefont {Boyle}\ \emph
  {et~al.}(2015{\natexlab{a}})\citenamefont {Boyle}, \citenamefont {Cossu},
  \citenamefont {Portelli},\ and\ \citenamefont {Yamaguchi}}]{Grid}%
  \BibitemOpen
  \bibfield  {author} {\bibinfo {author} {\bibfnamefont {P.}~\bibnamefont
  {Boyle}}, \bibinfo {author} {\bibfnamefont {G.}~\bibnamefont {Cossu}},
  \bibinfo {author} {\bibfnamefont {A.}~\bibnamefont {Portelli}}, \ and\
  \bibinfo {author} {\bibfnamefont {A.}~\bibnamefont {Yamaguchi}},\ }\href
  {https://github.com/paboyle/Grid} {\enquote {\bibinfo {title} {Grid},}\ }
  (\bibinfo {year} {2015}{\natexlab{a}})\BibitemShut {NoStop}%
\bibitem [{\citenamefont {Boyle}\ \emph
  {et~al.}(2015{\natexlab{b}})\citenamefont {Boyle}, \citenamefont {Yamaguchi},
  \citenamefont {Cossu},\ and\ \citenamefont {Portelli}}]{Boyle:2015tjk}%
  \BibitemOpen
  \bibfield  {author} {\bibinfo {author} {\bibfnamefont {P.}~\bibnamefont
  {Boyle}}, \bibinfo {author} {\bibfnamefont {A.}~\bibnamefont {Yamaguchi}},
  \bibinfo {author} {\bibfnamefont {G.}~\bibnamefont {Cossu}}, \ and\ \bibinfo
  {author} {\bibfnamefont {A.}~\bibnamefont {Portelli}},\ }\href {\doibase
  10.22323/1.251.0023} {\bibfield  {journal} {\bibinfo  {journal} {PoS}\
  }\textbf {\bibinfo {volume} {LATTICE2015}},\ \bibinfo {pages} {023} (\bibinfo
  {year} {2015}{\natexlab{b}})},\ \Eprint {http://arxiv.org/abs/1512.03487}
  {arXiv:1512.03487 [hep-lat]} \BibitemShut {NoStop}%
%%CITATION = ARXIV:1512.03487;%%
\bibitem [{\citenamefont {Pochinsky}(2008)}]{Pochinsky:2008zz}%
  \BibitemOpen
  \bibfield  {author} {\bibinfo {author} {\bibfnamefont {A.}~\bibnamefont
  {Pochinsky}},\ }\href {\doibase 10.22323/1.066.0040} {\bibfield  {journal}
  {\bibinfo  {journal} {PoS}\ }\textbf {\bibinfo {volume} {LATTICE2008}},\
  \bibinfo {pages} {040} (\bibinfo {year} {2008})}\BibitemShut {NoStop}%
%%CITATION = POSCI,LATTICE2008,040;%%
\bibitem [{\citenamefont {Pochinsky}\ \emph {et~al.}(2008)\citenamefont
  {Pochinsky} \emph {et~al.}}]{qlua}%
  \BibitemOpen
  \bibfield  {author} {\bibinfo {author} {\bibfnamefont {A.}~\bibnamefont
  {Pochinsky}} \emph {et~al.},\ }\href
  {https://usqcd.lns.mit.edu/w/index.php/QLUA} {\enquote {\bibinfo {title}
  {Qlua},}\ } (\bibinfo {year} {2008})\BibitemShut {NoStop}%
\bibitem [{\citenamefont {Wolff}(2004)}]{Wolff:2003sm}%
  \BibitemOpen
  \bibfield  {author} {\bibinfo {author} {\bibfnamefont {U.}~\bibnamefont
  {Wolff}} (\bibinfo {collaboration} {ALPHA}),\ }\href {\doibase
  10.1016/S0010-4655(03)00467-3, 10.1016/j.cpc.2006.12.001} {\bibfield
  {journal} {\bibinfo  {journal} {Comput.Phys.Commun.}\ }\textbf {\bibinfo
  {volume} {156}},\ \bibinfo {pages} {143} (\bibinfo {year} {2004})},\ \Eprint
  {http://arxiv.org/abs/hep-lat/0306017} {arXiv:hep-lat/0306017 [hep-lat]}
  \BibitemShut {NoStop}%
%%CITATION = HEP-LAT/0306017;%%
\bibitem [{\citenamefont {Fritzsch}\ and\ \citenamefont
  {Ramos}(2013)}]{Fritzsch:2013je}%
  \BibitemOpen
  \bibfield  {author} {\bibinfo {author} {\bibfnamefont {P.}~\bibnamefont
  {Fritzsch}}\ and\ \bibinfo {author} {\bibfnamefont {A.}~\bibnamefont
  {Ramos}},\ }\href {\doibase 10.1007/JHEP10(2013)008} {\bibfield  {journal}
  {\bibinfo  {journal} {JHEP}\ }\textbf {\bibinfo {volume} {1310}},\ \bibinfo
  {pages} {008} (\bibinfo {year} {2013})},\ \Eprint
  {http://arxiv.org/abs/1301.4388} {arXiv:1301.4388} \BibitemShut {NoStop}%
\bibitem [{\citenamefont {Carosso}\ \emph {et~al.}(2018)\citenamefont
  {Carosso}, \citenamefont {Hasenfratz},\ and\ \citenamefont
  {Neil}}]{Carosso:2018bmz}%
  \BibitemOpen
  \bibfield  {author} {\bibinfo {author} {\bibfnamefont {A.}~\bibnamefont
  {Carosso}}, \bibinfo {author} {\bibfnamefont {A.}~\bibnamefont {Hasenfratz}},
  \ and\ \bibinfo {author} {\bibfnamefont {E.~T.}\ \bibnamefont {Neil}},\
  }\href {\doibase 10.1103/PhysRevLett.121.201601} {\bibfield  {journal}
  {\bibinfo  {journal} {Phys. Rev. Lett.}\ }\textbf {\bibinfo {volume} {121}},\
  \bibinfo {pages} {201601} (\bibinfo {year} {2018})},\ \Eprint
  {http://arxiv.org/abs/1806.01385} {arXiv:1806.01385 [hep-lat]} \BibitemShut
  {NoStop}%
%%CITATION = ARXIV:1806.01385;%%
\bibitem [{\citenamefont {Hasenfratz}\ and\ \citenamefont
  {Witzel}(2019{\natexlab{b}})}]{Hasenfratz:2019hpg}%
  \BibitemOpen
  \bibfield  {author} {\bibinfo {author} {\bibfnamefont {A.}~\bibnamefont
  {Hasenfratz}}\ and\ \bibinfo {author} {\bibfnamefont {O.}~\bibnamefont
  {Witzel}},\ }\href@noop {} {\  (\bibinfo {year} {2019}{\natexlab{b}})},\
  \Eprint {http://arxiv.org/abs/1910.06408} {arXiv:1910.06408 [hep-lat]}
  \BibitemShut {NoStop}%
%%CITATION = ARXIV:1910.06408;%%
\bibitem [{\citenamefont {Ryttov}\ and\ \citenamefont
  {Shrock}(2017)}]{Ryttov:2017kmx}%
  \BibitemOpen
  \bibfield  {author} {\bibinfo {author} {\bibfnamefont {T.~A.}\ \bibnamefont
  {Ryttov}}\ and\ \bibinfo {author} {\bibfnamefont {R.}~\bibnamefont
  {Shrock}},\ }\href {\doibase 10.1103/PhysRevD.95.105004} {\bibfield
  {journal} {\bibinfo  {journal} {Phys. Rev.}\ }\textbf {\bibinfo {volume}
  {D95}},\ \bibinfo {pages} {105004} (\bibinfo {year} {2017})},\ \Eprint
  {http://arxiv.org/abs/1703.08558} {arXiv:1703.08558 [hep-th]} \BibitemShut
  {NoStop}%
%%CITATION = ARXIV:1703.08558;%%
\bibitem [{\citenamefont {Anderson}\ \emph {et~al.}(2017)\citenamefont
  {Anderson}, \citenamefont {Burns}, \citenamefont {Milroy}, \citenamefont
  {Ruprecht}, \citenamefont {Hauser},\ and\ \citenamefont {Siegel}}]{UCsummit}%
  \BibitemOpen
  \bibfield  {author} {\bibinfo {author} {\bibfnamefont {J.}~\bibnamefont
  {Anderson}}, \bibinfo {author} {\bibfnamefont {P.~J.}\ \bibnamefont {Burns}},
  \bibinfo {author} {\bibfnamefont {D.}~\bibnamefont {Milroy}}, \bibinfo
  {author} {\bibfnamefont {P.}~\bibnamefont {Ruprecht}}, \bibinfo {author}
  {\bibfnamefont {T.}~\bibnamefont {Hauser}}, \ and\ \bibinfo {author}
  {\bibfnamefont {H.~J.}\ \bibnamefont {Siegel}},\ }\href {\doibase
  10.1145/3093338.3093379} {\bibfield  {journal} {\bibinfo  {journal}
  {Proceedings of PEARC17}\ } (\bibinfo {year} {2017}),\
  10.1145/3093338.3093379}\BibitemShut {NoStop}%
\bibitem [{\citenamefont {Towns}\ \emph {et~al.}(2014)\citenamefont {Towns},
  \citenamefont {Cockerill}, \citenamefont {Dahan}, \citenamefont {Foster},
  \citenamefont {Gaither}, \citenamefont {Grimshaw}, \citenamefont {Hazlewood},
  \citenamefont {Lathrop}, \citenamefont {Lifka}, \citenamefont {Peterson},
  \citenamefont {Roskies}, \citenamefont {Scott},\ and\ \citenamefont
  {Wilkins-Diehr}}]{xsede}%
  \BibitemOpen
  \bibfield  {author} {\bibinfo {author} {\bibfnamefont {J.}~\bibnamefont
  {Towns}}, \bibinfo {author} {\bibfnamefont {T.}~\bibnamefont {Cockerill}},
  \bibinfo {author} {\bibfnamefont {M.}~\bibnamefont {Dahan}}, \bibinfo
  {author} {\bibfnamefont {I.}~\bibnamefont {Foster}}, \bibinfo {author}
  {\bibfnamefont {K.}~\bibnamefont {Gaither}}, \bibinfo {author} {\bibfnamefont
  {A.}~\bibnamefont {Grimshaw}}, \bibinfo {author} {\bibfnamefont
  {V.}~\bibnamefont {Hazlewood}}, \bibinfo {author} {\bibfnamefont
  {S.}~\bibnamefont {Lathrop}}, \bibinfo {author} {\bibfnamefont
  {D.}~\bibnamefont {Lifka}}, \bibinfo {author} {\bibfnamefont {G.~D.}\
  \bibnamefont {Peterson}}, \bibinfo {author} {\bibfnamefont {R.}~\bibnamefont
  {Roskies}}, \bibinfo {author} {\bibfnamefont {J.~R.}\ \bibnamefont {Scott}},
  \ and\ \bibinfo {author} {\bibfnamefont {N.}~\bibnamefont {Wilkins-Diehr}},\
  }\href {\doibase 10.1109/MCSE.2014.80} {\bibfield  {journal} {\bibinfo
  {journal} {Computing in Science \& Engineering}\ }\textbf {\bibinfo {volume}
  {16}},\ \bibinfo {pages} {62} (\bibinfo {year} {2014})}\BibitemShut {NoStop}%
\end{thebibliography}%
